\pdfoutput=1

\def\jpsi{{\ensuremath{J\kern-0.15em/\kern-0.15em\psi}}\xspace}
\def\psitwos{{\ensuremath{\psi(2S)}}\xspace}
\def\etac{{\ensuremath{\eta_{c}}}\xspace}
\def\chic{{\ensuremath{\chi_{c0,1,2}}}\xspace}
\def\chicone{{\ensuremath{\chi_{c1}}}\xspace}

\def\theX{{\ensuremath{\chi_{c1}(3872)}}\xspace}

\def\g{{\ensuremath{\gamma}}\xspace}

\def\D{{\ensuremath{D}}\xspace}
\def\Dstar{{\ensuremath{D^{*}}}\xspace}
\def\Dbar{{\ensuremath{\overline{D}}}\xspace}
\def\Dstarbar{{\ensuremath{\overline{D}^{*}}}\xspace}

\def\Dz{{\ensuremath{D^0}}\xspace}
\def\Dzb{{\ensuremath{\overline{D}^0}}\xspace}
\def\Dp{{\ensuremath{D^+}}\xspace}
\def\Dpm{{\ensuremath{D^\pm}}\xspace}
\def\Dstarz{{\ensuremath{D^{*0}}}\xspace}
\def\Dstarzb{{\ensuremath{\overline{D}^{*0}}}\xspace}
\def\Dstarp{{\ensuremath{D^{*+}}}\xspace}

\def\BF         {{\ensuremath{\mathcal{B}}}\xspace}
\def\BR         {\BF}
 
\def\to                 {\ensuremath{\rightarrow}\xspace}

\def\B     {\ensuremath{B}\xspace}
\def\Bbar    {{\ensuremath{\offsetoverline{\B}}}\xspace}
\def\Bp     {\ensuremath{\B^+}\xspace}                
\def\Bm     {\ensuremath{\B^-}\xspace}                
\def\Bd     {\ensuremath{\B^0}\xspace}                
\def\Bs     {\ensuremath{\B_s^0}\xspace}                
\def\Bc     {\ensuremath{\B_c^+}\xspace}

\mathchardef\PLambda="7103                 
\def\Lz          {{\ensuremath{\PLambda}}\xspace}
\def\Lb           {{\ensuremath{\Lz^0_\bquark}}\xspace}
 \def\Pp      {\ensuremath{p}\xspace}               
\def\proton      {{\ensuremath{\Pp}}\xspace}
\def\antiproton      {{\ensuremath{\overline{\Pp}}}\xspace}
         
  \def\PK      {\ensuremath{K}\xspace}                 
\def\kaon    {{\ensuremath{\PK}}\xspace}
\def\KS      {{\ensuremath{\kaon^0_{\mathrm{S}}}}\xspace}

\def\Kp      {{\ensuremath{\kaon^+}}\xspace}
\def\Km      {{\ensuremath{\kaon^-}}\xspace}

 \def\Ppi         {\ensuremath{\pi}\xspace}                
\def\pion   {{\ensuremath{\Ppi}}\xspace}
\def\pip    {{\ensuremath{\pion^+}}\xspace}
\def\pim    {{\ensuremath{\pion^-}}\xspace}
\def\piz    {{\ensuremath{\pion^0}}\xspace}

\def\Kpm     {{\ensuremath{\kaon^\pm}}\xspace}

\def\Dp      {{\ensuremath{\D^+}}\xspace}
\def\ep      {{\ensuremath{e^+}}\xspace}
\def\en      {{\ensuremath{e^-}}\xspace}
\def\Dm      {{\ensuremath{\D^-}}\xspace}
\def\Dpm     {{\ensuremath{\D^\pm}}\xspace}

\def\Bpm     {{\ensuremath{\B^\pm}}\xspace}

\newcommand{\nospaceunit}[1]{\ensuremath{\text{#1}}}   

\def\fb   {\ensuremath{\aunit{fb}}\xspace}
\def\invfb   {\ensuremath{\fb^{-1}}\xspace}

\def\cm{\ensuremath{\aunit{cm}}\xspace}
\def\mm{\ensuremath{\aunit{mm}}\xspace}
\def\fm{\ensuremath{\aunit{fm}}\xspace}
\def\mum{\ensuremath{\,\mu\nospaceunit{\text{m}}}\xspace}
\def\s{\ensuremath{\aunit{s}}\xspace}

\newcommand{\aunit}[1]{\ensuremath{\text{\,#1}}}  
\newcommand{\tev}{\aunit{Te\kern -0.1em V}\xspace}
\newcommand{\gev}{\aunit{Ge\kern -0.1em V}\xspace}
\newcommand{\gevc}{\gev}
\newcommand{\gevcc}{\gev}
\newcommand{\mev}{\aunit{Me\kern -0.1em V}\xspace}
\newcommand{\kev}{\aunit{ke\kern -0.1em V}\xspace}
\newcommand{\mevcc}{\mev}
\newcommand{\kevcc}{\kev}

\def\pt{\ensuremath{p_{\mathrm{T}}}}
\def\y{\ensuremath{y}}

\def\thebaroffset{0.18em}
\newcommand{\offsetoverline}[2][\thebaroffset]{\kern #1\overline{\kern -#1 #2}}%
\def\PD      {\ensuremath{D}\xspace}                 
\def\Dbar    {{\ensuremath{\offsetoverline{\PD}}}\xspace}
\def\Dstarb  {{\ensuremath{\Dbar{}^*}}\xspace}

 \def\Pc      {\ensuremath{c}\xspace}                 
                  \def\cquark    {{\ensuremath{\Pc}}\xspace}
 \def\cquarkbar {{\ensuremath{\overline \cquark}}\xspace}
 \def\PSigma      {\ensuremath{\Sigma}\xspace}   
 \mathchardef\PSigma="7106
\def\Sigmares    {{\ensuremath{\PSigma}}\xspace}
\def\Sigmac       {{\ensuremath{\Sigmares_\cquark}}\xspace}

\def\Lc         {{\ensuremath{\PLambda^+_c}}\xspace}

\mathchardef\PXi="7104
\def\Xires       {{\ensuremath{\PXi}}\xspace}
\def\Xic         {{\ensuremath{\Xires_\cquark}}\xspace}
\def\Xicp        {{\ensuremath{\Xires^+_\cquark}}\xspace}
\def\Xicz         {{\ensuremath{\Xires^0_\cquark}}\xspace}
\def\Xicprim         {{\ensuremath{\Xires^\prime_\cquark}}\xspace}

\def\Xiccpp      {\ensuremath{\Xires_{cc}^{++}}\xspace}            
\def\Tcc      {\ensuremath{T_{cc}^{+}}\xspace}  

\def\Bub     {{\ensuremath{\B^-}}\xspace}
\def\Bm      {{\ensuremath{\Bub}}\xspace}
\def\Bu      {{\ensuremath{\B^+}}\xspace}
\def\Bp      {{\ensuremath{\Bu}}\xspace}
\def\Ds      {{\ensuremath{\D^+_\squark}}\xspace}
\def\Dsp     {{\ensuremath{\D^+_\squark}}\xspace}
\def\Dsm     {{\ensuremath{\D^-_\squark}}\xspace}

\def\Lambdares   {{\ensuremath{\PLambda}}\xspace}

 \def\Ps      {\ensuremath{s}\xspace}      
\def\squark    {{\ensuremath{\Ps}}\xspace}
\def\squarkbar {{\ensuremath{\overline \squark}}\xspace}

 \def\Pb      {\ensuremath{b}\xspace}                 
\def\bquark    {{\ensuremath{\Pb}}\xspace}
\def\bquarkbar {{\ensuremath{\overline \bquark}}\xspace}

 \def\Pu      {\ensuremath{u}\xspace}                 
\def\uquark    {{\ensuremath{\Pu}}\xspace}
\def\uquarkbar {{\ensuremath{\overline \uquark}}\xspace}

 \def\Pd      {\ensuremath{d}\xspace}                 
\def\dquark    {{\ensuremath{\Pd}}\xspace}
\def\dquarkbar {{\ensuremath{\overline \dquark}}\xspace}

 \def\Pq      {\ensuremath{q}\xspace}                 
\def\quark     {{\ensuremath{\Pq}}\xspace}
\def\quarkbar  {{\ensuremath{\overline \quark}}\xspace}

 \mathchardef\PXi="7104

\def\Xires       {{\ensuremath{\PXi}}\xspace}
\def\Xibm         {{\ensuremath{\Xires^-_\bquark}}\xspace}
\def\antiproton  {{\ensuremath{\overline \proton}}\xspace}
\def\Bz      {{\ensuremath{\B^0}}\xspace}
\def\Bs      {{\ensuremath{\B^0_\squark}}\xspace}

 \def\PXi         {\ensuremath{\Xi}\xspace}                 \mathchardef\PXi="7104
\def\Xires       {{\ensuremath{\PXi}}\xspace}
\def\Xicz        {{\ensuremath{\Xires^0_\cquark}}\xspace}
\def\Xic        {{\ensuremath{\Xires_\cquark}}\xspace}

\def\Pcnew {\ensuremath{P_\cquark}\xspace}
\def\Pcsnew {\ensuremath{P_{\cquark\squark}}\xspace}

 \mathchardef\PUpsilon="7107

 \def\NR {non-resonant\xspace}
 \def\phsp {phasespace\xspace}

\documentclass[12pt,a4paper]{article}

\usepackage{ifthen} 

\usepackage{graphicx}  
\usepackage{color}
\newlength\typewidth
\newlength\lfskip
\newboolean{articletitles}
\setboolean{articletitles}{true} 

\usepackage[numbers]{natbib}

\usepackage[breaklinks]{hyperref}    
\usepackage[all]{hypcap} 
\usepackage{mciteplus}

\usepackage{amssymb}
\usepackage{amsmath} 

\usepackage{comment}

\usepackage{xspace} 

\begin{document}

\begin{titlepage}

\quad
\vspace*{3.0cm}

{\bf\boldmath\huge
\begin{center}
Exotic Hadrons at LHCb
\end{center}
}

\begin{center}
Daniel Johnson,$^1$ Ivan Polyakov,$^2$ Tomasz Skwarnicki,$^3$ and Mengzhen Wang$^4$
\bigskip\\
{\it\footnotesize
$ ^1$University of Birmingham, Birmingham, United Kingdom;\\email: daniel.johnson@cern.ch\\
$ ^2$European Organization for Nuclear Research (CERN), Geneva, Switzerland;\\e-mail: ivan.polyakov@cern.ch\\
$ ^3$Syracuse University, Syracuse, NY 13244, USA; \\e-mail: tskwarni@syr.edu\\
$ ^4$Sezione INFN di Milano, Milano, Italy;\\e-mail: mengzhen.wang@cern.ch\\
}
\quad\\
March 6, 2024
\end{center}

\vspace{\fill}

\begin{abstract}
\noindent

It has been five years since the data sample from the LHCb detector, the first experiment optimized for heavy-flavor physics studies at a hadronic collider, was completed. 
These data led to many major discoveries in exotic hadron spectroscopy, which we review in this article. We supplement the experimental results with a selection of phenomenological interpretations. As the upgraded LHCb detector is expected to collect a larger data sample starting in 2024, the near- and further-future potential of the LHCb program in exotic hadron physics is also discussed. 

\end{abstract}

\vspace*{1.1cm}
\vfill

\begin{center}
{\it Annu. Rev. Nucl. Part. Sci. 74: In press.} \href{https://doi.org/10.1146/annurev-nucl-102422-040628}{https://doi.org/10.1146/annurev-nucl-102422-040628}
\end{center}

\end{titlepage}

\newpage

\tableofcontents

\section{INTRODUCTION}
\label{sec:introduction}

Hadrons are bound states created by strong interactions, which couple objects carrying color charge: quarks and gluons. We have known about quarks -- elementary fermions with fractional electric charge and one unit of strong color charge -- since the mid nineteen-sixties when their existence was deduced from the observed mass spectrum of hadrons which could be explained using three light quark (hereafter $q$) flavors: up ($u$, electric charge $+2/3e$), down ($d$, electric charge $-1/3e$) and strange ($s$, electric charge $-1/3e$). All hadrons at that time could be explained as bound states either of a quark and an antiquark ($q\quarkbar$) or of three quarks ($qqq$), so called conventional mesons and baryons, respectively. The  properties of strong interactions which drove the observed light hadron mass spectrum were quark-flavor independence and the confinement of color charges. Combined with the similarity of the light quark masses ($m_q$), flavor independence accounted for the observed isospin symmetry ($m_u\approx m_d$) which, through inclusion of the strange quark ($m_s\approx m_{u,d}$), was extended to $SU(3)_{flavor}$ symmetry.  
Confinement limits the extent of unscreened color fields to about a size of a light hadron, $\sim1\fm$. When the lightest baryons i.e.\ nucleons, protons $(uud)$ and neutrons $(udd)$, are in close proximity, residual strong interactions can bind them into more extended structures, known as nuclei, $[(qqq)(qqq)...]$. Such nuclear interactions bear some resemblance to the residual electromagnetic interactions through which electrically neutral atoms form molecular structures.

It took another decade to develop a complete theory of strong interactions based on $SU(3)_{color}$ symmetry -- Quantum Chromo-Dynamics (QCD) -- which introduced the gluon ($g$), a massless elementary boson mediating color interactions. Unlike the photons of QED, gluons carry complex color charges, a unit of color and a unit of anticolor. Consequently, gluons could become hadron constituents and form so called hybrid mesons $(q\quarkbar g)$, hybrid baryons $(qqqg)$, or glueballs $(gg)$. Going beyond the well established conventional hadronic states, $(q\quarkbar)$ and $(qqq)$, hybrids are classified as ``exotic'' hadrons.  

Also falling into the broad category of ``exotic'' hadrons are those bound states comprising more than the minimal number of quarks. Such multiquark states can come in the form of either exotic mesons, e.g.\ tetraquarks $(qq\quarkbar\quarkbar)$, or exotic baryons, e.g.\ pentaquarks  $(qqqq\quarkbar)$. In fact, the very first quark-model proposals \cite{GellMann:1964nj,*Zweig:1981pd} speculated about their existence. QCD provided further motivation for multiquark states since quark (antiquark) pairs, $\{qq\}$, in antisymmetric color combinations act as a single antiquark (quark). Such pairs could thus become the building blocks of diquark tetraquarks, $(\{qq\}\{\quarkbar\quarkbar\})$, or diquark pentaquarks, 
$(\{qq\}\{qq\}\quarkbar)$. These hadrons would have compact sizes of the order of the confining scale. Different color schemes of compact multiquark hadrons have been also proposed, like that of so-called hadrocharmonium, in which a light hadron surrounds the heavy charmonium state in the center, both color-polarized, $(\{\cquark\cquarkbar\}\{q\quarkbar\})$, $(\{\cquark\cquarkbar\}\{qqq\})$. A further mechanism for the creation of multiquark structures arises if the nuclear type of binding is effective in baryon-meson and meson-meson combinations so as to produce ``molecular'' pentaquarks ($[(qqq)(q\quarkbar)]$) or tetraquarks ($[(q\quarkbar)(q\quarkbar)]$).
They would be loosely bound, with extended sizes due to two, nearly independent, confinement volumes. These various quark binding schemes could lead to different hadron families, or to hadrons with mixed features. 
 
Even after the passage of half a century since the formation of QCD, believed to be the exact theory of strong interactions, experimental studies of the hadron spectrum still attract great interest. Because the strong coupling constant becomes large in the confining domain that drives hadronic structures, perturbative methods -- so useful in QED applications -- are of limited use in QCD. The lattice QCD approach, using numerical methods applied to the QCD Lagrangian, has been very successful in reproducing the observed mass spectrum of stable (i.e. long-lived) conventional hadrons. Numerical simulations of unstable (e.g.\ excited mesons and baryons) or multiquark structures are in their infancy and are often limited by computational practicalities. Thus, instead of first-principle theoretical predictions, phenomenological models are used to predict the full hadron spectrum, introducing often contradictory assumptions and uncontrollable systematic uncertainties. This leaves 
many fundamental questions (Do gluons act as hadron constituents? Do compact and/or molecular multiquark hadrons exist? Do they mix with each other and with conventional hadrons? Are diquarks useful concepts for compact hadron substructure?) still to be answered experimentally.    
Therefore, hadron spectroscopy is one of the last frontiers within the Standard Model of interactions still largely driven by experimental studies.

Light hadron masses are often dominated by quark interaction energy, making their quark content less evident, especially given the presence of three different quark flavors of similar masses.
For excited states, this energy can be easily converted into the creation of additional $q\quarkbar$ pairs resulting in large decay widths. 
Constituent quarks are highly relativistic making quark-spin-dependent mass splittings large. 
The masses of charm ($c$, electric charge $+2/3e$) and bottom ($b$, electric charge $-1/3e$) quarks differ substantially and are much larger than their interaction energies, making their presence in hadrons obvious. These heavy quarks (hereafter $Q$) are nearly static in hadrons, suppressing their spin-dependent couplings. These large $Q$ masses lie at the heart of several effective theories simplifying their theoretical modeling. Furthermore, several states with lower excitation energies are below the threshold for easy hadron disintegration, resulting in narrow widths. These factors conspire to make heavy-quark hadrons particularly important for explorations of various hadronic structures. 

The LHCb experiment is the first operating at a hadronic collider to have been optimized for the study of hadrons containing charm and/or bottom quarks. Large production cross-sections at the highest energy proton-proton collider, combined with the high instantaneous luminosity of the Large Hadron Collider provide an unmatched source of \cquark- and \bquark-quarks, limited chiefly by the processing capabilities of the detector. The architecture of LHCb detector is optimized for an efficient detection of heavy-quark hadrons with simultaneous suppression of backgrounds arising from the large number of particles produced at the primary $pp$ interaction point. Not surprisingly, the LHCb experiment has produced a large number of discoveries in hadron spectroscopy, which we review in this article.

\section{THE LHCB EXPERIMENT}
\label{sec:lhcbexp}

\begin{figure}
    \centering
    \includegraphics[width=.8\textwidth]{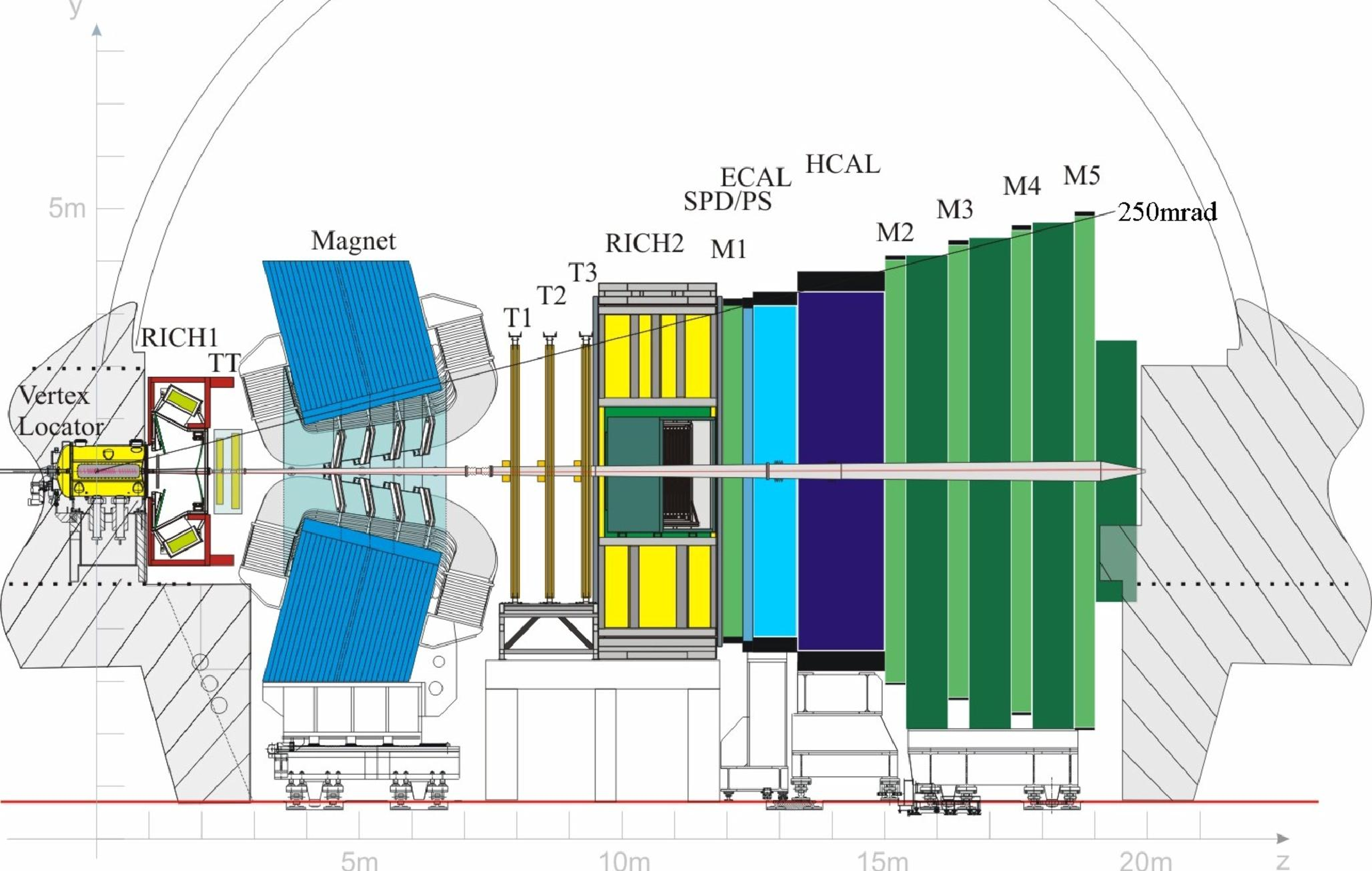}
    \caption{Schematic of the LHCb experiment at CERN~\cite{LHCb:2014set}.\label{fig:lhcbdetector}}
\end{figure}

The LHCb experiment was conceived for the study of heavy flavour hadrons at the LHC; most notably to examine the production and decay of objects containing heavy \bquark- or \cquark-quarks. The forward geometry of the LHCb detector underscores this objective, instrumenting only regions of high pseudorapidity -- between approximately 2 and 5 pseudorapidity units -- where \bquark- and \cquark-hadrons are copiously produced and can be detected in a relatively small solid angle, minimizing detector cost.   A schematic of the detector is shown in Figure~\ref{fig:lhcbdetector}.

In the endeavour to perform the most excellent studies of hadronic spectroscopy, three objectives must be achieved: mis-reconstruction effects must be kept to a minimum, the effect of limited experimental resolution must be minimised, and background processes – real and combinatorial – must be suppressed. The characteristics of the LHCb experiment which allow it to satisfy these demands are now discussed.

A hallmark of the LHCb detector is its excellent particle identification (PID) performance. For muons, this is achieved through the five ``stations'' of its muon system~\cite{Archilli:2013npa}, comprising multi-wire proportional chambers and gaseous electron multiplier detectors (M1-M5 in Fig.~\ref{fig:lhcbdetector}). The system efficiently vetoes hadronic backgrounds and achieves a muon trigger efficiency surpassing 95\%, of particular importance for those final states featuring a \jpsi meson as many discussed in this review do. Useful electron identification capability is delivered by combining information from the hadron and electromagnetic calorimeter systems. 

Central to many of the decays used in studies  of spectroscopy by the LHCb collaboration are final states that consist largely or entirely of as many as 9 light stable hadrons: pions, kaons, and protons. Without mitigation, the combinatorial background introduced through the mis-assignation of mass hypothesis, in particular for kaons and pions during reconstruction of such high-multiplicity final states would be very high. This is overcome principally using two Ring Imaging Cherenkov (RICH) detectors which collect the Cherenkov radiation emitted as charged particles traverse their radiator volumes, with the two RICH detectors accommodating particles in differing momentum ranges. Considering the whole momentum range from 2\gevc\footnote{convention $\hbar=c=1$ is adopted throughout the article} to 100\gevc, the efficiency for correctly identifying kaons is found to be around 95\% with a pion misidentification rate of around 10\%.

That the best experimental resolution should be targeted is obvious. Its consequences are particularly important for hadron spectroscopy in two areas. Firstly greater momentum resolution translates directly into better resolution in invariant-mass spectra. With improved mass resolution, signals emerge more significantly above broad background shapes, nearby partially reconstructed contributions are more distinctly separated, and clusters of states become more readily resolved (where their natural widths are not limiting), simplifying otherwise complicated fits to overlapping resonances. The performance of the LHCb experiment’s tracking system is essential here. It consists of a $55\mum$ pitch silicon-strip Vertex Locator, with 104 double-sided modules (one with strips oriented radially, the other azimuthally) arranged at discrete $z$, surrounding the interaction point and a further 4-layer, $183\mum$ pitch silicon microstrip detector just upstream of the warm dipole magnet (TT). The magnet has a bending power of 4Tm, and is followed by the final 3 layers of the tracking system (T1-T3). Two technologies are in use in each layer: the innermost region is instrumented once more with $183\mum$ silicon microstrips and the outer, experiencing lower particle flux, with straw tubes. The relative momentum uncertainty for tracks with momentum up to 300\gevc increases from 0.5\% below 20\gevc to around 0.8\% for tracks with momentum 100\gevc. Through comparison to a range of \cquark- and \bquark-hadrons whose masses are well known, the systematic uncertainty on the momentum scale is found to be around 0.03\%. The achieved relative mass resolution is around 0.5\% up to the $\PUpsilon(3S)$ resonance. Such excellent resolution explains, for instance, why fits to backgrounds from $B\to D^{*0} DK$ processes, where a neutral product of the $D^{*0}$ decay was not reconstructed, could be circumvented in the study of $\Bpm\to\Dp\Dm\Kpm$ decays discussed later. The processes were cleanly separated experimentally.

The second consequence of excellent experimental resolution is for the precise determination of impact parameters (IP). The LHCb experiment relies heavily on the fact that final state particles produced during the decays of relatively long-lived \bquark- and \cquark-hadrons -- each typically travelling several millimeters in the laboratory frame -- exhibit a large impact parameter with respect to the well-measured $pp$ collision point. Many of the online event filters select based on the presence of significantly displaced tracks. Strict impact parameter cuts continue to be employed offline to suppress combinatorial backgrounds formed when ``background'' particles produced promptly in the $pp$ collision are wrongly included in the reconstruction of a particular decay process. They are especially effective in decays of \bquark-particles that involve an intermediate open-charm hadron (\Dz or \Dpm). The LHCb experiment achieves an IP resolution of approximately $13\mum$ for the highest-momentum tracks, which deteriorates to around $80\mum$ for the lowest momentum tracks thanks to the greater role of multiple scattering. Compared to the scale of \bquark- and \cquark-hadron flight distances, this resolution gives an exceptional opportunity to suppress background from prompt processe.

Despite the low rate of $pp$ interactions per crossing of proton bunches in the LHCb -- only a little over 1 during Run 1 (2011--2012) and Run 2 (2015--2018) -- the abundance of soft hadronic production accompanying the hard scatter of an LHC $pp$ collision allows wide scope for combinatorial backgrounds where particles (mainly pions) are wrongly employed during the reconstruction of a particular process. Requirements on the minimum  impact parameters of included particles, enabled by the displaced decays of the original \bquark- and \cquark-hadrons in most studies of spectroscopy at LHCb, are critical to suppress such backgrounds. They are not universal, however, and significant progress has been made in studies of promptly produced objects, originating at a point indistinguishable from the $pp$ collision, including some of those discussed in this review.

A major challenge across the LHCb physics programme is the use of decay processes that involve one or more neutral particles; the relatively low efficiency for photon reconstruction and the non-$4\pi$ acceptance and incomplete reconstruction of the $pp$ interaction products the primary considerations. Although the reconstruction is challenging, noteworthy efforts have been made to investigate hadron spectroscopy using decays to final states involving neutrals.

The detectors built at the $e^+e^-$ B-factories, most recently Belle II, have also excellent heavy-hadron reconstruction efficiencies, which can even surpass the LHCb capabilities especially for high-multiplicity final states, the final states with photons or involving $K^0_s\to\pi^=\pi^-$ decays. However, the significance of the LHCb experiment is in achieving a good heavy-hadron detection at a high energy hadron-hadron collider. The strong production cross-sections provide overwhelming competitive advantage over the electromagnetic processes.  Furthermore, $b$-baryons, which are unreachable at the B-factories, are copiously produced at the LHC together with $b$-mesons, providing additional avenues for exotic hadron studies.

\section{THE $\theX$ STATE}
\label{sec:x3872}

The beginning of experimental results on exotic heavy quark spectroscopy dates two decades back, when the narrow $\theX$ state was discovered in 
$B^{+,0}\to\theX K^{+,0}$ (charge conjugate states are implied), 
$\theX\to\pi^+\pi^-\jpsi$, 
$\jpsi\to\ell^+\ell^-$ decays
by the Belle $e^+e^-\to B\Bbar$ experiment \cite{Choi:2003ue}.
Just $36\pm7$ signal events were observed in the data sample corresponding to an integrated luminosity of about 140\invfb. 
The observation was later confirmed by the BaBar experiment \cite{BaBar:2004oro},
and, by inclusive studies, dominated by the prompt production at the $p\antiproton$ Tevatron collider, by the CDF ($730\pm90$ signal events in 0.22 fb$^{-1}$ of data) \cite{CDF:2003cab}, and by the D0 \cite{D0Abazov:2004kp} experiments.
The state was intriguing since its mass coincided exactly with the threshold for $D^0\Dbar^{*0}$ decays. 
As if to underline the advantages of a high-energy $pp$ collider for studies of hadronic spectroscopy, it took a dataset initially corresponding to an integrated luminosity of just 0.035\invfb  for the LHCb collaboration to observe promptly produced $565\pm62$ $\theX$ events
at the LHC \cite{Aaij:2011sn}.
The ease of $\theX$ detection at LHCb reflects not only the increased heavy quark cross-section at the higher beam collision energies ($\sqrt{s}=7-13$ TeV LHC vs $2$ TeV Tevatron), but also, much more so, the higher detection efficiency of the forward LHCb detector as compared to the central detectors, which were optimized to high-$\pt$ physics rather than heavy flavor studies.

Settling the $\theX$ quantum numbers was an early, important contribution of the LHCb experiment. Previously, the CDF experiment narrowed them down to $J^{PC}=1^{++}$ or $2^{-+}$ in the sample of $2292\pm113$ $\theX$ events (0.78 fb$^{-1}$) \cite{CDF:2006ocq}.
The sensitivity was limited by high backgrounds and the lack of $\theX$ polarization in the inclusive studies. 
The Belle experiment, with $173\pm16$ signal events reconstructed in $B$ decays ($711$ fb$^{-1}$), which allowed for the $\theX$ polarization and small backgrounds, agreed with this $J^{PC}$ selection
\cite{Belle:2011vlx}.
With a 2011 dataset corresponding to an integrated luminosity of 1\invfb, the LHCb collaboration reconstructed $313\pm26$ signal events in \B-meson decays,
with a background level comparable to that of the $\ep\en$ experiments, thanks to the excellent \B-decay vertex separation from the primary $pp$ interaction point~\cite{LHCb:2013kgk}.
By performing a full five-dimensional angular analysis for the first time, it was possible to favor overwhelmingly the $1^{++}$ assignment over $2^{-+}$. 
Like the previous analyses, it was assumed that $\theX\to(\pi^+\pi^-)\jpsi$ decay proceeds via $S$-wave.
With the 2011-2012 dataset ($3$ fb$^{-1}$) and $1011\pm38$ signal events, the LHCb collaboration found no evidence for $D$-wave contributions, setting an upper limit of $<4\%$ at 95\%\ C.L. on their fraction, and firmly established $J^{PC}=1^{++}$ relatively to all other alternatives~\cite{LHCb:2015jfc}. 

Such quantum numbers are consistent with the $D^0\Dbar^{*0}$ molecular interpretation and compact tetraquark models, as well as the conventional $\chi_c(\rm 2\,^3P_1)$ meson hypothesis. 
However, the $\theX$ cannot be a pure charmonium state, which is best evidenced by the LHCb analysis of the dipion mass spectrum in $\theX\to\pi^+\pi^-\jpsi$ decays performed with the 2011-2018 data sample (9 fb$^{-1}$) and $6788\pm117$ signal events reconstructed in $B^+\to\theX K^+$ decays \cite{LHCb:2022jez}. The decay is dominated by the isospin violating $\theX\to\rho^0\jpsi$ decay, with the isospin-conserving $\theX\to\omega\jpsi$ decay contributing significantly  $(21.4\pm2.3\pm2.0)\%$, strongly amplified by $\rho^0-\omega$ interference.
From this LHCb determined the ratio of coupling constants, ${g_{\theX\to\rho^0\jpsi}}/{g_{\theX\to\omega\jpsi}} =
    0.29\pm0.04$,
    which is an order of magnitude larger than the amount of isopin violation determined for the pure $c\cquarkbar$ state,
    ${g_{\psi(2S)\to\piz\jpsi}}/{g_{\psi(2S)\to\eta\jpsi}}
=0.045\pm0.001$ \cite{LHCb:2022jez}.
Large isospin violation is naturally expected 
in models in which the $\theX$ state has a significant $D\Dbar^{*}$ component, 
either as constituents (i.e.\ in the \lq\lq molecular model") or generated dynamically in the decay~\cite{Tornqvist:2003na,*Tornqvist:2004qy,*Voloshin:2003nt,*Swanson:2003tb}
The proximity of the $\theX$ mass to the $D^0\Dbar^{*0}$ threshold, enhances such contributions over $D^+D^{*-}$ combinations, which are 8\mev heavier.
It has also been suggested in compact tetraquark models that two neutral states, expected in the diaquark model, could be degenerate and mix to form the \theX, giving rise to large isospin violation in $\theX$ decays~\cite{Terasaki:2007uv,*Maiani:2017kyi}.

The large number of reconstructed $\theX\to\pi^+\pi^-\jpsi$ decays coming from the displaced \bquark-hadron vertex allowed the LHCb collaboration to achieve the best sensitivity in seeking to measure the $\theX$ mass and width.
Two complementary approaches were taken: that of exclusive $B^+\to\theX K^+$ reconstruction \cite{LHCb:2020fvo}, which gives the best background suppression, and that of semi-inclusive $\theX$ reconstruction with a displaced vertex, which integrates over all \bquark-hadron species \cite{LHCb:2020xds}, yielding more signal ($15,630\pm380$ events in a dataset corresponding to an integrated luminosity of 3\invfb), but also resulting in a higher background level.
The mass determinations are consistent with those previous, and are the most precise to date \cite{ParticleDataGroup:2022pth}. The measured mass is still indistinguishable from the $D^0\Dbar^{*0}$ threshold. The width determinations, according to a Breit-Wigner parameterization, yielded values different from zero for the first time (combined: $1.19\pm0.21$ \mev \cite{ParticleDataGroup:2022pth}).
However, this parameterization has questionable value for a state coinciding with a coupled-channel threshold. Therefore, the LHCb collaboration also undertook an analysis using a Flatt\'e-like parameterization which contains parameters for the couplings to the
analyzed $\pi^+\pi^-\jpsi$ channel, as well as to the $D^0\Dbar^{*0}$,  $D^+\Dbar^{*-}$ thresholds, and the $\pi^+\pi^-\pi^0\jpsi$ channel constrained to the measurements made by the other experiments. The pole mass, the couplings to $\pi^+\pi^-\jpsi$ and to $D\Dbar^{*}$ channels (the latter assumed to be isospin symmetric), and $\Gamma_0$ parameter to catch all other contributions were free parameters when fitting the observed $\pi^+\pi^-\jpsi$ mass distributions.
The obtained underlying lineshape is much narrower, with FWHM$=0.22\pm0.07$ \mev (statistical error only), than the Breit-Wigner lineshape. However, the fit quality is only marginally improved reflecting difficulty in probing the lineshape when the detector mass resolution dominates over the lineshape effects by a large factor. The pole position, and therefore the binding energy, is poorly constrained because of its strong correlation to the other fit parameters. 
Furthermore, the obtained value of $\Gamma_0=1.4\pm0.4\mev$ was uncomfortably large, pointing to the importance of modes not explicitly represented in the lineshape model. In the future, increased statistics and including in the analysis other invariant-mass distributions, alongside that of the $\pi^+\pi^-\jpsi$ candidates, in a simultaneous coupled-channel fit will improve LHCb sensitivity to the $\theX$ mass and width parameters. 

Decays of the \theX to a \jpsi or \psitwos meson and a photon provide a valuable opportunity to understand its nature. The partial widths of these decays depend on the overlap of the $c\cquarkbar$ wave function in the \theX with that in the charmonium state.
Therefore the ratio of the corresponding branching fractions, \newline $R_{\psi\g} \equiv 
{\BR(\theX\to\psitwos\g)}/{\BR(\theX\to\jpsi\g)}$, should depend heavily on whether the \theX has $\D\Dstarbar$ molecule or charmonium components.

\begin{figure}[htbp]
\centering
\includegraphics[width=0.9\textwidth]{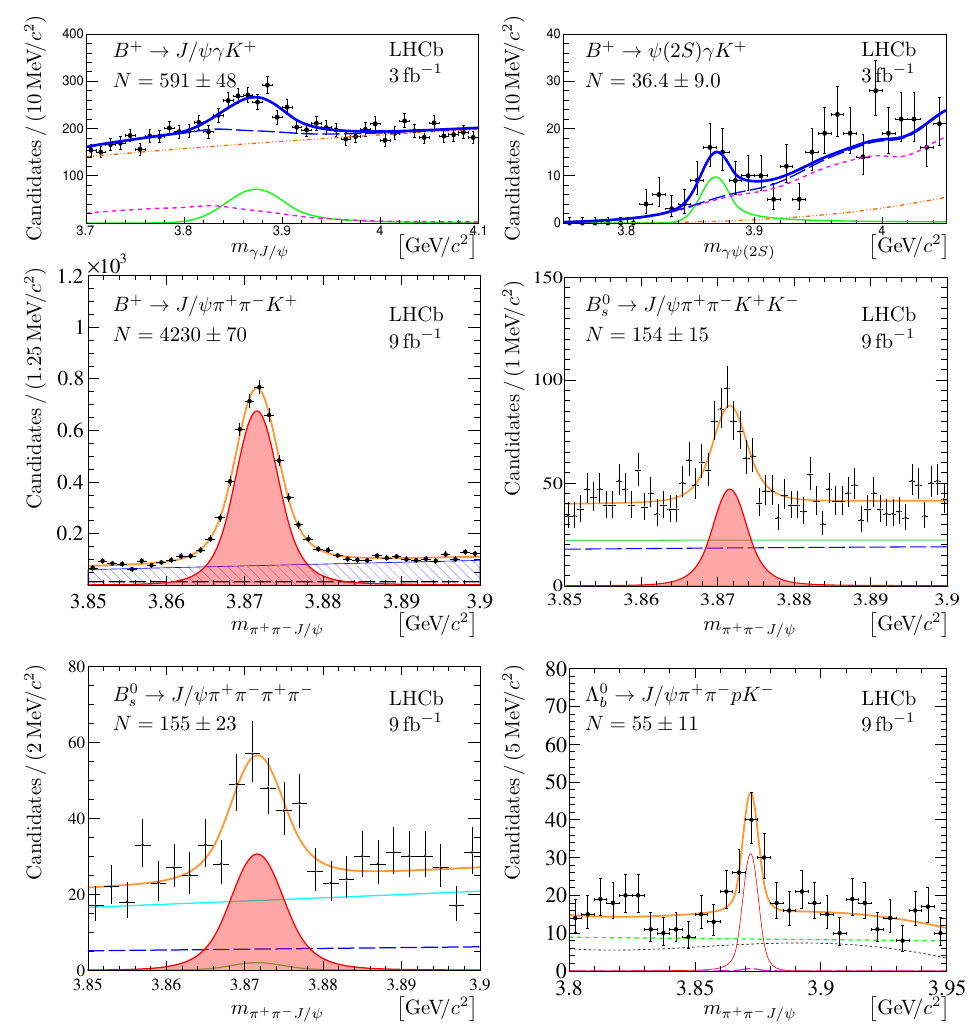}
\caption{The \theX signal in various $b$-hadron decays with corresponding yields: $\Bp\to\theX_{\jpsi\g}\Kp$, $\Bp\to\theX_{\psitwos\g}\Kp$~\cite{LHCb:2014jvf}, $\Bp\to\theX_{\jpsi\pip\pim}\Kp$~\cite{LHCb:2020fvo},  
$\Bs\to\theX_{\jpsi\pip\pim}\Kp\Km$~\cite{LHCb:2020coc},
$\Bs\to\theX_{\jpsi\pip\pim}\pip\pim$~\cite{LHCb:2023reb},
$\Lb\to\theX_{\jpsi\pip\pim}\proton\Km$~\cite{LHCb:2019imv}.
}
\label{fig:x3872_1}
\end{figure}

The BaBar collaboration has reported evidence for the $\theX\to\psitwos\g$ decay and measured $R_{\psi\g}=3.4\pm1.4$~\cite{BaBar:2008flx}. In contrast, the Belle collaboration found no significant $\theX\to\psitwos\g$ signal and reported an upper limit $R_{\psi\g}<2.1~(90\%~{\rm C.L.})$~\cite{Belle:2011wdj}.
Using a data sample corresponding to an integrated luminosity of 3\invfb, collected during 2011 and 2012, the LHCb collaboration obtained evidence for the $\theX\to\psitwos\g$ decay in a sample of $B^{+}\to\theX K^{+}$ decays with a
yield of $36.4\pm9.0$ candidates and a
significance of 4.4 standard deviations (see Fig.~\ref{fig:x3872_1}). 
The ratio $R_{\psi\g}$ was measured to be $2.46\pm0.64\pm0.29$~\cite{LHCb:2014jvf}, being compatible with both the preceding measurements by the BaBar and Belle collaborations. 
The measured value exceeds estimates of ${\cal{O}}(10^{-3})$ for the \theX as a pure $\D\Dstarbar$ molecular state~\cite{Swanson:2004cq,Dong:2009uf,Ferretti:2014xqa}, thus indicating the presence of a charmonium component.
However, it is not in conflict with a predominantly molecular nature for the \theX~\cite{Guo:2014taa} and can be explained with only a $5-12\%$ admixture of charmonium~\cite{Dong:2009uf} or even less~\cite{Takeuchi:2016hat}.

In a more recent analysis by the BESIII collaboration, no signal of the decay $\theX\to\psitwos\g$ was found and an upper limit on $R_{\psi\g}$ was set at $0.59~(90\%~CL)$~\cite{BESIII:2020nbj}. This result is in tension with the measurements by the LHCb and BaBar collaborations. A future analysis from the LHCb collaboration, incorporating the larger Run 2 dataset, will elucidate it.

To study \theX production in decays of various \bquark hadrons the $\theX\to\pip\pim\jpsi$ decay mode was used. 
In addition to the well established $\Bp\to\theX\Kp$~\cite{LHCb:2020fvo} decay, 
the new decays $\Bs\to\theX\Kp\Km$~\cite{LHCb:2020coc}, $\Bs\to\theX\pip\pim$~\cite{LHCb:2023reb} and $\Lb\to\theX\proton\Km$~\cite{LHCb:2019imv} were observed for the first time (see Fig.~\ref{fig:x3872_1}).
It was found that $(38.9\pm4.9)\%$ of $\Bs\to\theX\Kp\Km$ decays and $(58\pm15)\%$ of $\Lb\to\theX\proton\Km$ decays proceed via the $\phi$ and $\PLambda(1520)$ intermediate resonances, respectively.
The $\Bs\to\theX\pip\pim$ decay was found to be dominated by $f_{0}(980)$ and $f_{0}(1500)$ resonance components.
Branching fractions of these decay modes relative to those in analogous decays, replacing the \theX candidate with a conventional \psitwos $c\cquarkbar$ state, including the subsequent  $X\to\pip\pim\jpsi$ decay, were measured to be
\begin{align*}
    \mathcal{R}_{\Bp\to{X}\Kp} & = (3.69 \pm 0.07 \pm 0.06) \times 10^{-2} ~, & 
    \mathcal{R}_{\Lb\to{X}\proton\Km} & = (5.4\pm1.1\pm0.2) \times 10^{-2} ~, \\
    \mathcal{R}_{\Bs\to{X}\phi} & = (2.42\pm0.23\pm0.07) \times 10^{-2} ~, & 
    \mathcal{R}_{\Bs\to{X}\pip\pim} & = (6.8\pm1.1\pm0.2) \times 10^{-2} ~,
\end{align*}
where $\mathcal{R}_{\B\to X Y}$ represents the corresponding ratio of branching fractions \newline ${\BR(\B\to\theX Y)/\BR(\B\to\psitwos Y)}$.
The ratio for $\Bs\to{X}\phi$ decays is consistent with, but more precise than, the value of $(2.21 \pm 0.29 \pm 0.17) \times 10^{-2}$ reported earlier by the CMS collaboration~\cite{PhysRevLett.125.152001}. 
The large variation of the ${\cal R}$ ratio between production modes may reflect a tetraquark component in the $\theX$ state \cite{Maiani:2020zhr}.

Just as for \bquark decays, the \theX prompt production cross-section 
was measured relative to that of \psitwos using the $\pip\pim\jpsi$ decay mode. 
In $pp$ collisions at ${\sqrt{s}=7}$~TeV the product of the \theX production cross-section and the branching ratio into $\pip\pim\jpsi$ was measured for \theX candidates with rapidity between 2.5 and 4.5 and transverse momentum between 5 and 20\gevc (both promptly and via \bquark decays). It was found to be $5.4 \pm 1.3 \pm 0.8$~nb~\cite{Aaij:2011sn}.
In an analysis of $pp$ collisions at $\sqrt{s}=8\tev$ and 13\tev the double-differential cross-sections of the \theX state relative to that of the \psitwos meson were measured as a function of transverse momentum (\pt) and rapidity (\y) separately for promptly produced \theX and those from $b$-decays~\cite{LHCb:2021ten}. 
The ratios, integrated over the kinematic region $4<\pt<20~\gevc$ and $2.0<\y<4.5$, were
\begin{align*}
    R^{\mathrm{8\,TeV}}_{\mathrm{prompt}} & = (7.6 \pm 0.5 \pm 0.9) \times 10^{-2} ~, &
    R^{\mathrm{8\,TeV}}_{\mathrm{non-prompt}} & = (4.6 \pm 0.4 \pm 0.5) \times 10^{-2} ~,\\
    R^{\mathrm{13\,TeV}}_{\mathrm{prompt}} & = (7.6 \pm 0.3 \pm 0.6) \times 10^{-2} ~, {\textrm{ and}}&
    R^{\mathrm{13\,TeV}}_{\mathrm{non-prompt}} & = (4.4 \pm 0.2 \pm 0.4) \times 10^{-2} ~.
\end{align*}

\begin{figure}[tbp]
\centering
\includegraphics[width=\textwidth]{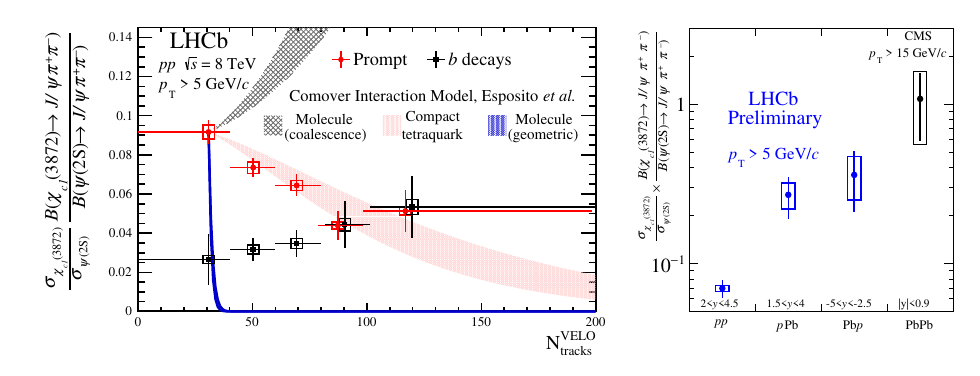}
\caption{Ratio of \theX and \psitwos cross-sections multiplied by branching fractions of decay to $\jpsi\pip\pim$ (left) in $pp$ collisions depending on event multiplicity
\cite{LHCb:2020sey}
and (right) in $pp$, $p$Pb and PbPb collisions \cite{LHCb:2022ixc}.}
\label{fig:x3872_2}
\end{figure}

In $pp$ collisions at $\sqrt{s}=8$~TeV the
production of \theX and \psitwos was also studied as a function of charged particle multiplicity~\cite{LHCb:2020sey} (see Fig.~\ref{fig:x3872_2}). Prompt production of the \theX relative to the \psitwos was found to decrease with event multiplicity while for \theX and \psitwos particles produced via $b$-decays no significant dependence on multiplicity was found, as expected for production at the isolated secondary vertex. 
Possible explanations involve break-up of the \theX, extended in size by its tetraquark component, by rescaterring with comoving partons. However, the exact interpretation is still unclear~\cite{Esposito_2021,Braaten:2020iqw}.

For the first time, production of the \theX was measured in $p$Pb collisions at ${\sqrt{s_{NN}}=8.16}$~TeV in both forward ($1.5<\y<4$,~$p$Pb) and backward ($-5<\y<-2.5$,~Pb$p$) rapidity regions~\cite{LHCb:2022ixc}. Measurements of cross-sections multiplied by branching fraction of decay to $\jpsi\pip\pim$ measured relative to that of the \psitwos yield the ratios 
\begin{align*}
    R^{p\mathrm{Pb}} & = 0.27 \pm 0.08 \pm 0.05 ~, \textrm{ and} &
    R^{\mathrm{Pb}p} & = 0.36 \pm 0.15 \pm 0.11.
\end{align*}
These are an order of magnitude larger than in $pp$ collisions
and closer to the value of ${1.08\pm0.49\pm0.52}$ measured by CMS in PbPb collisions at significantly higher transverse momentum~\cite{CMS:2021znk}. 
In this ratio, effects from modification of the nuclear
parton distribution function largely cancel.
Given that in $pp$ collisions this ratio decreases with multiplicity, one may suggest that as the collision system size -- or density -- increases, the \theX production becomes dominated by a new mechanisms, such as quark coalescence~\cite{LHCb:2022ixc}.
However, the experimental errors in all the heavy-ion collision results are very large. More data are needed before these effects are clearly established.

\section{PENTAQUARK STATES}
\label{sec:pentaquarks}

Multiquark states consisting of four quarks and one antiquark have been predicted since the inception of the quark model, and decades of efforts have been made to search for these pentaquarks experimentally. Over the first 50 years, candidates came and went, but none were widely-accepted.  Some of these states had both exotic and conventional interpretations and are the subject of continuing debate, such as the $\Lz(1405)$ particle which can be thought of as either a $\overline{K}N$ bound state or as a conventional baryon with mass shifted due to the couplings to virtual channels~\cite{Isgur:1978xj}. Others were not confirmed in subsequent experiments having improved statistical precision. One such, classic, example is the $\it{\Theta}(1540)^+$ particle \cite{LEPS:2003wug, CLAS:2003yuj, CLAS:2003wfm}. 

During the past decade, the LHCb collaboration has reported observations and evidences of several pentaquark candidates that are charmonium-like, containing a $\cquark\cquarkbar$ pair. Various exclusive \bquark hadron decays were studied, and intermediate states emerged in strong decays to a charmonium meson and a light baryon~\cite{Aaij:2015tga, LHCb:2019kea, LHCb:2020jpq,  LHCb:2021chn, LHCb:2022ogu}. Their decay signatures indicate a minimal quark content consisting of a $\cquark\cquarkbar$ quark pair and three light quarks, placing them firmly outside the household of conventional baryons.

\subsection{Pentaquark candidates in beauty baryon decays}
The first observation of charmonium-like pentaquark candidates originated in a study of the $\Lb\to\jpsi\proton\Km$ decay. Using a data sample corresponding to an integrated luminosity of $3\invfb$, the LHCb collaboration reported the observation of two $\jpsi\proton$ states 
($\cquark\cquarkbar\uquark\uquark\dquark$)
in an amplitude analysis~\cite{Aaij:2015tga}. One of these states, referred to as $\Pcnew(4450)^+$, has a mass around 4450\mev. It contributes an obvious peaking structure in the $\jpsi\proton$ invariant-mass spectrum with a width too narrow to be explained by reflections from conventional $\Lz^*(\to\proton\Km)$ resonances~\cite{Aaij:2016phn}.
The other broad $\Pcnew(4380)^+$ state (width of about 200 \mev) was
 not directly visible in the $\jpsi\proton$ mass spectrum, but was needed in the amplitude model to reach a good description of the multi-dimensional dataset. As this amplitude model is now obsolete
 \cite{LHCb:2019kea,Wang:2020giv}, the question of whether broad pentaquark states are needed for a good description of these data remains.

\begin{figure}[tb]
    \centering
    \includegraphics[width=0.88\linewidth]{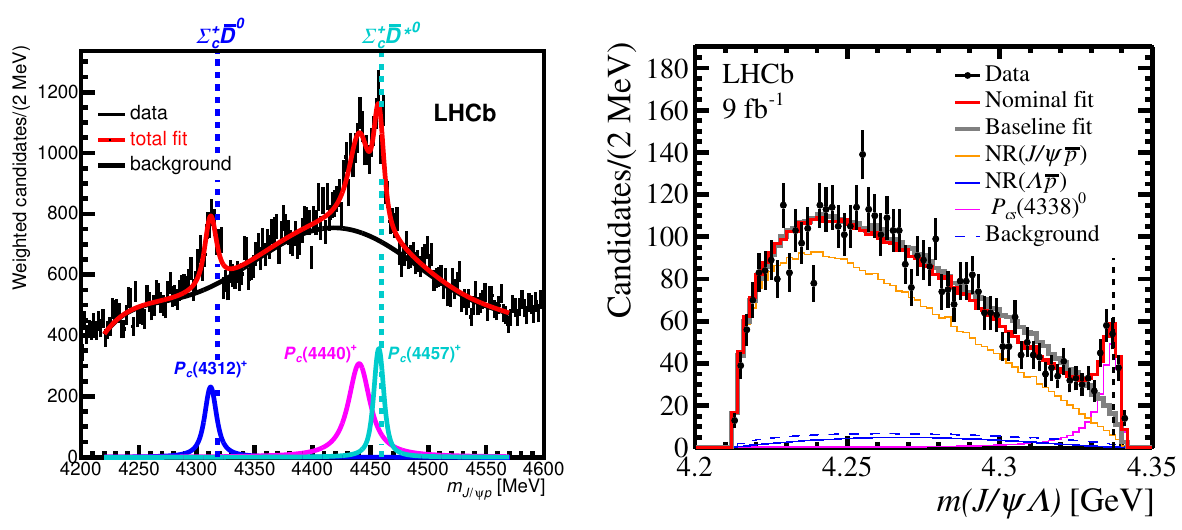}
    \caption{Pentaquark candidates observed by LHCb. The left figure  shows the peaking states $\Pcnew(4312)^+, \Pcnew(4440)^+$ and $\Pcnew(4457)^+$ observed in a $\jpsi\proton$ mass-spectrum analysis of the $\Lb\to\jpsi\proton\Km$ data sample \cite{LHCb:2019kea}. Black dots with error bars show the data distribution, where single-event weights are applied to suppress the contribution from the conventional $\Lz^*$ components. Peaking positions of these $\Pcnew$ states are close to the mass thresholds of $\Sigma_c^+ \Dbar^{(*)0}$ combinations, which are highlighted using vertical dashed lines in the figure. The right figure shows the $\jpsi\Lambdares$ mass projection of the $\Bm\to\jpsi\Lambdares\antiproton$ amplitude analysis where the pentaquark candidate $\Pcsnew(4338)^0$ was observed \cite{LHCb:2022ogu}. The $\Pcsnew(4338)^0$ peaking position is close to the mass threshold of the $\Xicp D^-$ combination, which is highlighted using a vertical black dashed line.}
 \label{fig:pentaquark}
\end{figure}

By adding data collected by the LHCb collaboration between 2015 and 2018 at a higher $pp$ collision energy, with a total integrated luminosity of 6\invfb, and with a further improvement in the event selection, a nine-fold increase of the sample size was reached in the reanalysis of $\Lb\to\jpsi\proton\Km$ decays~\cite{LHCb:2019kea}. 
A one-dimensional $\jpsi\proton$ mass-spectrum analysis was performed.  
With the improved statistical precision, the $\jpsi\proton$ structure at 4450\mev was found to be composed of two narrower peaks close to one another, and a new peaking state at 4312\mev was also uncovered, as shown in Fig.~\ref{fig:pentaquark}. 
Masses, widths and relative fractions of these new narrow peaking states are listed in Tab.~\ref{tab:2019Pc_massWidth}. 
The $\Pcnew(4312)^+$ state lies right below the mass threshold of the $\Sigmac\Dbar$ baryon-meson combination, and has a statistical significance of 7.3$\sigma$. The overlapping narrow states $\Pcnew(4440)^+$ and $\Pcnew(4457)^+$ are close to the mass threshold of the $\Sigmac\Dstarb$ combination, and the statistical significance of the two-peak hypothesis against the $\Pcnew(4450)^+$ single-peak hypothesis is 5.4$\sigma$.

The narrow widths of the $\Pcnew(4312)^+$, $\Pcnew(4440)^+$ and $\Pcnew(4457)^+$ states, in spite of their large decay \phsp, call for a decay-width suppression mechanism. In the tightly bound multiquark model, it is questionable whether a potential barrier, e.g.\ an angular momentum barrier~\cite{Ali:2019npk},  can provide enough spatial separation between the $\cquark$ and $\cquarkbar$ quarks for an effective width suppression. The loosely-bound meson-baryon ``molecular'' model provides a more natural interpretation of these narrow structures, since $\cquark$ and $\cquarkbar$ are separated by much larger distances, making it harder for them to recombine to form the $\jpsi$ meson. The binding energy of hadronic molecules cannot be large, usually predicted to be at $O(10\mev)$ level~\cite{Liu:2019tjn} on par with nucleon binding, thus their masses are expected to be near the mass thresholds of the meson-baryon combinations, which is consistent with the observed $\Pcnew$ peak positions. 
The observed mass splitting between the $\Pcnew(4312)$ and the other two narrow $\Pcnew$ states cannot be reproduced with comparable precision in the compact pentaquark model when attributed to the difference between quark orbital angular momenta~\cite{Ali:2019npk}.
The existence of $\Sigma_c \Dbar^{(*)}$ hadronic molecules was predicted \cite{Yang:2011wz, Wu:2010jy, Wu:2012md, Karliner:2015ina} before the first pentaquark observation~\cite{Aaij:2015tga}, and continues to be the most popular theoretical interpretation~\cite{Liu:2019tjn, Chen:2019asm, Guo:2019fdo, He:2019ify, Huang:2019jlf, Shimizu:2019ptd, Guo:2019kdc, Xiao:2019mvs, Meng:2019ilv, Wu:2019rog, Shen:2019evi, Xiao:2019gjd, Voloshin:2019aut, Sakai:2019qph, Wang:2019hyc, Yamaguchi:2019seo, Xu:2019zme, PavonValderrama:2019nbk, Peng:2019wys, Liu:2019zvb, Pan:2019skd, Burns:2019iih}. It was suggested that a triangle singularity could generate near-threshold peaks in the $\jpsi\proton$ mass spectrum~\cite{Guo:2015umn, Meissner:2015mza, Liu:2015fea, Mikhasenko:2015vca}, but LHCb has demonstrated that the observed peaks are too narrow to be explained in that way (Supplemental Material of Ref.~\cite{LHCb:2019kea}).
The hadrocharmonium model can accommodate the $\Pcnew(4440)^+$ and $\Pcnew(4457)^+$ ($\Pcnew(4312)^+$) as quasi $\psi(2S)p$ ($\chi_{c0}p$) states \cite{Eides:2019tgv}.
When taking this hypothesis as input to fix its free parameters,
seven more states are predicted in the same mass range (quasi $\eta_c(2S)p$, $\chi_{c1}p$, $h_cp$, $\chi_{c2}p$ states), which have not been observed.

A multidimensional amplitude analysis of $\Lb\to\jpsi\proton\Km$ decays using the  combined LHCb Run 1 and 2 datasets has not been published. 
Such an analysis could provide additional inputs for understanding the nature of these $\Pcnew$ states by measuring their spin-parity quantum numbers and investigating the possible contributions from broad $\jpsi\proton$ states.
The helicity amplitude formulation used in Ref.~\cite{Aaij:2015tga} was later corrected to properly align proton helicities from conventional ($pK^-$) and exotic ($\jpsi p$) hadron contributions \cite{Wang:2020giv}.
Some of the ongoing efforts of the amplitude analysis are documented in Ref.~\cite{Wang:2806799}, where several outstanding challenges have been identified.
The huge data sample and complicated multi-dimensional fit function with a large number of free parameters require the use of high-speed processors, like GPUs, to perform the amplitude fits within a reasonable timeframe. The small relative production fractions for the narrow $\Pcnew$ states in $\Lb\to\jpsi\proton\Km$ decays, as summarized in Table~\ref{tab:2019Pc_massWidth}, result in a low tolerance for mis-modeling of the rest of the components contributing to the decay amplitude. 
Better external inputs concerning the properties of conventional hadrons ($\Lz$ excitations) contributing to the $\Lb$ decay would be helpful, especially in the poorly known high $\proton \Km$ mass region. Potential $\Sigmac^{(*)}\Dbar^{(*)}$ coupled-channel effects  can have a significant impact on the interpretation of the properties of $\Pcnew$ states~\cite{Wang:2806799}. A joint analysis of $\Lb\to\jpsi\proton\Km$ and $\Lb\to\Sigmac^{(*)}\Dbar^{(*)}\Km$ data using a coupled-channel lineshape model will be useful alongside the addition of data collected with the upgraded LHCb detector.

\begin{table}[tbp]
\centering
\caption{Masses and widths of narrow pentaquarks observed by the LHCb collaboration in the $\Lb\to\jpsi\proton\Km$ data sample~\cite{LHCb:2019kea}.}
\label{tab:2019Pc_massWidth}
\begin{tabular}{cccc} 
State & mass [$\mev$] & width [$\mev$]  & relative fraction [\%]\\
\hline
$\Pcnew(4312)^+$ & $4311.9 \pm 0.7 ^{+6.8}_{-0.6}$ & $9.8 \pm 2.7 \pm ^{+3.7}_{-4.5}$ & $0.30 \pm 0.07 ^{+0.34}_{-0.09}$ \\
$\Pcnew(4440)^+$ & $4440.3 \pm 1.3 ^{+4.1}_{-4.7}$ & $20.6 \pm 4.9 \pm ^{+8.7}_{-10.1}$ & $1.11 \pm 0.33 ^{+0.22}_{-0.10}$ \\
$\Pcnew(4457)^+$ & $4457.3 \pm 0.6 ^{+4.1} _{-1.7}$ & $6.4 \pm 2.0 ^{+5.7}_{-1.9}$ & $0.53 \pm 0.16 ^{+0.15}_{-0.13}$  \\ \hline
\end{tabular}
\end{table}

Triggered by the pentaquark-candidate observations made in the $\Lb\to\jpsi\proton\Km$ decay~\cite{Aaij:2015tga}, two additional amplitude analyses were performed to search for pentaquarks in similar \bquark baryon decays. In 2016, the LHCb collaboration released the amplitude analysis result of the Cabibbo-suppressed decay $\Lb\to\jpsi\proton\pim$ using data collected in 2011 and 2012, which supported the existence of pentaquark candidates contributing to the $\jpsi\proton$ system~\cite{Aaij:2016ymb}.  Using the larger $pp$ dataset collected between 2011 and 2018, corresponding to an integrated luminosity of $9\invfb$, an amplitude analysis of approximately 1750 $\Xibm\to\jpsi\Lambdares\Km$ decays was carried out and evidence was found for a $\Pcsnew(4459)^0$ particle decaying to the $\jpsi\Lambdares$ final state. Based on a relativistic Breit-Wigner lineshape description, its mass and width were measured to be $4458.8 \pm 2.9 ^{+4.7}_{-1.1}\mev$ and $17.3 \pm 6.5 ^{+8.0} _{-5.7}\mev$, respectively, behaving as a narrow peaking state just below the mass threshold of the $\Xicz\Dstarz$ combination. Inspired by the observation of the $\Pcnew(4440)^+$ and $\Pcnew(4457)^+$ states close to the $\Sigmac\Dstarb$ mass threshold, the two-peak hypothesis of the $\Pcsnew(4459)^0$ particle was tested, but it could  be neither confirmed nor rejected using the current dataset.

\subsection{Pentaquark candidates in beauty meson decays}

Decays of \B mesons to $\jpsi\proton$ or $\jpsi\Lambdares$ particles and an additional light-flavor baryon offer an unique environment to precisely study properties of hidden-charm pentaquark candidates. These decays benefit from smaller contributions from light-flavor conventional resonances decaying into baryon-baryon final states, compared to the situation of the $\Lb\to\jpsi\proton\Km$~\cite{LHCb:2019kea} and $\Xibm\to\jpsi\Lambdares\Km$~\cite{LHCb:2020jpq} decays whose amplitudes are dominated by conventional resonances in $\proton\Km$ and $\Lambdares\Km$ systems.  In 2019, the LHCb collaboration reported the first observation of the decay $\Bs\to\jpsi\proton\antiproton$~\cite{LHCb:2019rmd} using a $pp$ dataset corresponding to an integrated luminosity of $5.2\invfb$. An enhancement of its measured branching fraction compared to the predicted value~\cite{Hsiao:2014tda} hinted at unexpected exotic resonances contributing to this decay channel. 
The intermediate structures in the $\Bs\to\jpsi\proton\antiproton$ decay were investigated using the LHCb data collected between 2011 and 2018, corresponding to an integrated luminosity of $9\invfb$~\cite{LHCb:2021chn}. An excess at about 4.3\gev was seen in the $\jpsi\proton$ invariant-mass spectrum, 
indicating the potential contributions of pentaquark candidates.  An amplitude analysis was performed to extract the properties of the possible state, and showed evidence for a $\jpsi\proton$ structure, denoted as $\Pcnew(4337)^+$, with a mass and width of $4337 ^{+7}_{-4} \pm 2 \mev$ and $29 ^{+26} _{-12} \pm 14 \mev$, respectively~\cite{LHCb:2021chn}.
The significance of the new state is $3.1\sigma$ and, as such, more data are needed to verify the existence of this state. 
The contribution from the $\Pcnew(4312)^+$ state to the $\Bs\to\jpsi\proton\antiproton$ decay was also tested, and no evidence was found.

The latest observation of a hidden-charm pentaquark candidate was found in the $\Bm\to\jpsi\Lambdares\antiproton$~\cite{LHCb:2022ogu} decay. 
An amplitude analysis was applied to a sample of $\Bm\to\jpsi\Lambdares\antiproton$ decays, using a helicity-based formalism optimized for analyzing three-body decays~\cite{JPAC:2019ufm}. An observation of a $\jpsi\Lambdares$ intermediate structure was reported with a statistical significance higher than 15$\sigma$. This  ($\cquark \cquarkbar \squark\uquark\dquark$) pentaquark candidate, denoted by $\Pcsnew(4338)^0$, was measured to have a Breit-Wigner mass of $4338.2 \pm 0.7 \pm 0.4 \mev$ and a width of $7.0 \pm 1.2 \pm 1.3 \mev$. 
The amplitude analysis favored a $J^P=1/2^-$ assignment. The hypothesis of $J=1/2$ was established and the $J^P=1/2+$ assignment was excluded with 90\% confidence.
Its mass -- right at 
the $\Xic\Dbar$ threshold -- and narrow width of $7.0\pm1.2\pm1.3$ \mev,
along with its $J^P=1/2^-$ spin-parity , support a molecular interpretation.
This makes the $\Pcsnew(4338)^0$ state a likely $\Xicz\Dbar^0$ analog 
of the ${\Pcnew}^+$ state in the $\Sigma_{\cquark}^0\Dbar^0$ system. 
Furthermore, as for the $\Pcsnew(4459)^0$ case, a $\Xicprim\Dbar$
state is expected, predicting a total of three additional strange
pentaquarks~\cite{Karliner:2022erb}.

\section{HIDDEN CHARM TETRAQUARK CANDIDATES}
\label{sec:hiddencharm}
  
One of the earliest heavy-quark tetraquark candidates was the $Z_c(4430)^+$ in its decay to the $\psi(2S)\pi^+$ final state, from a study of around two thousand  $\Bbar^0\to\psi(2S)\pip\Km$ decays by the Belle collaboration~\cite{Choi:2007wga}. 
It was the first explicitly exotic tetraquark structure, but the road to its confirmation at 
$e^+e^-$ machines was not always smooth~\cite{BaBar:2008bxw}.
The LHCb experiment reconstructed 25,000 of the \B decays in a dataset corresponding to an integrated luminosity of 3\invfb. By means of a four-dimensional amplitude analysis, they confirmed significant resonant activity with spin-parity $J^P=1^+$ in the 
$\psi(2S)\pi^+$ channel in the mass region of interest~\cite{LHCb:2014zfx}. The result was consistent with the amplitude analysis results by Belle \cite{Chilikin:2013tch}. When interpreted as one Breit-Wigner resonance, the $Z_c(4430)^+$ mass and widths are $4478\pm18$\mev
and $181\pm31$\mev, respectively~\cite{ParticleDataGroup:2022pth}.
The LHCb analysis indicated also a significant structure in the $0^-$ or $1^+$ $\psi(2S)\pip$ wave at lower mass.
The Belle experiment analyzed 32,000 $\Bbar^0\to\jpsi\pip\Km$ decays and claimed a significant,
and very broad ($370\pm70^{+\phantom{0}70}_{-132}$ \mev),
$1^+$ $Z_c(4200)^+$ resonance in the amplitude model, in addition to the $Z_c(4430)^+$ resonance which had been included with fixed mass and width~\cite{Belle:2014nuw}.
LHCb also demonstrated, with high confidence, the presence of exotic hadron structures
in $\Bbar^0\to\psi(2S)\pip\Km$ and
$\Bbar^0\to \jpsi\pip\Km$ decays
using a semi-model-independent approach~\cite{LHCb:2014zfx,LHCb:2015sqg,LHCb:2019maw}. Without dependence upon an amplitude model, this showed 
that these channels cannot be described
only using conventional kaon excitations decaying to the $\pip\Km$ final state. 
However, such an approach does not provide a complete picture of possible exotic contributions. 

The LHCb data collected between 2011 and 2018, corresponding to an integrated luminosity of 9\invfb, yields 2.3 million and 140,000  reconstructed $\Bbar\to X \pip\Km$ candidates where $X$ is a $\jpsi$ or a $\psi(2S)$ states respectively (72 and 6 fold increases, respectively). 
No amplitude analyses of the two decays in the LHCb dataset have yet been published, but unofficial results are available in a form of Ph.D. thesis~\cite{Beiter:2854734}.
A large number (6-10) of relatively broad, exotic $\psi(nS)\pip$ and $\psi(nS)\Km$ states of various $J^P$ are required for a good description of the data. Conceivably,  such states can be explained by means of direct color couplings among the four quarks of a compact tetraquark system. In such states, the $c$ and $\cquarkbar$ share the same confinement volume and can relatively easily fuse to form the $\psi(nS)$ states. This is in contrast with ``molecular" charmed meson-antimeson pairs where decays to the $\psi(nS)$ states are suppressed. There is no clear evidence for production of the likely molecular, narrow $Z_c(3900)^+$ and $Z_c(4020)^+$ states in $B^+\to\psi(nS)\pip\Km$ decays~\cite{ParticleDataGroup:2022pth}.   
Comparing the two amplitude analyses, except for the parameters of the $Z_c(4200)^+$, the masses, widths and exact $J^P$ composition of resonances do not align well~\cite{Beiter:2854734}. This may be due to the fact that coupled channel effects, though expected, have not yet been accounted for in the amplitude fits. Direct three-body decays have been also neglected.  
Until more satisfactory amplitude models have been developed, with better correspondence between the $\jpsi$ and $\psi(2S)$ decays, discussion of the related exotic states should be approached with care.
This applies especially to the $1^+$ wave (including the $Z_c(4430)^+$ and $Z_c(4200)^+$ structures), which requires the largest number of poles to describe the LHCb data, and at the same time is expected to couple to $D\Dbar^*$ and $D^*\Dbar^*$ channels in an $S$-wave. 
Improvements to the amplitude models applied to the large LHCb $B\to\psi(nS)\pi^-K^+$ data sets are being pursued, thus updated results are expected to be published in the future.

\begin{figure}[tb]
    \centering
    \includegraphics[width=\textwidth]{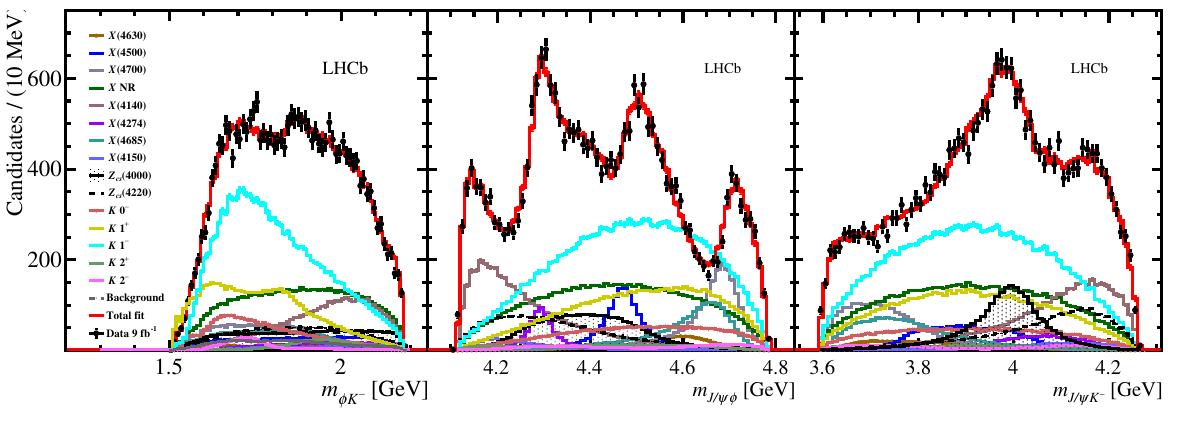}
    \caption{
    Distributions of the $\phi\Km$ (left), $\jpsi\phi$ (middle) and $\jpsi\Km$ (right) invariant masses of $\Bm\to\jpsi\phi\Km$ data sample collected by LHCb during Run 1 and 2 \cite{LHCb:2021uow}.  Black dots with error bars show the data distribution. Projections of the amplitude fit are overlaid. Contributions from different intermediate structures are presented using different line colors, as indicated in the legend of the figure. Four peaks clearly visible in the $\jpsi\phi$ mass spectrum are described using the $X(4140)$, $X(4274)$, $X(4500)$ and $X(4700)$ states, which were previously observed using LHCb Run 1 data~\cite{LHCb:2016nsl, LHCb:2016axx}. A broad peak at $4\gev$  in the $\jpsi\Kp$ mass spectrum corresponds to the $Z_{cs}(4000)^-$ tetraquark candidate. The amplitude model includes also the  $X(4150), X(4630), X(4685)$ and $Z_{cs}(4220)^-$ exotic states and conventional kaon excitations, which are needed to reach a good description of the multi-dimensional data. }
    \label{fig:B2JpsiPhiK}
\end{figure}

In 2009, the CDF experiment reported an evidence of a very narrow ($\Gamma=12\pm9$ \mev) near-threshold $X(4140)$ structure decaying to the $\jpsi\phi$ final state in $58\pm10$ reconstructed  $\Bm\to\jpsi\phi\Km$ decays. This report, using a dataset corresponding to an integrated luminosity of 2.7\invfb of $\proton\antiproton$ Tevatron data~\cite{CDF:2009jgo}, triggered a series of further investigations into the nature of resonances contributing to the decay~\cite{Cheng-Ping:2009sgk, CDF:2011pep, LHCb:2012wyi, BaBar:2014wwp, CMS:2013jru, D0:2015nxw, D0:2013jvp}. A charmonium state at this mass is expected to have a large decay width to open-charm pair, thus the narrow width and decay to $\jpsi\phi$ pointed to the exotic nature of the $X(4140)$ particle~\cite{CDF:2009jgo}. 
The first result from LHCb data collected in 2010 (corresponding to an integrated $pp$ luminosity of just 0.37\invfb) involved the study of $382\pm22$ reconstructed $\Bm\to\jpsi\phi\Km$ decays and did not confirm the existence of the narrow $X(4140)$ state \cite{LHCb:2012wyi}. 
The CMS collaboration reconstructed $2480\pm160$ of the decays in a dataset corresponding to an integrated luminosity of 5.2\invfb and observed the narrow ($28\pm24$ \mev) $X(4140)$ state with high significance~\cite{CMS:2013jru}. 
Results from the CDF \cite{CDF:2011pep}, CMS \cite{CMS:2013jru} and D0 \cite{D0:2013jvp} collaborations also indicated a second, relatively narrow $\jpsi\phi$ state in $4274-4328$ \mev range. All these results were obtained by performing simple fits to the  $\jpsi\phi$ invariant-mass distirbutions with significant assumptions about the background shapes in the signal regions.
The first six-dimensional amplitude analysis of $\Bm\to\jpsi\phi\Km$ decays, modeling the backgrounds as reflections of various kaon excitations, $K^{*-}\to\phi\Km$, was performed by the LHCb collaboration using around 4,000 reconstructed decays. Only data collected during 2011 and 2012 data-taking were included, corresponding to an integrated luminosity of 3\invfb~\cite{LHCb:2016nsl, LHCb:2016axx}. Four prominent $\jpsi\phi$ resonances were needed for a good description of the data, including the $X(4140)$ threshold structure, albeit with significantly larger width ($83\pm30$ \mev) than obtained by the other experiments without amplitude analysis. Its quantum numbers were determined to be $J^{PC}=1^{++}$. The second
$\jpsi\phi$ structure, $X(4274)$, was confirmed and assigned the same quantum numbers.
Two further new states, $X(4500)$ and $X(4700)$, were found with $0^{++}$ quantum numbers. 
The amplitude analysis was updated to the full 2011-2018 LHCb dataset in 2021, in which around 24,000 $\Bm\to\jpsi\phi\Km$ decays were reconstructed with improved purity~\cite{LHCb:2021uow}. 
The amplitude fit projections onto subsystem masses are shown in Fig.~\ref{fig:B2JpsiPhiK}. The four broad $\jpsi\phi$ mass peaks indicate the contributions from the $X(4140)$, $X(4274)$, $X(4500)$ and $X(4700)$ particles. A broad $\jpsi\Km$ mass peak at about $4\gev$ suggests a contribution from an explicitly exotic $\jpsi\Km$ structure -- a tetraquark candidate with minimal quark content ($\cquark\cquarkbar u\squarkbar$). From the amplitude analysis this $J^P=1^+$ $Z_{cs}(4000)^-$ state has a mass of $4003 \pm 6 ^{+\phantom{1}4} _{-14}$ \mev and a width of $131 \pm 15 \pm 26$ 
 \mev.
This is likely a different state than the narrow ($\Gamma\sim 8-13$ \mev) $Z_{cs}(3985)^{-,0}$ state observed by the BES III experiment in decays to $D_s^-D^{(*)}$.
Properties of all the potentially exotic components in the LHCb amplitude model are summarized in Table~\ref{tab:B2JpsiphiK_Run12_PropertiesSummary}. In addition to all four $X\to\jpsi\phi$ states observed using the 2011-2012 data~\cite{LHCb:2016nsl, LHCb:2016axx},   there are several additional exotic candidates which do not generate a visible peaking state in two-body final state invariant-mass spectra (Fig.~\ref{fig:B2JpsiPhiK}), but are found to significantly improve the fit quality. They include the $2^{-+}$ $X(4150)$, $1^{++}$ $X(4685)$ and $1^{-+}$ $X(4630)$ states decaying to the $\jpsi\phi$ final state, and the $1^{+}$ $Z_{cs}(4220)^-$ state decaying into the $\jpsi\Km$ final state.

\begin{table}[tbp]
\centering
\caption{Properties of exotic hadrons contributing to the $\Bm\to\jpsi\phi\Km$ decay measured in Ref.~\cite{LHCb:2021uow}.}
\label{tab:B2JpsiphiK_Run12_PropertiesSummary}
\begin{tabular}{cccr@{\:$\pm$\:}lr@{\:$\pm$\:}lr@{\:$\pm$\:}l}
\hline
$J^P$ & \multicolumn{1}{c}{Component} &Significance&            \multicolumn{2}{c}{$M_0$\,[MeV]}   & \multicolumn{2}{c}{$\Gamma_0$\,[MeV]}      & \multicolumn{2}{c}{FF\,[\%]}          \\
\hline
 $2^-$ &$X(4150)$            & 4.8 $\sigma$  & $4146$&$18\pm33$ & $135$&$28\,_{-\,30}^{+\,59}$   & $2.0$&$0.5\,_{-\,1.0}^{+\,0.8}$ \\
\hline
$1^-$  &$X(4630)$            & 5.5 $\sigma$  & $4626$&$ 16\,_{-\,110}^{+\,\phantom{0}18}$ & $174$&$27\,_{-\,\phantom{0}73}^{+\,134}$  & $2.6$&$0.5\,_{-\,1.5}^{+\,2.9}$ \\
\hline
$0^+$ &$X(4500)$           & 20 $\sigma$  & $4474$&$3 \pm3 $ & $77$&$6\,_{-\,\phantom{0}8}^{+\,10}$ &  $5.6$&$0.7\,_{-\,0.6}^{+\,2.4}$ \\
 &$X(4700)$           & 17 $\sigma$  & $4694$&$ 4 \,_{-\,\phantom{0}3}^{+\,16}$ & $87$&$8\,_{-\,\phantom{0}6}^{+\,16}$ &  $8.9$&$1.2\,_{-\,1.4}^{+\,4.9}$ \\
\hline
$1^+$ &$X(4140)$           & 13 $\sigma$  & $4118$&$ 11 \,_{-\,36}^{+\,19}$ & $162$&$21\,_{-\,49}^{+\,24}$ & $17$&$3\,_{-\,\phantom{1}6}^{+\,19}$ \\
 &$X(4274)$           & 18 $\sigma$  & $4294$&$4 \,_{-\,6}^{+\,3}$  & $53$&$5\pm5$   & $2.8$&$0.5\,_{-\,0.4}^{+\,0.8}$ \\
 &$X(4685)$           & 15 $\sigma$  & $4684$&$ 7\,_{-\,16}^{+\,13}$  & $126$&$15 \,_{-\,41}^{+\,37}$ & $7.2$&$1.0 \,_{-\,2.0}^{+\,4.0}$ \\
\hline\hline
$1^+$  &$Z_{cs}(4000)^-$          & 15 $\sigma$ & $4003$&$ 6\,_{-\,14}^{+\,\phantom{0}4}$  & $131$&$15\pm26$  & $9.4$&$2.1\pm3.4$ \\
  &$Z_{cs}(4220)^-$          & 5.9 $\sigma$  & $4216$&$24\,_{-\,30}^{+\,43}$ & $233$&$52\,_{-\,73}^{+\,97}$ & $10$&$4\,_{-\,\phantom{0}7}^{+\,10}$ \\ \hline
\end{tabular}
\end{table}

After the observation of the $Z_{cs}(4000)^-$ tetraquark candidate~\cite{LHCb:2021uow}, the LHCb collaboration also performed an amplitude analysis 
of $\Bz\to\jpsi\phi\KS$ decays~\cite{LHCb:2023hxg}.
Due to the lower reconstruction efficiency of the $\KS$ compared to $\Km$ particles with the LHCb detector, the $\Bz\to\jpsi\phi\KS$ signal yield is one order of magnitude lower than the $\Bm\to\jpsi\phi\Km$ signal yield, which makes it hard to perform a completely independent amplitude analysis of the $\Bz$ decays. Assuming isospin symmetry, a simultaneous amplitude analysis was performed of both the $\Bz\to\jpsi\phi\KS$ and $\Bm\to\jpsi\phi\Km$ decays, where most of lineshape and coupling parameters were shared between these two channels, except for those related to the $Z_{cs}(4000)^-$ and $Z_{cs}(4000)^0$ states. Evidence at the level of $4\sigma$ was seen for a $Z_{cs}(4000)^0$ state, having a mass and width consistent with that of the $Z_{cs}(4000)^-$ particle~\cite{LHCb:2023hxg}. 

A simple fit to the high-mass peaking structure led the LHCb collaboration to conclude the presence of  a $\jpsi\phi$ state with a mass of $4741\pm6\pm6$ \mev and width of $53\pm15\pm11$ \mev in a study of $\Bs\to\jpsi\phi\pip\pim$ decays~\cite{LHCb:2020coc}.

While $Z_{cs}^-\to\jpsi\Km$ states are minimally four-quark objects, the $X\to\jpsi\phi$ states could either be conventional $(\cquark\cquarkbar)$ states,  $(\cquark\cquarkbar\squark\squarkbar)$ tetraquarks or $(\cquark\cquarkbar g)$ hybrids.
The $X(4140)$ state, which was initially claimed to have a very narrow width \cite{CDF:2009jgo,CDF:2011pep}, has a width of $162\pm21^{+24}_{-49}$\mev as determined by the latest LHCb analysis. Nevertheless, its high fit fraction in the \phsp suppressed region makes it a probable $(\cquark\cquarkbar\squark\squarkbar)$ candidate. The quantum number pattern of the observed $\jpsi\phi$ states does not fit the expectations for charmonia, making it likely that the compact tetraquark dynamics is involved, possibly mixing in with high radial excitations of the $(\cquark\cquarkbar)$ states. Like for the $\Bm\to\psi(nS)\pip\Km$ decays discussed above, the amplitude analysis of $\Bm\to\jpsi\phi\Km$ performed by the LHCb experiment did not include coupled-channel terms from $D^{(*)}_s\Dbar^{(*)}_{(s)}$ pairs, nor three-body contributions, thus the exact spectral decomposition of the obvious $\jpsi\phi$ and $\jpsi\Km$ mass structures may contain model-dependent biases.    

Recently, the LHCb collaboration has detected $\Bm \to \Dsp \Dsm \Km$ decays for the first time, reconstructing roughly 360 candidates, and performed an amplitude analysis~\cite{LHCb:2022dvn}. A significant (much more than $5\sigma$) near-threshold peak  is seen in the $\Dsp \Dsm$ mass spectrum, which can be modelled as a $0^{++}$  $X(3960)$ state at $3956 \pm 5 \pm 10$ \mev with rather narrow width of $43 \pm 13 \pm 8$ \mev~\cite{LHCb:2022aki}. There is also a possible $0^{++}$ state at $4133\pm6\pm6$ \mev,  with a relatively narrow width ($\Gamma=67\pm17\pm7$ \mev), generating  
 the $\Dsp \Dsm$ invariant-mass dip via interference with a sizeable \NR component.  
An alternative way to describe the mass dip is to use K-Matrix lineshape for the $X(3960)$ including the coupling to the $\jpsi\phi$ channel. The $X(3960)$ state is too narrow to be a $\rm 3^3P_0$ charmonium state.


\section{CHARMING AND STRANGE TETRAQUARK CANDIDATES}
\label{sec:charmingstrange}

Conventional states involving a charm and an anti-strange quark are well established, but no evidence for corresponding states including a charm and strange quark had arisen before 2020. Bound states containing both a charm and a strange quark would be manifestly exotic, having minimal 4-quark content in order to form a color-singlet state. Experimental context aside, precious few suggestions had been made for their existence even in the theoretical literature, prior to that point~\cite{Molina:2010tx,Liu:2016ogz,Cheng:2020nho}.

Experimental evidence for bound states containing both a charm and a strange quark first emerged in studies published by the LHCb collaboration~\cite{LHCb:2020bls,LHCb:2020pxc} involving $\Bpm\to\Dp\Dm K^{\pm}$ decays, with the charm mesons 
reconstructed in their decays to the $\Km\pip\pip$ final state. 
Care was taken to ensure that the combinatorial backrounds that often plague high-multiplicity, fully hadronic final states (the feed-down from 3 kaon or 4 pion charmed-meson decays) were heavily suppressed. The LHCb particle-identification system was effective at reducing backgrounds from pion-kaon misidentification, and the particular topology of the decay -- two long-lived charged $D$ mesons -- allowed effectively all background from charmless and single-charm decays to be removed.
Two methods were used to analyse the three-body \phsp of the \Bpm decay. 

The first approach~\cite{LHCb:2020bls} built on the expectation that conventional resonances would only appear in the $\Dp\Dm$ channel, not in the $\Dp\Km$ ($\cquark\squark\uquarkbar \dquarkbar$) or doubly-charged $\Dp\Kp$ ($\cquark\squarkbar\uquark \dquarkbar$) channels. A semi-model-independent treatment, similar to that previously used in the analysis of other three-body \bquark hadron decays featuring only one two-body decay chain in which  conventional resonances were expected~\cite{BaBar:2008bxw,LHCb:2014zfx,LHCb:2015sqg,Aaij:2016phn}. The approach involved assessing how well the data in entire \phsp could be described assuming the presence of resonances in $\Dp\Dm$. The \phsp was divided into slices of 
 $\Dp\Dm$ invariant mass and the helicity angle was decomposed in terms of the orthogonal basis of Legendre Polynomials. The spin component of the amplitude for contributing resonances would be expected to give rise to a helicity structure that could be described using a finite set of Legendre Polynomials, the size of which would correspond to the maximum spin of the $D^+D^-$ resonances in that slice. A limit on the order of the Legendre Polynomials available for the decomposition was introduced, motivated by the fact that very high-spin $(>3)$ intermediate states are unlikely to be formed during the decay of a pseudoscalar particle to three more pseudoscalars. A recombination of the descriptions obtained in each slice allows a test of how well the \phsp has been described, as well as the possibility to test the success of that description in different projections of the \phsp. Poor agreement was observed, where the disagreement was most marked when the $\Dp\Km$ projection of the \phsp was considered, particularly in the region of invariant mass near $2.9\gevcc$.

\begin{figure}[tbp]
\centering
\includegraphics[width=\textwidth]{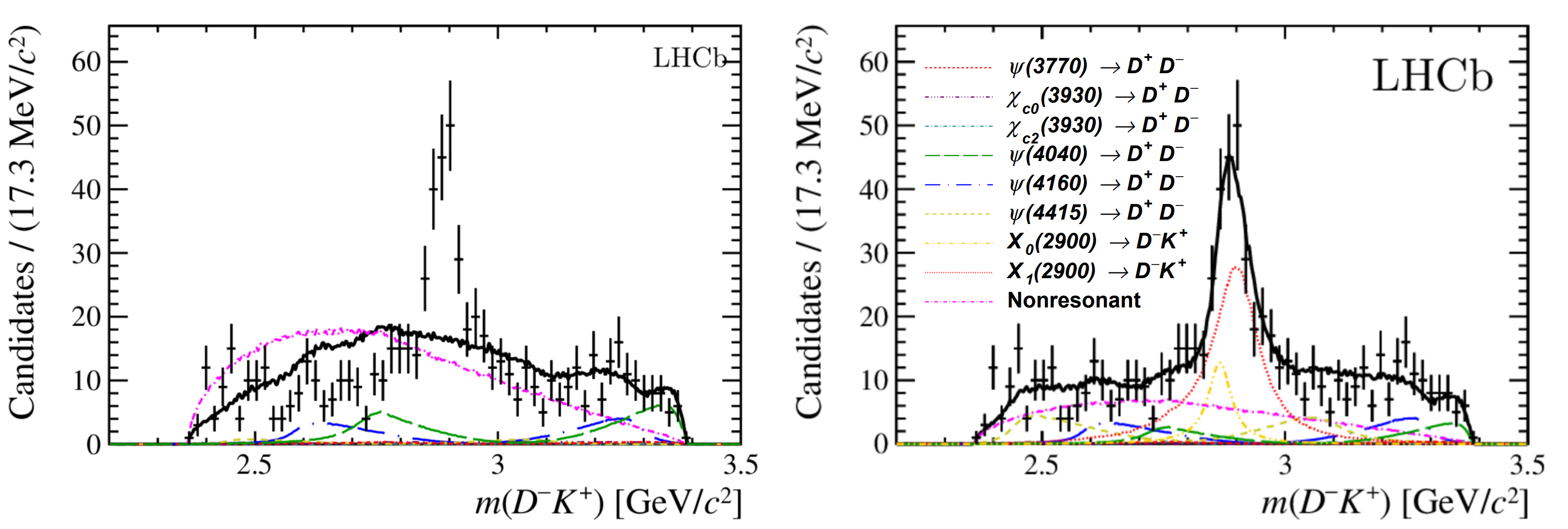}
\caption{Invariant mass distributions of $\Dm\Kp$ pairs in the decay $\Bp\to\Dp\Dm\Kp$ (black points)~\cite{LHCb:2020pxc}. In the left plot, the data are fit with an amplitude model (solid black line) corresponding to the second approach described in the text. This consists of resonances only in the $\Dp\Dm$ channel, and individual model components are indicated by colored lines. Severe discrepancy is observed around $2.9\gevcc$. In the right plot, resonant components are added to the model in the $\Dm\Kp$ channel, and an adequate description is achieved.\label{fig:Bp2DpDmKp_modindep_order4}}
\end{figure}

To characterize the exotic hadron contributions, the second approach involved a full amplitude analysis of the three-body \phsp~\cite{LHCb:2020pxc}. An amplitude was constructed consisting of known charmonium resonances (modelled by a Relativistic Breit-Wigner function in invariant mass with helicity structure corresponding to the resonance spin) and a \NR component. Complex coefficients multiplying each resonance were free to vary, but only a rather poor description of the data was achieved. Only when two new amplitudes were added in the $D^+K^-$ spectrum, with quantum numbers $J^P = 0^+$ and $1^-$, was an adequate description obtained, as illustrated in Fig.~\ref{fig:Bp2DpDmKp_modindep_order4}. The two new amplitudes corresponded to 
$T_{cs0}(2900)^0$ and $T_{cs1}(2900)^0$
resonant components having masses of $2866\pm7\pm2\mevcc$ and $2904\pm5\pm1\mevcc$ and widths of $57\pm12\pm4\mevcc$ and $110\pm11\pm4\mevcc$ for the $0^+$ and $1^-$ states, respectively.

The interpretation of the two new amplitudes in the $D^+K^-$ channel has not been easy. Much speculation focused on the proximity of the new states to the $\Dbar^*K^*$ and $\Dbar_1  K$ thresholds: $2902\mevcc$ and $2913\mevcc$, respectively~\cite{Burns:2020xne}. Compact tetraquark explanations often yield states with expected masses well below the slightly lighter observed $0^+$ state~\cite{Guo:2021mja}. Difficulty has been encountered in accounting for both the $0^+$ and $1^-$ states within a single $\Dbar^*K^*$ or $\Dbar_1 K$ molecular picture though, treated separately, more progress has been made~\cite{Molina:2022jcd}. Attempts have, equally, been made to accommodate the structures as the products of coupled-channel kinematic effects arising from rescattering processes~\cite{Liu:2020orv}. 

In order to make progress in understanding the unexpected structures, a close partnership between experiment and theory may prove fruitful. The first study by the LHCb collaboration employed data taken between 2011 and 2018, but an increase in the size of the dataset available to pursue these particular $\Bp$ decays of at least an order of magnitude can be anticipated during Run 3 and Run 4. Furthermore LHCb will not be alone in acquiring datasets of interest to engage these processes: the Belle II collaboration is well-placed to make important contributions here, too, with their higher absolute track reconstruction efficiency making up in high multiplicity final states for much lower $\Bpm$ production rates. A significant opportunity exists for experimentalists and theorists to implement new theoretical models early on in the analysis of ever-growing data samples.

The study of other decays is also highly likely to shed light on the nature of the $T_{cs}(2900)^0$ structures. In particular, the results of a future study of $\Bp\to\Dp\Km\pip$ decays~\cite{LHCb:2020bls,LHCb:2020pxc} and a future first revealing of the LHCb data for $\Bd\to\Dp\Dzb\Km$ decays are obvious candidates, both sharing the channel of interest in the discovery study.

\begin{figure}[htbp]
\centering
\includegraphics[width=.6\textwidth]{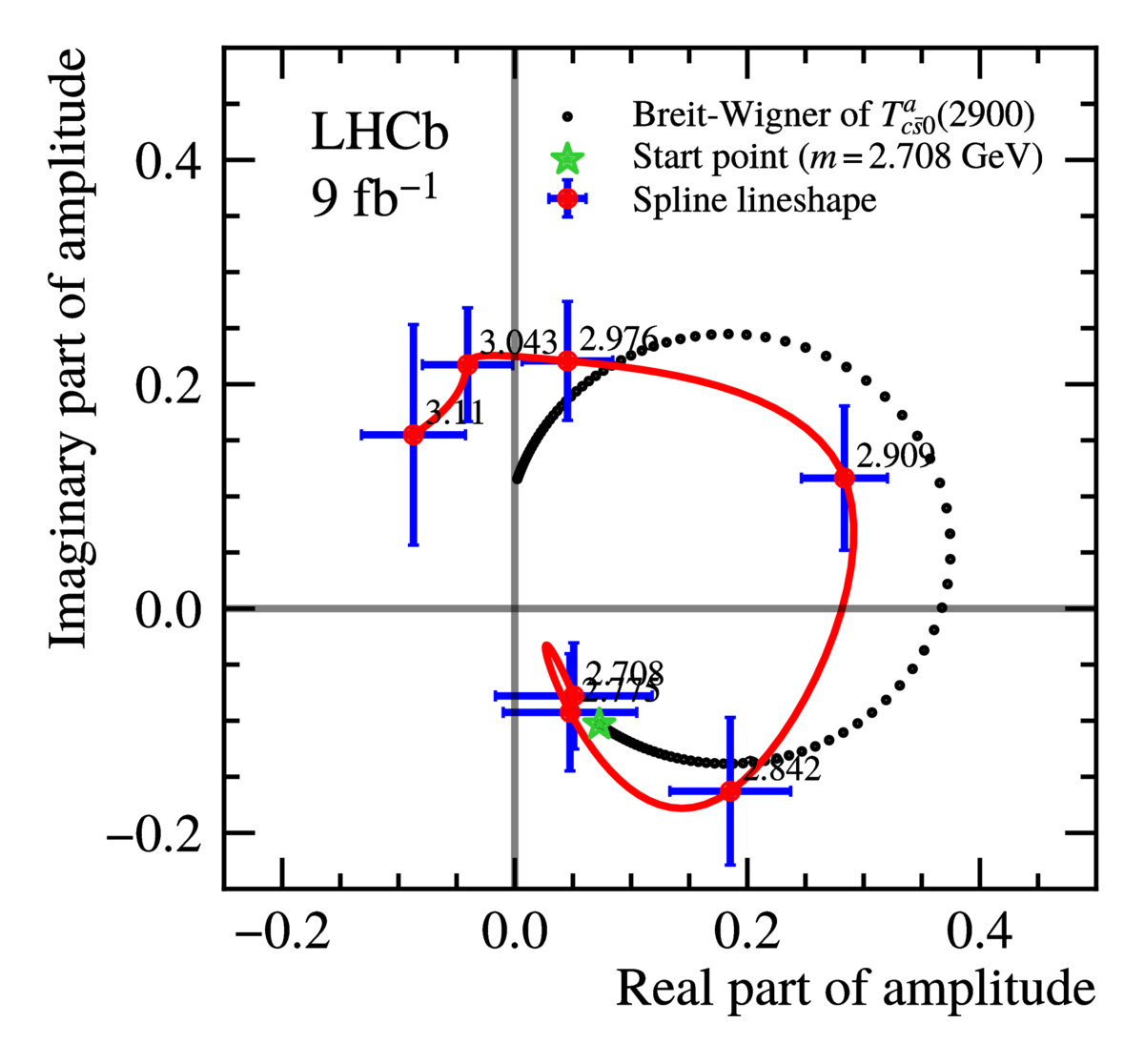}
\caption{Argand diagram, showing the complex coefficients associated to a spline function used to describe the S-wave component in the $\Dsm\pip$ channel \cite{LHCb:2022sfr}. The amplitude follows a trajectory consistent with that expected for a resonant contribution in that S-wave channel (approximately circular counter-clockwise mass evolution).\label{fig:Tcsbarargand}}
\end{figure}

Although at face value the study of decays involving a $\Dsp\pim$ channel 
($\cquark\squarkbar \uquarkbar \dquark$) seems only indirectly connected to that of $\Bp\to\Dp\Dm\Kp$, a remarkable parallel emerged, leading to the discovery of the $T_{c \squarkbar 0}(2900)$ states.
The LHCb collaboration undertook amplitude analyses of the decays $\Bd\to\Dzb\Dsp\pim$ and $\Bp\to\Dm\Dsp\pip$, process in which conventional $(\quark\quarkbar)$ resonances would only be expected in the $\Dzb\pim$ or $\Dm\pip$ channels~\cite{LHCb:2022sfr,LHCb:2022lzp}. The two datasets were fit simultaneously, where isospin relationships were imposed between corresponding amplitudes in the two channels. The Dalitz plot visualisation of the decay \phsp clearly illustrates the similarity of the resonant structure. 
Resonant amplitudes, modelled again by relativistic Breit-Wigner functions, were introduced corresponding to various excited open-charm states, and a quasi-model-independent spline function was used to model the \NR $S$-wave contribution. The model fails to describe the \phsp well, particularly in the region around a mass of $2.9\gevcc$ in the $\Dsp\pim$ and $\Dsm\pip$ spectra in the $\Bd\to\Dzb\Dsp\pim$ and $\Bp\to\Dm\Dsp\pip$ decays, respectively. A good fit quality is achieved through the addition of two new $\Dsp\pim$ states, sharing a complex coefficient, mass, and width, according to isospin symmetry expectations. The best fit was achieved with a spin-parity, $J^P=0^+$, and the fitted mass and width of the new states, named $T_{c\squarkbar 0}^a(2900)^0$ and $T_{c\squarkbar 0}^a(2900)^{++}$, are $2909\pm 10\mevcc$ and $134\pm19\mevcc$ 
respectively. An encouraging confirmation of the resonant character of the new state was obtained through the use of a spline function to model the S-wave component in the $\Dsm\pip$ channel. Plotting the amplitude at the knots of the spline function, it is observed in Fig.~\ref{fig:Tcsbarargand} that the amplitude follows a trajectory consistent with that expected for a resonance in the channel, as a function of $\Dsm\pip$ invariant mass.

It is interesting that the earlier study of the $\Dp\Kp$ channel $\Bp\to\Dp\Dm\Kp$ decay did not lead to the need for introduction of a doubly-charged contribution, though the sample size was considerably smaller in that case~\cite{LHCb:2020bls,LHCb:2020pxc}. In actual fact an excess was observed at a mass of $2.9\gevcc$ in the $\Dp\Kp$ channel in that analysis but it was adequately described as the reflection of the spin-2 $\chi_{c2}(3930)$ component of the amplitude model. It will be of interest whether the description remains satisfactory as the larger Run 3 dataset is brought into view. As for the the states discovered in the $\Dp\Dm\Kp$ final state, the use of single-charm decays is likely to be of considerable value here, in this case $\Bm\to\Dsp\Km\pim$ where the $\Dsp\pim$ channel appears once more.

The claim for observation of a new isospin pair of states, with one electrically-neutral and one doubly charged, was a striking one. The relationship between the earlier-discussed $T_{cs}(2900)^0$ structures and the new $T_{c\squarkbar 0}^a(2900)$ states is non-obvious; aside from the differing quark content, their widths differ significantly also. Interpretation of the new states followed a similar pattern to that of the $T_{cs}(2900)^0$ states, with models ranging from compact tetraquarks to molecular models~\cite{Lian:2023cgs,Agaev:2022eyk}, once more noting the proximity of the $2.9\gevcc$ structure to the $D^*K^*$ threshold and to that of the $D_s^*\rho$ combination~\cite{Molina:2022jcd}. Attempts have also been made to provide a unified description of the $\Dp\Km$ and $\Ds\pi$ states, as part of a new $SU(3)$ flavour 6-plet, though the absence of a corresponding non-strange $D^*(2900)^{0}$ state is notable~\cite{Dmitrasinovic:2023eei}.

\section{HIDDEN DOUBLE CHARM TETRAQUARK CANDIDATES}
\label{sec:hiddendoublecharm}

If the discoveries outlined in Sec.~\ref{sec:charmingstrange} had been made earlier, those tempted to apply a molecular interpretation to the excesses observed near meson-pair thresholds might have been less surprised by evidence of a new structure near threshold in the invariant mass spectrum of promptly produced pairs of $\jpsi$ mesons~\cite{LHCb:2020bwg}. It was, however, the searches for tetraquark states comprising four heavy $b$ quarks, ($\bquark\bquark\bquarkbar\bquarkbar$), decaying to $(\Upsilon(1S)\to\mu^+\mu^-)\mu^+\mu^-$~\cite{LHCb:2018uwm,CMS:2020qwa}, which had captured attention of experimentalists before the LHCb collaboration turned to look for ($\cquark\cquark\cquarkbar\cquarkbar$) states.
Such states might be expected to decay into a pair of \jpsi (or other charmonium) mesons, each composed of a $c\cquarkbar$ pair. Predictions for such states were already in place in the range from $5.8\gevcc$ to $7.4\gevcc$~\cite{Iwasaki:1976cn,Chao:1980dv,Ader:1981db,Li:1983ru,Berezhnoy:2011xn,Wu:2016vtq,Karliner:2016zzc,Barnea:2006sd,Debastiani:2017msn,Liu:2019zuc,Chen:2016jxd,Wang:2019rdo,Bedolla:2019zwg,Lloyd:2003yc,Chen:2020lgj,Wang:2018poa,Anwar:2017toa}.

A search for tetraquark states decaying to the $(\jpsi\to\mu^+\mu^-)(\jpsi\to\mu^+\mu^-)$ final state was undertaken by the LHCb collaboration using the full LHCb 2011--2018 dataset (9~$\invfb$). The sample preparation was relatively straightforward, consisting of momentum thresholds for momentum tracks, standard muon-ID, and requirement that pairs of oppositely charged muon tracks form a reasonable vertex with an invariant mass from 3\gevcc to 3.2\gevcc. A further requirement that the \jpsi candidates be consistent with originating at the primary $pp$ collision point suppressed the background from \B-decays, thanks to the displaced production point for candidates produced via those processes. The possibility for muon tracks to be used twice in reconstructing the $\jpsi\jpsi$ pair was excluded simply by requiring a significant angular separation for the tracks. 

The resulting sample contained a large fraction of correlated $\jpsi$ candidates. 
Finally, the invariant mass spectrum of the \jpsi pair could be considered, with improved resolution thanks to a kinematic refit with mass-constraints on each \jpsi candidate. As resonant \jpsi pair production is associated with a single parton scattering (SPS) as opposed to the independent \jpsi production of a double-parton scattering (DPS), and as such SPS interactions are characterised by a higher transverse momentum for the \jpsi pair, it was easy to suppress the DPS background. The residual DPS production component was identified by exploiting the fact that it typically inhabits the high \jpsi pair invariant mass region. The di-\jpsi spectrum displayed two peculiar features: firstly a broad near-threshold structure and a second, narrower structure around $6.9\gevcc$ (often referred to as the $X(6900)$).
Both appeared to grow in significance as the transverse momentum of the \jpsi pair increased and neither appeared to originate in efficiency structures, themselves found to be rather flat as a function of invariant mass.

\begin{figure}[htbp]
\centering
\includegraphics[width=.8\textwidth]{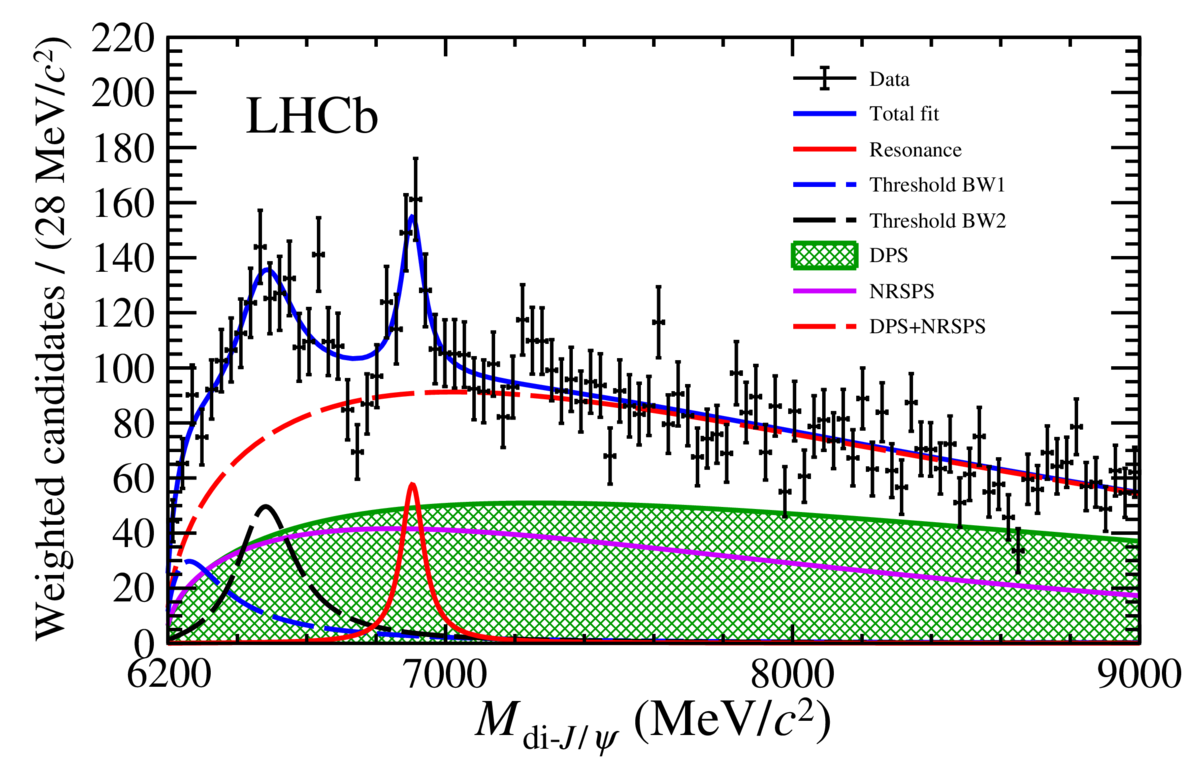}
\caption{Fit to the $\jpsi\jpsi$ invariant-mass spectrum according to the first of two models, where a single relativistic Breit-Wigner function is used to describe the higher-mass excess and two more are used to describe the near-threshold region~\cite{LHCb:2020bwg}.\label{fig:Tpsipsi_model1}}
\end{figure}

The interpretation of the structures is complicated, and affects the functions applied in a fit to the di-\jpsi spectrum. Two models achieved a satisfactory description of the data. The first, shown in Fig.~\ref{fig:Tpsipsi_model1}, comprised a single Breit-Wigner (BW) function corresponding to the higher-mass, narrower structure, and two more BW functions to describe the near-threshold shape. A second model allowed for interference between the component describing the \NR SPS production and the near-threshold resonance. The first model yields a measurement for the higher-mass sate of $6905\pm11\mevcc$ and width of $80\pm19\mev$, whereas the second produced a slightly lower mass of  $6886\pm11\mevcc$ and width of $168\pm33\mevcc$. 

Since no hadron identification is needed, and all LHC detectors are equipped with muon detectors, the $\jpsi\jpsi$ final state is also accessible to the ATLAS and CMS experiments \cite{ATLAS:2023bft,CMS:2023owd}. The lower absolute reconstruction efficiencies, due to the higher muon transverse-momentum thresholds, are offset by the higher integrated luminosities (135-140 $\invfb$, Run 2) obtained by these central detectors collecting their data at the maximal LHC instantaneous luminosity.
The LHCb discovery prompted these collaborations to analyze their $\mu^+\mu^-\mu^+\mu^-$ samples near the double-charmonium threshold with a superior sensitivity. The 6.9 GeV structure was confirmed by both experiments. In addition to the $\jpsi\jpsi$ structure, the ATLAS collaboration has a $4.7\sigma$ evidence for a $\jpsi\psi(2S)$ mass structure in the similar mass range \cite{ATLAS:2023bft}. 
The CMS collaboration found a clear evidence for the lower-mass state at 6.6~\gev and a $4.1\sigma$ evidence for a peak at 7.3~\gev~\cite{CMS:2023owd}. 

Once more, the nature of the observed four-quark mass structure is unclear. Decomposition of the observed mass spectrum into resonant contributions is assumption-dependent, though it is likely that multiple resonances play a role. Their widths are of the order 100~\mev or more, unlike for the tetraquark and pentaquark candidates -- good candidates for the ``molecular'' states -- discussed in the previous sections. Both of these observations point to compact tetraquark dynamics. However, coupled-channel effects (e.g.\ from $\jpsi\psi(2S)$) can also play a role~\cite{Dong:2020nwy}.
The addition of the new, larger Run 3 dataset should shed light on some of these questions.

\section{NARROW DOUBLE CHARM TETRAQUARK}
\label{sec:doublecharm}

In 2017 the LHCb collaboration observed a \Xiccpp state in its decay to the $\Lc\Km\pip\pip$ final state and having a mass around 3621\mev~\cite{LHCb:2017iph}. 
The signal consisted of $313\pm33$ candidates and had overwhelming significance in Run 2 data corresponding to an integrated luminosity of $1.7~\invfb$.
Later analysis of the full Run~2 dataset -- approximately 3 times larger -- confirmed observation of the state in the $\Lc\Km\pip\pip$ and $\Xicp\pip$ decay modes with a combined yield exceeding two thousand candidates~\cite{LHCb:2019epo}. 
Together with the observation of hundreds of signal candidates for the $X(6900)$ state decaying to $\jpsi\jpsi$ (see Sec.~\ref{sec:hiddendoublecharm}), this indicated that the LHCb data sample could be used to search for even more rare hadronic states containing two charm quarks.
This included tetraquark states formed by two charm quarks and two light anti-quarks. They have been awaited since the 1980's~\cite{Ader:1981db,BALLOT1983449} as candidates for a tightly-bound exotic hadron state. 
The measured mass of the \Xiccpp baryon, allowed to refine models of binding of doubly-heavy diquark  and implied that the mass of the $(cc\uquarkbar \dquarkbar)$ tetraquark should be close to the sum of \D and \Dstar masses~\cite{Karliner:2017qjm} and therefore might be narrow providing a clear experimental signature.
Also, in analogy to the number of narrow tetraquark states seen in $\D^{(*)}\Dbar^{(*)}$ spectra near the corresponding mass thresholds (for instance the \theX, or $Z_c(3900)$), one might expect similar \lq\lq molecular" states near the $\D^{(*)}\D^{(*)}$ thresholds. 
At the same time, no consensus regarding these states existed and theoretical predictions for the isospin-zero ground state $(cc\uquarkbar\dquarkbar)$ varied within a $\pm250\mev$ range around $\Dz\Dstarp$ mass threshold.

\begin{figure}[tbp]
\centering
\includegraphics[width=.73\textwidth]{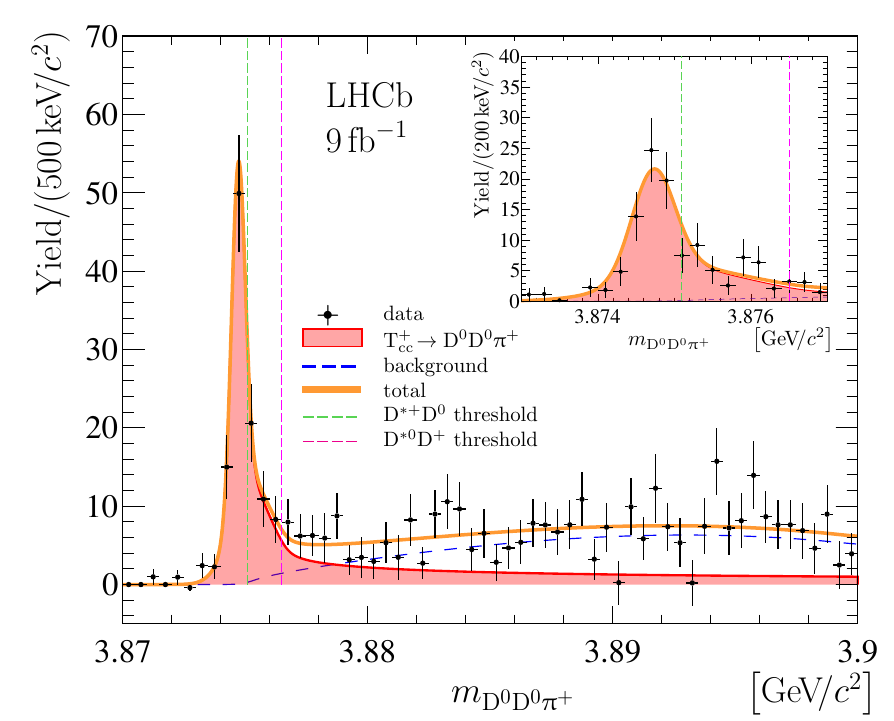}
\caption{Fit to $\Dz\Dz\pip$ invariant mass with the \Tcc signal described as a unitarized Breit-Wigner function on top of \NR background described modelled by a \phsp term multiplied by a polynomial, all convolved with detector resolution~\cite{LHCb:2021vvq}.
} 
\label{fig:Tcc}
\end{figure}

Using the full Run~1 and~2 LHCb data sample, the mass spectra of various $\D\D(\pip)$ combinations, where \D represents \Dz or \Dp produced directly in $pp$ collisions, were analyzed~\cite{LHCb:2021vvq,LHCb:2021auc}. Peaking structures were observed in the $\Dz\Dz$, $\Dz\Dp$ and $\Dz\Dz\pip$ mass spectra near the corresponding mass thresholds. 
The peak in the $\Dz\Dz\pip$ distribution appeared to lie a fraction of $1\mev$ below the $\Dz\Dstarp$ threshold. It was narrower than the peak in $\Dz\Dp$ channel and, as such, was the most natural candidate for the predicted ground $(cc\uquarkbar\dquarkbar)$ tetraquark state, \Tcc.
Given the close proximity to the threshold, a simple Breit-Wigner function
could not be used for the description of its shape.
Instead a unitarised Breit–Wigner model was constructed, taking into account coupled-channel effects and relying on two assumptions. The first assumption was that the observed state should decay strongly in the $\Dz\Dstarp$ and $\Dp\Dstarz$ channels and, subsequently, to $\Dz\Dz\pip$, $\Dz\Dp\piz$ and $\Dz\Dp\g$ final states, with equal couplings.
Secondly, it was assumed to have isospin $I=0$ and quantum numbers $J^P=1^+$, in accordance with the theoretical expectation for the \Tcc ground state. 
In the derivation of the subsequent \Tcc line shape, unitarity is ensured by accounting for the energy dependence of the $\Dz\Dz\pip$, $\Dz\Dp\g$ and $\Dz\Dp\piz$ decay widths.
Analyticity is ensured by accounting for the imaginary part of the total width via Kramer-Kronig relations.
The resulting function for \Tcc lineshape happens to depend on only one parameter, the mass position of the peak, given that the decay coupling to $\D\Dstar$ is large enough. 
The function is then convolved with a detector resolution of around $400\kevcc$, which is an order of magnitude larger than the natural \Tcc width calculated in this model.  The \NR $\Dz\Dstarp$ background in the fit to the data is assigned a \phsp model.
The resulting model provided a good description of the data as shown in Fig.~\ref{fig:Tcc}. The mass of the state relative to the $\Dz\Dstarp$ threshold was measured to be 
$\delta m_{\Tcc} = -359 \pm 40 ^{+9}_{-6} \kevcc$.
For this value of the mass, the model calculation gives a width of $\Gamma = 48 \pm 2 ^{+\phantom{1}0}_{-14} \kev$, where the first width uncertainty is due to that on the mass measurement, and the second is due to the uncertainty in the coupling parameter of the \Tcc to $\D\Dstar$, $|g|$. While the lower limit reflects the experimental constraints, the upper one does not and is entirely a feature of the model, namely that the width of the \Tcc saturates at high values of the coupling $|g|$.

Using the same \Tcc decay model, with the \Tcc mass value determined from the fit to the $\Dz\Dz\pip$ channel, the expected \Tcc signal shapes in the $\Dz\Dz$, $\Dz\Dp$ and $\Dz\pip$ mass distributions from \Tcc decays to the $\Dz\Dz\pip$, $\Dz\Dp\piz$ and $\Dz\Dp\g$ final states were obtained. 
In all cases a good agreement with the data distributions was observed when fitting their amplitudes on top of the \NR backgrounds.
Together with non-observation of any peaking structures in the $\Dp\Dp$ and $\Dp\Dz\pip$ mass spectra, this provides additional support for the assumed \Tcc decay model, and in particular, for the assumption that the discovered state is an isoscalar. In addition, the dependence of \Tcc production on its transverse momentum and event track multiplicity was studied. It was compared to the production of $\Dz\Dz$ pairs with mass in the $3.75<m_{\Dz\Dz}<3.87\gevcc$ range, expected to have a large contribution from double-parton scattering, and of $\Dz\Dzb$ pairs with the mass below 3.87\gevcc, given the similarity to charmonium production. While the $\theX$ production was found to be suppressed with respect to $\psitwos$ production at high multiplicities (see Fig.~\ref{fig:x3872_2}) \cite{LHCb:2020sey}, no such suppression was observed for the \Tcc state. This suggests that different mechanisms might play role in their production. However, given the low statistics, no firm conclusions could be made.

The \Tcc state is very close to the $\Dz\Dstarp$ threshold (see Fig.~\ref{fig:Tcc}), reminiscent the \theX state being close to the $\Dz\Dstarzb$ threshold.
As for the \theX, it is natural to assume that the \Tcc is dominantly a loosely-bound $\D\Dstar$ molecule which, in fact, matches the theoretical predictions based on the non-relativistic constituent quark model made a long time before the \Tcc discovery~\cite{Janc:2004qn}. 
However, since compact tetraquark models predict  that the \Tcc mass also be in this range, 
 both configurations might play a role.
At the same time, a good description of the $\Dz\Dz\pip$ mass spectra by the model assuming only $\Tcc\to\D\Dstar$ decays, discussed above, does not leave much room for \NR $\Tcc\to\D\D\pion(\g)$ decay modes, which are expected to be sizeable for a compact tetraquark.
Compared to the \Tcc, the \theX has an order of magnitude larger width, it decays to the lighter charmonium states, and it can contain an admixture of the conventional charmonium state itself.
This makes the \Tcc a much simpler state for theoretical description and allows for less ambiguous interpretation of its experimentally measured properties. Hence it has a great potential for helping to identify the relative roles of compact tetraquark and ``molecular'' configurations.

\section{FUTURE PROSPECTS}
\label{sec:future}

During Run~1 and~2 the LHCb experiment successfully operated at instantaneous luminosities reaching ${{\cal{L}} \sim 4\times 10^{32} \cm^{-2} \s^{-1}}$ and  
produced excellent results.
To cope with higher instantaneous luminosity, while maintaining or improving detection efficiency, a major upgrade of the detector has been installed during Long~Shutdown~2~\cite{LHCb:2023hlw}.
The new detector is now in  place. 

Increased luminosity will result in signal rates exceeding 1~MHz, which previously was the maximum output rate of the hardware (`level-0') trigger of the original detector.
If a similar hardware trigger had been used, it would have resulted in a large loss of the signal efficiency devaluing the gain from the higher luminosity, especially for fully hadronic channels.
To overcome this limitation, the level-0 trigger was eliminated, thus for the first time introducing a full-software trigger and readout of the entire detector at the visible LHC beam-beam collision rate of 30 MHz. This allows to discriminate signals from the backgrounds based on full event reconstruction. Hence trigger efficiencies are kept as high as in Run~1 and Run 2 or, in case of fully hadronic channels, are even doubled.
This new approach required migration of the first stage of the partial event reconstruction to a farm of GPU-equipped compute nodes and speeding up the rest of event reconstruction via efficient use of memory allocations and multi-threading.
To cope with the increased number of beam-beam interactions per beam crossing, and thus detector occupancy, the experiment's tracking system was completely upgraded. A new vertex detector is made of an array of silicon pixel detectors, replacing the silicon strips used before the upgrade. The inner radius was decreased from $8.2\mm$ to $5.1\mm$.
The silicon-strip tracker in front of the dipole magnet has a much improved readout segmentation and also extends closer to the beam pipe. The tracker behind the magnet, which used to be a combination of silicon-strips in the inner part and straw-tubes in the outer part was replaced by scintillating fibers. 
All sensors, including new photo-detectors in the RICH detectors (multi-anode PMTs instead of HPDs), are now read out at the 40~MHz beam crossing frequency.
The new nominal luminosity should reach ${{\cal{L}} \sim 2\times 10^{33} \cm^{-2} \s^{-1}}$ and a total integrated luminosity of around $50\invfb$ is expected by the end of Run~4 of the LHC (including Run~1-2). Together with improved trigger efficiency, 
the total yields in most physics channels will be an order of magnitude larger than in Run~1-2.
This opens vast opportunities for extension of exotic hadron studies.

For various $b$ decays to final states containing a \jpsi or \psitwos decaying to $\mu^+\mu^-$ signal yields will reach 
millions of events.
For final states with other types of charmonium states decaying to hadrons, like \etac or \chic decaying to $\proton\antiproton$, which have relatively low branching fractions and lower trigger efficiencies, signal yields will reach up to hundreds of thousands events.
This will allow for precise characterisation of already-discovered tetraquark and pentaquark states with hidden charmonium. 
At the same time, signal yields for $b$ decays to final states with double open charm ($\D\Dbar$, $\Lc\Dbar$ { etc.}) will reach tens of thousands of events allowing for observation of tetraquarks and pentaquarks in these modes. That will open the way to constrain coupled-channel effects in their various decay modes
and, eventually, a joint description of corresponding decay amplitudes via e.g.\ K-matrix formalism.

In relation to exploration of the \theX state, the larger dataset would allow precise determination  of relative branching fractions of its various decays and access to new suppressed decay modes ($\chicone\pip\pim$, $\proton\antiproton$ {etc.}). 
Moreover, decays of other exotic states to the \theX may be discovered.
Improved studies of the \theX mass lineshape in selected decay channels will elucidate the nature of the \theX.
With the SMOG system~\cite{SMOG} 
capable of injecting various noble gases into the volume of LHCb vertex detector, 
studies of the \theX production
can be extended to a larger number of colliding systems.

Run~3 and~4 will also make it possible to do detailed study of doubly-charmed exotic states, namely the \Tcc and resonances in $\jpsi\jpsi$ system, and to search for analogous states with beauty quarks. They will also increase prospects for studies of exotic states with hidden beauty, so far observed only in $e^+e^-$ collisions, and searches for other types of exotic states like tetra/pentaquarks decaying only weakly or hadrons formed by six quarks.

To exploit the High Luminosity LHC (HL-LHC), a further upgrade (`Upgrade~II') is planned to be for Long~Shutdown~4~\cite{LHCbCollaboration:2776420}. 
After the second upgrade, the detector will operate at a maximum instantaneous luminosity of 
${{\cal{L}} \sim 1.5\times 10^{34} \cm^{-2} \s^{-1}}$,
with a plan to integrate a dataset corresponding to an integrated luminosity of $300\invfb$ by the end of data collection, thus increasing statistics in most channels by another order of magnitude. 
The LHCb experiment
will be capable of studying exotic hadrons with charm in decays of any \bquark-hadron up to $\Bc$ meson, and in almost all conceivable final states,
which otherwise will not be accessible in any other experiment foreseen in the coming decades.
For a number of exotic states, double-differential measurements of their production cross-sections will be performed, together with their dependence on collision environment.
Potentially, the obtained experimental knowledge will reach a volume and quality  to allow a profound understanding of nature of the multiquark states and will help to improve knowledge of non-perturbative QCD.

While exotic hadron spectroscopy has a bright future in the LHCb program, other experimental programs have an important role to play in investigating exotic hadrons which may not be reachable at the LHC, like e.g.\ $\psi(4230)$ or $Z_b^{+,0}$ states \cite{Lebed:2022vfu}.

\section{CONCLUSIONS}
\label{sec:conclusions}

The first generation of the LHCb experiment, which collected data over the past decade (Run 1 and 2 at the LHC), had a profound impact on exotic hadron spectroscopy. 
This was achieved thanks to its unique capabilities (Sec.~\ref{sec:lhcbexp}) to efficiently trigger on, and suppress backgrounds in, final states containing heavy charm and bottom quarks at the LHC. The resulting heavy-quark dataset will be unrivalled for decades to come. 
It demonstrated that many previously known exotic hadron candidates can be best studied at such a high-energy hadron collider, in spite of the experimental complications stemming from the large number of particles produced in such environment.
The LHCb collaboration's achievements in studies of the \theX particle (Sec.~\ref{sec:x3872}) are excellent examples including the definite determination of its quantum numbers, reaching sensitivity to observe its natural lineshape for the first time, providing precise clarification of the isopin-violating nature of its discovery mode $\theX\to\pi^+\pi^-\jpsi$, and uncovering evidence for unusual particle-multiplicity dependence of its prompt production.

The improved sensitivity helped to demonstrate that many hidden-charm exotic hadrons previously observed in $B$ meson decays, like the $Z_c(4430)^+\to\psi(2S)\pi^+$ and $X(4140)\to\jpsi\phi$, were only the tip of the iceberg of a rich spectrum of similar states with various quantum numbers (Sec.~\ref{sec:hiddencharm}).
Exotic $\jpsi K^+$ resonances in the same $B$ meson decays have been observed for the first time. None of these states are very narrow, qualitatively fitting the expectations of compact tetraquark model, $(\cquark\cquarkbar q\overline{q})$, in which there is no effective mechanism to suppress $\cquark$ and $\cquarkbar$ finding each other to form $\psi(2S)$ or $\jpsi$.
Coupled-channel effects still need to be taken into account in the amplitude analyses used to extract their spectra, before the spectral decomposition of these resonances is to be fully trusted. 
So far, there is no evidence for comparatively narrow \lq\lq molecular" tetraquark states, like the $Z_c(3900)^{+,0}\to\jpsi\pi^{+,0}$ so well established in $e^+e^-$ production.

In contrast, among the most striking discoveries by the LHCb experiment are the relatively narrow pentaquark states $P_c^+\to\jpsi p$ and $P_{cs}^0\to\jpsi\Lambda$, matching the expectations for \lq\lq molecular" states near the charmed-baryon+charmed-meson mass thresholds (Sec.~\ref{sec:pentaquarks}). Whether broader pentaquark states, matching compact pentaquark spectroscopy, exist is an open experimental question. These hidden-charm pentaquarks have been observed in $\bquark$ baryon decays, which are not reachable at the $e^+e^-$ $B-$factories.

Less clear is the origin of the charmed-and-strange tetraquark candidates discovered by the LHCb experiment in $B$ meson decays (Sec.~\ref{sec:charmingstrange}). The BelleII experiment has good prospects to contribute here. 

The other frontier opened by the LHCb experiment concerns $(\cquark\cquark\cquarkbar\cquarkbar)$ states observed in prompt productions, and decaying to vector-charmonium pairs (Sec.~\ref{sec:hiddendoublecharm}).
Since the latter are detected the in $(\mu^+\mu^-)(\mu^+\mu^-)$ final state, the ATLAS and CMS experiments have good sensitivity to them. They both confirmed the mass structures discovered by the LHCb collaboration and contributed additional evidence for the states. These structures are not unlikely to be due to compact tetraquark dynamics, though coupled-channel effects may also play a role.

Last but not least, the LHCb experiment also discovered the narrowest exotic hadron known to date, the $T_{cc}^+\to D^0D^0\pi^+$ (Sec.~\ref{sec:doublecharm}), carrying two units of charm $(\cquark\cquark\uquarkbar\dquarkbar)$.
Having some similarities with the \theX state, due to the proximity to the $\D\Dstar$ threshold (or $\D\Dstarb$ in the case of \theX), it is also a much simpler state, from both an experimental and theoretical point of view, since the two heavy quarks cannot form a conventional meson. 
It thus provides better prospects for understanding the relative roles of compact-tetraquark and ``molecular'' configurations.

The full potential of the already-accumulated LHCb data has not yet been realized. More experimentally-difficult final states, including neutral particles ($\gamma$, $\pi^0$, $\eta$, $\omega$) are being analyzed and results should become available in the coming years. Prompt production of exotic hadrons, which comes with difficult-to-control high backgrounds, may also lead to interesting new observations. Central (semi)exclusive production of clean low-multiplicity final state has also a potential to be fully explored. 

To conclude this article, it became very clear from the LHCb discoveries of several narrow pentaquarks near the baryon-meson mass thresholds with heavy quarks inside, that \lq\lq molecular" type of bindings are at work. More of such states are likely to be discovered in the future. This fits well with the narrow heavy tetraquark states discovered earlier at the $e^+e^-$ colliders near meson-meson thresholds. There are many other observations of broader exotic hadron structures, which don't fit the molecular model, which point to presence of compact color bindings. However, because of experimental difficulties of resolving broader states, as well as large theoretical uncertainties in modelling such systems, it is harder to draw strong conclusions. It is also quite possible that both types of bindings contribute simultaneously to the formation of the observed structures. 
More effort is needed on both experimental and theoretical sides to clarify if compact multiquark hadrons really exist, and how the two types of bindings interplay with each other. 
The order-of-magnitude increase in the data samples expected to be collected using the upgraded LHCb experiment over the next decade, and the further order-of-magnitude increase expected by the Upgrade II of LHCb (Sec.~\ref{sec:future})
will help this goal.

\section*{DISCLOSURE STATEMENT}
The authors are not aware of any affiliations, memberships, funding, or
financial holdings that might be perceived as affecting the objectivity of
this review. 

\section*{ACKNOWLEDGMENTS}
This work was supported by the National Science Foundation (USA) Award Number 2102879, the Science and Technology Facilities Council (UK) [grant reference ST/W004305/1] and the 2022/2023 INFN Research Grant Program (Italy) [Announcement n.23591].

\bibliographystyle{LHCb}
\bibliography{ExoticHadronsAtLHCb-arxiv}

\ifx\mcitethebibliography\mciteundefinedmacro
\PackageError{LHCb.bst}{mciteplus.sty has not been loaded}
{This bibstyle requires the use of the mciteplus package.}\fi
\providecommand{\href}[2]{#2}
\begin{mcitethebibliography}{100}
\mciteSetBstSublistMode{n}
\mciteSetBstMaxWidthForm{subitem}{\alph{mcitesubitemcount})}
\mciteSetBstSublistLabelBeginEnd{\mcitemaxwidthsubitemform\space}
{\relax}{\relax}

\bibitem{GellMann:1964nj}
M.~Gell-Mann, \ifthenelse{\boolean{articletitles}}{\emph{{A schematic model of
  baryons and mesons}},
  }{}\href{http://dx.doi.org/10.1016/S0031-9163(64)92001-3}{Phys.\ Lett.\
  \textbf{8} (1964) 214}\relax
\mciteBstWouldAddEndPuncttrue
\mciteSetBstMidEndSepPunct{\mcitedefaultmidpunct}
{\mcitedefaultendpunct}{\mcitedefaultseppunct}\relax
\EndOfBibitem
\bibitem{Zweig:1981pd}
G.~Zweig, \ifthenelse{\boolean{articletitles}}{\emph{{An SU$_3$ model for
  strong interaction symmetry and its breaking}}, }{} 1964.
\newblock { {version 1:
  \href{http://cds.cern.ch/record/352337/files/CERN-TH-401.pdf}{CERN-TH-401}};
  {version 2:
  \href{http://cds.cern.ch/record/570209/files/CERN-TH-412.pdf}{CERN-TH-412}}
  }\relax
\mciteBstWouldAddEndPuncttrue
\mciteSetBstMidEndSepPunct{\mcitedefaultmidpunct}
{\mcitedefaultendpunct}{\mcitedefaultseppunct}\relax
\EndOfBibitem
\bibitem{LHCb:2014set}
LHCb, R.~Aaij {\em et~al.}, \ifthenelse{\boolean{articletitles}}{\emph{{LHCb
  Detector Performance}},
  }{}\href{http://dx.doi.org/10.1142/S0217751X15300227}{Int.\ J.\ Mod.\ Phys.\
  A \textbf{30} (2015), no.~07 1530022},
  \href{http://arxiv.org/abs/1412.6352}{{\normalfont\ttfamily
  arXiv:1412.6352}}\relax
\mciteBstWouldAddEndPuncttrue
\mciteSetBstMidEndSepPunct{\mcitedefaultmidpunct}
{\mcitedefaultendpunct}{\mcitedefaultseppunct}\relax
\EndOfBibitem
\bibitem{Archilli:2013npa}
F.~Archilli {\em et~al.},
  \ifthenelse{\boolean{articletitles}}{\emph{{Performance of the Muon
  Identification at LHCb}},
  }{}\href{http://dx.doi.org/10.1088/1748-0221/8/10/P10020}{JINST \textbf{8}
  (2013) P10020}, \href{http://arxiv.org/abs/1306.0249}{{\normalfont\ttfamily
  arXiv:1306.0249}}\relax
\mciteBstWouldAddEndPuncttrue
\mciteSetBstMidEndSepPunct{\mcitedefaultmidpunct}
{\mcitedefaultendpunct}{\mcitedefaultseppunct}\relax
\EndOfBibitem
\bibitem{Choi:2003ue}
Belle, S.~K. Choi {\em et~al.},
  \ifthenelse{\boolean{articletitles}}{\emph{{Observation of a narrow
  charmonium-like state in exclusive $B^\pm\to K^\pm \pi^+ \pi^- \jpsi$
  decays}}, }{}\href{http://dx.doi.org/10.1103/PhysRevLett.91.262001}{Phys.\
  Rev.\ Lett.\  \textbf{91} (2003) 262001},
  \href{http://arxiv.org/abs/hep-ex/0309032}{{\normalfont\ttfamily
  arXiv:hep-ex/0309032}}\relax
\mciteBstWouldAddEndPuncttrue
\mciteSetBstMidEndSepPunct{\mcitedefaultmidpunct}
{\mcitedefaultendpunct}{\mcitedefaultseppunct}\relax
\EndOfBibitem
\bibitem{BaBar:2004oro}
BaBar, B.~Aubert {\em et~al.},
  \ifthenelse{\boolean{articletitles}}{\emph{{Study of the $B \to J/\psi K^-
  \pi^+ \pi^-$ decay and measurement of the $B \to X(3872) K^-$ branching
  fraction}}, }{}\href{http://dx.doi.org/10.1103/PhysRevD.71.071103}{Phys.\
  Rev.\ D \textbf{71} (2005) 071103},
  \href{http://arxiv.org/abs/hep-ex/0406022}{{\normalfont\ttfamily
  arXiv:hep-ex/0406022}}\relax
\mciteBstWouldAddEndPuncttrue
\mciteSetBstMidEndSepPunct{\mcitedefaultmidpunct}
{\mcitedefaultendpunct}{\mcitedefaultseppunct}\relax
\EndOfBibitem
\bibitem{CDF:2003cab}
CDF, D.~Acosta {\em et~al.},
  \ifthenelse{\boolean{articletitles}}{\emph{{Observation of the narrow state
  $X(3872) \to J/\psi \pi^+ \pi^-$ in $\bar{p}p$ collisions at $\sqrt{s} =
  1.96$\tev}}, }{}\href{http://dx.doi.org/10.1103/PhysRevLett.93.072001}{Phys.\
  Rev.\ Lett.\  \textbf{93} (2004) 072001},
  \href{http://arxiv.org/abs/hep-ex/0312021}{{\normalfont\ttfamily
  arXiv:hep-ex/0312021}}\relax
\mciteBstWouldAddEndPuncttrue
\mciteSetBstMidEndSepPunct{\mcitedefaultmidpunct}
{\mcitedefaultendpunct}{\mcitedefaultseppunct}\relax
\EndOfBibitem
\bibitem{D0Abazov:2004kp}
{D0}, V.~M. Abazov {\em et~al.},
  \ifthenelse{\boolean{articletitles}}{\emph{{Observation and properties of the
  $X(3872)$ decaying to $J/\psi \pi^+ \pi^-$ in $p \bar p$ collisions at
  $\sqrt{s}= 1.96$\tev}},
  }{}\href{http://dx.doi.org/10.1103/PhysRevLett.93.162002}{{Phys.\ Rev.\
  Lett.\ } \textbf{93} (2004) 162002},
  \href{http://arxiv.org/abs/hep-ex/0405004}{{\normalfont\ttfamily
  arXiv:hep-ex/0405004}}\relax
\mciteBstWouldAddEndPuncttrue
\mciteSetBstMidEndSepPunct{\mcitedefaultmidpunct}
{\mcitedefaultendpunct}{\mcitedefaultseppunct}\relax
\EndOfBibitem
\bibitem{Aaij:2011sn}
LHCb, R.~Aaij {\em et~al.},
  \ifthenelse{\boolean{articletitles}}{\emph{{Observation of $X(3872) $
  production in $pp$ collisions at $\sqrt{s}=7\tev$}},
  }{}\href{http://dx.doi.org/10.1140/epjc/s10052-012-1972-7}{Eur.\ Phys.\ J.\
  \textbf{C72} (2012) 1972},
  \href{http://arxiv.org/abs/1112.5310}{{\normalfont\ttfamily
  arXiv:1112.5310}}\relax
\mciteBstWouldAddEndPuncttrue
\mciteSetBstMidEndSepPunct{\mcitedefaultmidpunct}
{\mcitedefaultendpunct}{\mcitedefaultseppunct}\relax
\EndOfBibitem
\bibitem{CDF:2006ocq}
CDF, A.~Abulencia {\em et~al.},
  \ifthenelse{\boolean{articletitles}}{\emph{{Analysis of the quantum numbers
  $J^{PC}$ of the $X(3872)$}},
  }{}\href{http://dx.doi.org/10.1103/PhysRevLett.98.132002}{Phys.\ Rev.\ Lett.\
   \textbf{98} (2007) 132002},
  \href{http://arxiv.org/abs/hep-ex/0612053}{{\normalfont\ttfamily
  arXiv:hep-ex/0612053}}\relax
\mciteBstWouldAddEndPuncttrue
\mciteSetBstMidEndSepPunct{\mcitedefaultmidpunct}
{\mcitedefaultendpunct}{\mcitedefaultseppunct}\relax
\EndOfBibitem
\bibitem{Belle:2011vlx}
Belle, S.-K. Choi {\em et~al.},
  \ifthenelse{\boolean{articletitles}}{\emph{{Bounds on the width, mass
  difference and other properties of $X(3872) \to \pi^+ \pi^- J/\psi$ decays}},
  }{}\href{http://dx.doi.org/10.1103/PhysRevD.84.052004}{Phys.\ Rev.\ D
  \textbf{84} (2011) 052004},
  \href{http://arxiv.org/abs/1107.0163}{{\normalfont\ttfamily
  arXiv:1107.0163}}\relax
\mciteBstWouldAddEndPuncttrue
\mciteSetBstMidEndSepPunct{\mcitedefaultmidpunct}
{\mcitedefaultendpunct}{\mcitedefaultseppunct}\relax
\EndOfBibitem
\bibitem{LHCb:2013kgk}
LHCb, R.~Aaij {\em et~al.},
  \ifthenelse{\boolean{articletitles}}{\emph{{Determination of the X(3872)
  meson quantum numbers}},
  }{}\href{http://dx.doi.org/10.1103/PhysRevLett.110.222001}{Phys.\ Rev.\
  Lett.\  \textbf{110} (2013) 222001},
  \href{http://arxiv.org/abs/1302.6269}{{\normalfont\ttfamily
  arXiv:1302.6269}}\relax
\mciteBstWouldAddEndPuncttrue
\mciteSetBstMidEndSepPunct{\mcitedefaultmidpunct}
{\mcitedefaultendpunct}{\mcitedefaultseppunct}\relax
\EndOfBibitem
\bibitem{LHCb:2015jfc}
LHCb, R.~Aaij {\em et~al.}, \ifthenelse{\boolean{articletitles}}{\emph{{Quantum
  numbers of the $X(3872)$ state and orbital angular momentum in its $\rho^0
  J\psi$ decay}}, }{}\href{http://dx.doi.org/10.1103/PhysRevD.92.011102}{Phys.\
  Rev.\ D \textbf{92} (2015), no.~1 011102},
  \href{http://arxiv.org/abs/1504.06339}{{\normalfont\ttfamily
  arXiv:1504.06339}}\relax
\mciteBstWouldAddEndPuncttrue
\mciteSetBstMidEndSepPunct{\mcitedefaultmidpunct}
{\mcitedefaultendpunct}{\mcitedefaultseppunct}\relax
\EndOfBibitem
\bibitem{LHCb:2022jez}
LHCb, R.~Aaij {\em et~al.},
  \ifthenelse{\boolean{articletitles}}{\emph{{Observation of sizeable
  \ensuremath{\omega} contribution to ${\theX\to\pi^+\pi^-\jpsi}$ decays}},
  }{}\href{http://dx.doi.org/10.1103/PhysRevD.108.L011103}{Phys.\ Rev.\ D
  \textbf{108} (2023), no.~1 L011103},
  \href{http://arxiv.org/abs/2204.12597}{{\normalfont\ttfamily
  arXiv:2204.12597}}\relax
\mciteBstWouldAddEndPuncttrue
\mciteSetBstMidEndSepPunct{\mcitedefaultmidpunct}
{\mcitedefaultendpunct}{\mcitedefaultseppunct}\relax
\EndOfBibitem
\bibitem{Tornqvist:2003na}
N.~A. Tornqvist, \ifthenelse{\boolean{articletitles}}{\emph{{Comment on the
  narrow charmonium state of Belle at 3871.8-MeV as a deuson}},
  }{}\href{http://arxiv.org/abs/hep-ph/0308277}{{\normalfont\ttfamily
  arXiv:hep-ph/0308277}}\relax
\mciteBstWouldAddEndPuncttrue
\mciteSetBstMidEndSepPunct{\mcitedefaultmidpunct}
{\mcitedefaultendpunct}{\mcitedefaultseppunct}\relax
\EndOfBibitem
\bibitem{Tornqvist:2004qy}
N.~A. Tornqvist, \ifthenelse{\boolean{articletitles}}{\emph{{Isospin breaking
  of the narrow charmonium state of Belle at 3872-MeV as a deuson}},
  }{}\href{http://dx.doi.org/10.1016/j.physletb.2004.03.077}{Phys.\ Lett.\
  \textbf{B590} (2004) 209},
  \href{http://arxiv.org/abs/hep-ph/0402237}{{\normalfont\ttfamily
  arXiv:hep-ph/0402237}}\relax
\mciteBstWouldAddEndPuncttrue
\mciteSetBstMidEndSepPunct{\mcitedefaultmidpunct}
{\mcitedefaultendpunct}{\mcitedefaultseppunct}\relax
\EndOfBibitem
\bibitem{Voloshin:2003nt}
M.~B. Voloshin, \ifthenelse{\boolean{articletitles}}{\emph{{Interference and
  binding effects in decays of possible molecular component of X(3872)}},
  }{}\href{http://dx.doi.org/10.1016/j.physletb.2003.11.014}{Phys.\ Lett.\
  \textbf{B579} (2004) 316},
  \href{http://arxiv.org/abs/hep-ph/0309307}{{\normalfont\ttfamily
  arXiv:hep-ph/0309307}}\relax
\mciteBstWouldAddEndPuncttrue
\mciteSetBstMidEndSepPunct{\mcitedefaultmidpunct}
{\mcitedefaultendpunct}{\mcitedefaultseppunct}\relax
\EndOfBibitem
\bibitem{Swanson:2003tb}
E.~S. Swanson, \ifthenelse{\boolean{articletitles}}{\emph{{Short range
  structure in the X(3872)}},
  }{}\href{http://dx.doi.org/10.1016/j.physletb.2004.03.033}{Phys.\ Lett.\
  \textbf{B588} (2004) 189},
  \href{http://arxiv.org/abs/hep-ph/0311229}{{\normalfont\ttfamily
  arXiv:hep-ph/0311229}}\relax
\mciteBstWouldAddEndPuncttrue
\mciteSetBstMidEndSepPunct{\mcitedefaultmidpunct}
{\mcitedefaultendpunct}{\mcitedefaultseppunct}\relax
\EndOfBibitem
\bibitem{Terasaki:2007uv}
K.~Terasaki, \ifthenelse{\boolean{articletitles}}{\emph{{A new tetra-quark
  interpretation of $X(3872)$}},
  }{}\href{http://dx.doi.org/10.1143/PTP.118.821}{Prog.\ Theor.\ Phys.\
  \textbf{118} (2007) 821},
  \href{http://arxiv.org/abs/0706.3944}{{\normalfont\ttfamily
  arXiv:0706.3944}}\relax
\mciteBstWouldAddEndPuncttrue
\mciteSetBstMidEndSepPunct{\mcitedefaultmidpunct}
{\mcitedefaultendpunct}{\mcitedefaultseppunct}\relax
\EndOfBibitem
\bibitem{Maiani:2017kyi}
L.~Maiani, A.~D. Polosa, and V.~Riquer,
  \ifthenelse{\boolean{articletitles}}{\emph{{A Theory of X and Z Multiquark
  Resonances}}, }{}\href{http://arxiv.org/abs/1712.05296}{{\normalfont\ttfamily
  arXiv:1712.05296}}\relax
\mciteBstWouldAddEndPuncttrue
\mciteSetBstMidEndSepPunct{\mcitedefaultmidpunct}
{\mcitedefaultendpunct}{\mcitedefaultseppunct}\relax
\EndOfBibitem
\bibitem{LHCb:2020fvo}
LHCb, R.~Aaij {\em et~al.}, \ifthenelse{\boolean{articletitles}}{\emph{{Study
  of the $\psi_2(3823)$ and $\chi_{c1}(3872)$ states in ${B^+ \rightarrow
  \left( \jpsi\pi^+\pi^-\right)K^+}$ decays}},
  }{}\href{http://dx.doi.org/10.1007/JHEP08(2020)123}{JHEP \textbf{08} (2020)
  123}, \href{http://arxiv.org/abs/2005.13422}{{\normalfont\ttfamily
  arXiv:2005.13422}}\relax
\mciteBstWouldAddEndPuncttrue
\mciteSetBstMidEndSepPunct{\mcitedefaultmidpunct}
{\mcitedefaultendpunct}{\mcitedefaultseppunct}\relax
\EndOfBibitem
\bibitem{LHCb:2020xds}
LHCb, R.~Aaij {\em et~al.}, \ifthenelse{\boolean{articletitles}}{\emph{{Study
  of the lineshape of the $\chi_{c1}(3872)$ state}},
  }{}\href{http://dx.doi.org/10.1103/PhysRevD.102.092005}{Phys.\ Rev.\ D
  \textbf{102} (2020), no.~9 092005},
  \href{http://arxiv.org/abs/2005.13419}{{\normalfont\ttfamily
  arXiv:2005.13419}}\relax
\mciteBstWouldAddEndPuncttrue
\mciteSetBstMidEndSepPunct{\mcitedefaultmidpunct}
{\mcitedefaultendpunct}{\mcitedefaultseppunct}\relax
\EndOfBibitem
\bibitem{ParticleDataGroup:2022pth}
Particle Data Group, R.~L. Workman {\em et~al.},
  \ifthenelse{\boolean{articletitles}}{\emph{{Review of Particle Physics}},
  }{}\href{http://dx.doi.org/10.1093/ptep/ptac097}{PTEP \textbf{2022} (2022)
  083C01}\relax
\mciteBstWouldAddEndPuncttrue
\mciteSetBstMidEndSepPunct{\mcitedefaultmidpunct}
{\mcitedefaultendpunct}{\mcitedefaultseppunct}\relax
\EndOfBibitem
\bibitem{BaBar:2008flx}
BaBar, B.~Aubert {\em et~al.},
  \ifthenelse{\boolean{articletitles}}{\emph{{Evidence for $X(3872) \to
  \psi_{2S} \gamma$ in $B^\pm \to X_{3872} K^\pm$ decays, and a study of $B \to
  c \bar{c} \gamma K$}},
  }{}\href{http://dx.doi.org/10.1103/PhysRevLett.102.132001}{Phys.\ Rev.\
  Lett.\  \textbf{102} (2009) 132001},
  \href{http://arxiv.org/abs/0809.0042}{{\normalfont\ttfamily
  arXiv:0809.0042}}\relax
\mciteBstWouldAddEndPuncttrue
\mciteSetBstMidEndSepPunct{\mcitedefaultmidpunct}
{\mcitedefaultendpunct}{\mcitedefaultseppunct}\relax
\EndOfBibitem
\bibitem{Belle:2011wdj}
Belle, V.~Bhardwaj {\em et~al.},
  \ifthenelse{\boolean{articletitles}}{\emph{{Observation of $X(3872)\to J/\psi
  \gamma$ and search for $X(3872)\to\psi'\gamma$ in B decays}},
  }{}\href{http://dx.doi.org/10.1103/PhysRevLett.107.091803}{Phys.\ Rev.\
  Lett.\  \textbf{107} (2011) 091803},
  \href{http://arxiv.org/abs/1105.0177}{{\normalfont\ttfamily
  arXiv:1105.0177}}\relax
\mciteBstWouldAddEndPuncttrue
\mciteSetBstMidEndSepPunct{\mcitedefaultmidpunct}
{\mcitedefaultendpunct}{\mcitedefaultseppunct}\relax
\EndOfBibitem
\bibitem{LHCb:2014jvf}
LHCb, R.~Aaij {\em et~al.},
  \ifthenelse{\boolean{articletitles}}{\emph{{Evidence for the decay
  $X(3872)\rightarrow\psi(2S)\gamma$}},
  }{}\href{http://dx.doi.org/10.1016/j.nuclphysb.2014.06.011}{Nucl.\ Phys.\ B
  \textbf{886} (2014) 665},
  \href{http://arxiv.org/abs/1404.0275}{{\normalfont\ttfamily
  arXiv:1404.0275}}\relax
\mciteBstWouldAddEndPuncttrue
\mciteSetBstMidEndSepPunct{\mcitedefaultmidpunct}
{\mcitedefaultendpunct}{\mcitedefaultseppunct}\relax
\EndOfBibitem
\bibitem{Swanson:2004cq}
E.~S. Swanson, \ifthenelse{\boolean{articletitles}}{\emph{{Molecular
  interpretation of the X(3872)}}, }{} in {\em {32nd International Conference
  on High Energy Physics}}, pp.~1037--1039, 10, 2004.
\newblock \href{http://arxiv.org/abs/hep-ph/0410284}{{\normalfont\ttfamily
  arXiv:hep-ph/0410284}}.
\newblock
  doi:~\href{http://dx.doi.org/10.1142/9789812702227\_0206}{10.1142/9789812702227\_0206}\relax
\mciteBstWouldAddEndPuncttrue
\mciteSetBstMidEndSepPunct{\mcitedefaultmidpunct}
{\mcitedefaultendpunct}{\mcitedefaultseppunct}\relax
\EndOfBibitem
\bibitem{Dong:2009uf}
Y.~Dong, A.~Faessler, T.~Gutsche, and V.~E. Lyubovitskij,
  \ifthenelse{\boolean{articletitles}}{\emph{{$J/\psi\gamma$ and $\psi(2S)
  \gamma$ decay modes of the X(3872)}},
  }{}\href{http://dx.doi.org/10.1088/0954-3899/38/1/015001}{J.\ Phys.\
  \textbf{G38} (2011) 015001},
  \href{http://arxiv.org/abs/0909.0380}{{\normalfont\ttfamily
  arXiv:0909.0380}}\relax
\mciteBstWouldAddEndPuncttrue
\mciteSetBstMidEndSepPunct{\mcitedefaultmidpunct}
{\mcitedefaultendpunct}{\mcitedefaultseppunct}\relax
\EndOfBibitem
\bibitem{Ferretti:2014xqa}
J.~Ferretti, G.~Galat\`a, and E.~Santopinto,
  \ifthenelse{\boolean{articletitles}}{\emph{{Quark structure of the $X(3872)$
  and $\chi_b(3P)$ resonances}},
  }{}\href{http://dx.doi.org/10.1103/PhysRevD.90.054010}{Phys.\ Rev.\ D
  \textbf{90} (2014), no.~5 054010},
  \href{http://arxiv.org/abs/1401.4431}{{\normalfont\ttfamily
  arXiv:1401.4431}}\relax
\mciteBstWouldAddEndPuncttrue
\mciteSetBstMidEndSepPunct{\mcitedefaultmidpunct}
{\mcitedefaultendpunct}{\mcitedefaultseppunct}\relax
\EndOfBibitem
\bibitem{Guo:2014taa}
F.-K. Guo {\em et~al.}, \ifthenelse{\boolean{articletitles}}{\emph{{What can
  radiative decays of the X(3872) teach us about its nature?}},
  }{}\href{http://dx.doi.org/10.1016/j.physletb.2015.02.013}{Phys.\ Lett.\
  \textbf{B742} (2015) 394},
  \href{http://arxiv.org/abs/1410.6712}{{\normalfont\ttfamily
  arXiv:1410.6712}}\relax
\mciteBstWouldAddEndPuncttrue
\mciteSetBstMidEndSepPunct{\mcitedefaultmidpunct}
{\mcitedefaultendpunct}{\mcitedefaultseppunct}\relax
\EndOfBibitem
\bibitem{Takeuchi:2016hat}
S.~Takeuchi, M.~Takizawa, and K.~Shimizu,
  \ifthenelse{\boolean{articletitles}}{\emph{{Radiative Decays of the $X(3872)$
  in the Charmonium-Molecule Hybrid Picture}},
  }{}\href{http://dx.doi.org/10.7566/JPSCP.17.112001}{JPS Conf.\ Proc.\
  \textbf{17} (2017) 112001},
  \href{http://arxiv.org/abs/1602.04297}{{\normalfont\ttfamily
  arXiv:1602.04297}}\relax
\mciteBstWouldAddEndPuncttrue
\mciteSetBstMidEndSepPunct{\mcitedefaultmidpunct}
{\mcitedefaultendpunct}{\mcitedefaultseppunct}\relax
\EndOfBibitem
\bibitem{BESIII:2020nbj}
BESIII, M.~Ablikim {\em et~al.},
  \ifthenelse{\boolean{articletitles}}{\emph{{Study of Open-Charm Decays and
  Radiative Transitions of the $X(3872)$}},
  }{}\href{http://dx.doi.org/10.1103/PhysRevLett.124.242001}{Phys.\ Rev.\
  Lett.\  \textbf{124} (2020), no.~24 242001},
  \href{http://arxiv.org/abs/2001.01156}{{\normalfont\ttfamily
  arXiv:2001.01156}}\relax
\mciteBstWouldAddEndPuncttrue
\mciteSetBstMidEndSepPunct{\mcitedefaultmidpunct}
{\mcitedefaultendpunct}{\mcitedefaultseppunct}\relax
\EndOfBibitem
\bibitem{LHCb:2020coc}
LHCb, R.~Aaij {\em et~al.}, \ifthenelse{\boolean{articletitles}}{\emph{{Study
  of $B^0_s \to \jpsi \pi^+\pi^-K^+K^-$ decays}},
  }{}\href{http://dx.doi.org/10.1007/JHEP02(2021)024}{JHEP \textbf{02} (2021)
  024}, \href{http://arxiv.org/abs/2011.01867}{{\normalfont\ttfamily
  arXiv:2011.01867}}, [Erratum: JHEP 04, 170 (2021)]\relax
\mciteBstWouldAddEndPuncttrue
\mciteSetBstMidEndSepPunct{\mcitedefaultmidpunct}
{\mcitedefaultendpunct}{\mcitedefaultseppunct}\relax
\EndOfBibitem
\bibitem{LHCb:2023reb}
LHCb, R.~Aaij {\em et~al.},
  \ifthenelse{\boolean{articletitles}}{\emph{{Observation of the
  $B^0_s\rightarrow \chi_{c1}(3872)\pi^+\pi^-$ decay}},
  }{}\href{http://dx.doi.org/10.1007/JHEP07(2023)084}{JHEP \textbf{07} (2023)
  084}, \href{http://arxiv.org/abs/2302.10629}{{\normalfont\ttfamily
  arXiv:2302.10629}}\relax
\mciteBstWouldAddEndPuncttrue
\mciteSetBstMidEndSepPunct{\mcitedefaultmidpunct}
{\mcitedefaultendpunct}{\mcitedefaultseppunct}\relax
\EndOfBibitem
\bibitem{LHCb:2019imv}
LHCb, R.~Aaij {\em et~al.},
  \ifthenelse{\boolean{articletitles}}{\emph{{Observation of the
  $\Lambda_b^0\rightarrow \chi_{c1}(3872)pK^-$ decay}},
  }{}\href{http://dx.doi.org/10.1007/JHEP09(2019)028}{JHEP \textbf{09} (2019)
  028}, \href{http://arxiv.org/abs/1907.00954}{{\normalfont\ttfamily
  arXiv:1907.00954}}\relax
\mciteBstWouldAddEndPuncttrue
\mciteSetBstMidEndSepPunct{\mcitedefaultmidpunct}
{\mcitedefaultendpunct}{\mcitedefaultseppunct}\relax
\EndOfBibitem
\bibitem{PhysRevLett.125.152001}
CMS Collaboration, A.~M. Sirunyan {\em et~al.},
  \ifthenelse{\boolean{articletitles}}{\emph{Observation of the
  ${B}_{s}^{0}\ensuremath{\rightarrow} \mathit{X}(3872)\ensuremath{\phi}$
  decay}, }{}\href{http://dx.doi.org/10.1103/PhysRevLett.125.152001}{Phys.\
  Rev.\ Lett.\  \textbf{125} (2020) 152001},
  \href{http://arxiv.org/abs/2005.04764}{{\normalfont\ttfamily
  arXiv:2005.04764}}\relax
\mciteBstWouldAddEndPuncttrue
\mciteSetBstMidEndSepPunct{\mcitedefaultmidpunct}
{\mcitedefaultendpunct}{\mcitedefaultseppunct}\relax
\EndOfBibitem
\bibitem{Maiani:2020zhr}
L.~Maiani, A.~D. Polosa, and V.~Riquer,
  \ifthenelse{\boolean{articletitles}}{\emph{{$X(3872)$ tetraquarks in $B$ and
  $B_s$ decays}},
  }{}\href{http://dx.doi.org/10.1103/PhysRevD.102.034017}{Phys.\ Rev.\ D
  \textbf{102} (2020), no.~3 034017},
  \href{http://arxiv.org/abs/2005.08764}{{\normalfont\ttfamily
  arXiv:2005.08764}}\relax
\mciteBstWouldAddEndPuncttrue
\mciteSetBstMidEndSepPunct{\mcitedefaultmidpunct}
{\mcitedefaultendpunct}{\mcitedefaultseppunct}\relax
\EndOfBibitem
\bibitem{LHCb:2021ten}
LHCb, R.~Aaij {\em et~al.},
  \ifthenelse{\boolean{articletitles}}{\emph{{Measurement of
  \ensuremath{\chi}$_{c1}$(3872) production in proton-proton collisions at $
  \sqrt{s} $ = 8 and 13\tev}},
  }{}\href{http://dx.doi.org/10.1007/JHEP01(2022)131}{JHEP \textbf{01} (2022)
  131}, \href{http://arxiv.org/abs/2109.07360}{{\normalfont\ttfamily
  arXiv:2109.07360}}\relax
\mciteBstWouldAddEndPuncttrue
\mciteSetBstMidEndSepPunct{\mcitedefaultmidpunct}
{\mcitedefaultendpunct}{\mcitedefaultseppunct}\relax
\EndOfBibitem
\bibitem{LHCb:2020sey}
LHCb, R.~Aaij {\em et~al.},
  \ifthenelse{\boolean{articletitles}}{\emph{{Observation of Multiplicity
  Dependent Prompt $\chi_{c1}(3872)$ and $\psi(2S)$ Production in $pp$
  Collisions}},
  }{}\href{http://dx.doi.org/10.1103/PhysRevLett.126.092001}{Phys.\ Rev.\
  Lett.\  \textbf{126} (2021), no.~9 092001},
  \href{http://arxiv.org/abs/2009.06619}{{\normalfont\ttfamily
  arXiv:2009.06619}}\relax
\mciteBstWouldAddEndPuncttrue
\mciteSetBstMidEndSepPunct{\mcitedefaultmidpunct}
{\mcitedefaultendpunct}{\mcitedefaultseppunct}\relax
\EndOfBibitem
\bibitem{Esposito_2021}
A.~Esposito {\em et~al.}, \ifthenelse{\boolean{articletitles}}{\emph{{The
  nature of X(3872) from high-multiplicity pp collisions}},
  }{}\href{http://dx.doi.org/10.1140/epjc/s10052-021-09425-w}{Eur.\ Phys.\ J.\
  C \textbf{81} (2021) 669},
  \href{http://arxiv.org/abs/2006.15044}{{\normalfont\ttfamily
  arXiv:2006.15044}}\relax
\mciteBstWouldAddEndPuncttrue
\mciteSetBstMidEndSepPunct{\mcitedefaultmidpunct}
{\mcitedefaultendpunct}{\mcitedefaultseppunct}\relax
\EndOfBibitem
\bibitem{Braaten:2020iqw}
E.~Braaten, L.-P. He, K.~Ingles, and J.~Jiang,
  \ifthenelse{\boolean{articletitles}}{\emph{{Production of $X(3872)$ at High
  Multiplicity}},
  }{}\href{http://dx.doi.org/10.1103/PhysRevD.103.L071901}{Phys.\ Rev.\ D
  \textbf{103} (2021), no.~7 L071901},
  \href{http://arxiv.org/abs/2012.13499}{{\normalfont\ttfamily
  arXiv:2012.13499}}\relax
\mciteBstWouldAddEndPuncttrue
\mciteSetBstMidEndSepPunct{\mcitedefaultmidpunct}
{\mcitedefaultendpunct}{\mcitedefaultseppunct}\relax
\EndOfBibitem
\bibitem{LHCb:2022ixc}
LHCb, R.~Aaij {\em et~al.},
  \ifthenelse{\boolean{articletitles}}{\emph{{Modification of
  $\chi{_{c1}(3872)}$ and $\psi{(2S)}$ production in $p$Pb collisions at
  $\sqrt{s_{NN}} = 8.16$\tev}}, }{}
\newblock \href{http://cds.cern.ch/record/2807146}{LHCb-CONF-2022-001}\relax
\mciteBstWouldAddEndPuncttrue
\mciteSetBstMidEndSepPunct{\mcitedefaultmidpunct}
{\mcitedefaultendpunct}{\mcitedefaultseppunct}\relax
\EndOfBibitem
\bibitem{CMS:2021znk}
CMS, A.~M. Sirunyan {\em et~al.},
  \ifthenelse{\boolean{articletitles}}{\emph{{Evidence for X(3872) in Pb-Pb
  Collisions and Studies of its Prompt Production at $\sqrt
  {s_{NN}}$=5.02\tev}},
  }{}\href{http://dx.doi.org/10.1103/PhysRevLett.128.032001}{Phys.\ Rev.\
  Lett.\  \textbf{128} (2022), no.~3 032001},
  \href{http://arxiv.org/abs/2102.13048}{{\normalfont\ttfamily
  arXiv:2102.13048}}\relax
\mciteBstWouldAddEndPuncttrue
\mciteSetBstMidEndSepPunct{\mcitedefaultmidpunct}
{\mcitedefaultendpunct}{\mcitedefaultseppunct}\relax
\EndOfBibitem
\bibitem{Isgur:1978xj}
N.~Isgur and G.~Karl, \ifthenelse{\boolean{articletitles}}{\emph{{P Wave
  Baryons in the Quark Model}},
  }{}\href{http://dx.doi.org/10.1103/PhysRevD.18.4187}{Phys.\ Rev.\
  \textbf{D18} (1978) 4187}\relax
\mciteBstWouldAddEndPuncttrue
\mciteSetBstMidEndSepPunct{\mcitedefaultmidpunct}
{\mcitedefaultendpunct}{\mcitedefaultseppunct}\relax
\EndOfBibitem
\bibitem{LEPS:2003wug}
LEPS, T.~Nakano {\em et~al.},
  \ifthenelse{\boolean{articletitles}}{\emph{{Evidence for a narrow ${S = +1}$
  baryon resonance in photoproduction from the neutron}},
  }{}\href{http://dx.doi.org/10.1103/PhysRevLett.91.012002}{Phys.\ Rev.\ Lett.\
   \textbf{91} (2003) 012002},
  \href{http://arxiv.org/abs/hep-ex/0301020}{{\normalfont\ttfamily
  arXiv:hep-ex/0301020}}\relax
\mciteBstWouldAddEndPuncttrue
\mciteSetBstMidEndSepPunct{\mcitedefaultmidpunct}
{\mcitedefaultendpunct}{\mcitedefaultseppunct}\relax
\EndOfBibitem
\bibitem{CLAS:2003yuj}
CLAS, S.~Stepanyan {\em et~al.},
  \ifthenelse{\boolean{articletitles}}{\emph{{Observation of an exotic ${S =
  +1}$ baryon in exclusive photoproduction from the deuteron}},
  }{}\href{http://dx.doi.org/10.1103/PhysRevLett.91.252001}{Phys.\ Rev.\ Lett.\
   \textbf{91} (2003) 252001},
  \href{http://arxiv.org/abs/hep-ex/0307018}{{\normalfont\ttfamily
  arXiv:hep-ex/0307018}}\relax
\mciteBstWouldAddEndPuncttrue
\mciteSetBstMidEndSepPunct{\mcitedefaultmidpunct}
{\mcitedefaultendpunct}{\mcitedefaultseppunct}\relax
\EndOfBibitem
\bibitem{CLAS:2003wfm}
CLAS, V.~Kubarovsky {\em et~al.},
  \ifthenelse{\boolean{articletitles}}{\emph{{Observation of an exotic baryon
  with ${S = +1}$ in photoproduction from the proton}},
  }{}\href{http://dx.doi.org/10.1103/PhysRevLett.92.032001}{Phys.\ Rev.\ Lett.\
   \textbf{92} (2004) 032001},
  \href{http://arxiv.org/abs/hep-ex/0311046}{{\normalfont\ttfamily
  arXiv:hep-ex/0311046}}, [Erratum: Phys.Rev.Lett. 92, 049902 (2004)]\relax
\mciteBstWouldAddEndPuncttrue
\mciteSetBstMidEndSepPunct{\mcitedefaultmidpunct}
{\mcitedefaultendpunct}{\mcitedefaultseppunct}\relax
\EndOfBibitem
\bibitem{Aaij:2015tga}
LHCb, R.~Aaij {\em et~al.},
  \ifthenelse{\boolean{articletitles}}{\emph{{Observation of $J/\psi p$
  Resonances Consistent with Pentaquark States in $\Lambda_b^0 \to J/\psi K^-
  p$ Decays}},
  }{}\href{http://dx.doi.org/10.1103/PhysRevLett.115.072001}{Phys.\ Rev.\
  Lett.\  \textbf{115} (2015) 072001},
  \href{http://arxiv.org/abs/1507.03414}{{\normalfont\ttfamily
  arXiv:1507.03414}}\relax
\mciteBstWouldAddEndPuncttrue
\mciteSetBstMidEndSepPunct{\mcitedefaultmidpunct}
{\mcitedefaultendpunct}{\mcitedefaultseppunct}\relax
\EndOfBibitem
\bibitem{LHCb:2019kea}
LHCb, R.~Aaij {\em et~al.},
  \ifthenelse{\boolean{articletitles}}{\emph{{Observation of a narrow
  pentaquark state, $P_c(4312)^+$, and of two-peak structure of the
  $P_c(4450)^+$}},
  }{}\href{http://dx.doi.org/10.1103/PhysRevLett.122.222001}{Phys.\ Rev.\
  Lett.\  \textbf{122} (2019), no.~22 222001},
  \href{http://arxiv.org/abs/1904.03947}{{\normalfont\ttfamily
  arXiv:1904.03947}}\relax
\mciteBstWouldAddEndPuncttrue
\mciteSetBstMidEndSepPunct{\mcitedefaultmidpunct}
{\mcitedefaultendpunct}{\mcitedefaultseppunct}\relax
\EndOfBibitem
\bibitem{LHCb:2020jpq}
LHCb, R.~Aaij {\em et~al.},
  \ifthenelse{\boolean{articletitles}}{\emph{{Evidence of a $J/\psi\Lambda$
  structure and observation of excited $\Xi^-$ states in the $\Xi^-_b \to
  J/\psi\Lambda K^-$ decay}},
  }{}\href{http://dx.doi.org/10.1016/j.scib.2021.02.030}{Sci.\ Bull.\
  \textbf{66} (2021) 1278},
  \href{http://arxiv.org/abs/2012.10380}{{\normalfont\ttfamily
  arXiv:2012.10380}}\relax
\mciteBstWouldAddEndPuncttrue
\mciteSetBstMidEndSepPunct{\mcitedefaultmidpunct}
{\mcitedefaultendpunct}{\mcitedefaultseppunct}\relax
\EndOfBibitem
\bibitem{LHCb:2021chn}
LHCb, R.~Aaij {\em et~al.},
  \ifthenelse{\boolean{articletitles}}{\emph{{Evidence for a new structure in
  the $J/\psi p$ and $J/\psi \bar{p}$ systems in $B_s^0 \to J/\psi p \bar{p}$
  decays}}, }{}\href{http://dx.doi.org/10.1103/PhysRevLett.128.062001}{Phys.\
  Rev.\ Lett.\  \textbf{128} (2022), no.~6 062001},
  \href{http://arxiv.org/abs/2108.04720}{{\normalfont\ttfamily
  arXiv:2108.04720}}\relax
\mciteBstWouldAddEndPuncttrue
\mciteSetBstMidEndSepPunct{\mcitedefaultmidpunct}
{\mcitedefaultendpunct}{\mcitedefaultseppunct}\relax
\EndOfBibitem
\bibitem{LHCb:2022ogu}
LHCb, R.~Aaij {\em et~al.},
  \ifthenelse{\boolean{articletitles}}{\emph{{Observation of a $\jpsi\PLambda$
  Resonance Consistent with a Strange Pentaquark Candidate in $\Bm\to
  \jpsi\Lambda\antiproton$ Decays}},
  }{}\href{http://dx.doi.org/10.1103/PhysRevLett.131.031901}{Phys.\ Rev.\
  Lett.\  \textbf{131} (2023), no.~3 031901},
  \href{http://arxiv.org/abs/2210.10346}{{\normalfont\ttfamily
  arXiv:2210.10346}}\relax
\mciteBstWouldAddEndPuncttrue
\mciteSetBstMidEndSepPunct{\mcitedefaultmidpunct}
{\mcitedefaultendpunct}{\mcitedefaultseppunct}\relax
\EndOfBibitem
\bibitem{Aaij:2016phn}
LHCb, R.~Aaij {\em et~al.},
  \ifthenelse{\boolean{articletitles}}{\emph{{Model-independent evidence for
  $J/\psi p$ contributions to ${\Lambda_b^0\to J/\psi p K^-}$ decays}},
  }{}\href{http://dx.doi.org/10.1103/PhysRevLett.117.082002}{Phys.\ Rev.\
  Lett.\  \textbf{117} (2016) 082002},
  \href{http://arxiv.org/abs/1604.05708}{{\normalfont\ttfamily
  arXiv:1604.05708}}\relax
\mciteBstWouldAddEndPuncttrue
\mciteSetBstMidEndSepPunct{\mcitedefaultmidpunct}
{\mcitedefaultendpunct}{\mcitedefaultseppunct}\relax
\EndOfBibitem
\bibitem{Wang:2020giv}
M.~Wang {\em et~al.}, \ifthenelse{\boolean{articletitles}}{\emph{{A novel
  method to test particle ordering and final state alignment in helicity
  formalism}}, }{}\href{http://dx.doi.org/10.1088/1674-1137/abf139}{Chin.\
  Phys.\ C \textbf{45} (2021), no.~6 063103},
  \href{http://arxiv.org/abs/2012.03699}{{\normalfont\ttfamily
  arXiv:2012.03699}}\relax
\mciteBstWouldAddEndPuncttrue
\mciteSetBstMidEndSepPunct{\mcitedefaultmidpunct}
{\mcitedefaultendpunct}{\mcitedefaultseppunct}\relax
\EndOfBibitem
\bibitem{Ali:2019npk}
A.~Ali and A.~Y. Parkhomenko,
  \ifthenelse{\boolean{articletitles}}{\emph{{Interpretation of the narrow
  $J/\psi p$ Peaks in $\Lambda_b \to J/\psi p K^-$ decay in the compact diquark
  model}}, }{}\href{http://dx.doi.org/10.1016/j.physletb.2019.05.002}{Phys.\
  Lett.\ B \textbf{793} (2019) 365},
  \href{http://arxiv.org/abs/1904.00446}{{\normalfont\ttfamily
  arXiv:1904.00446}}\relax
\mciteBstWouldAddEndPuncttrue
\mciteSetBstMidEndSepPunct{\mcitedefaultmidpunct}
{\mcitedefaultendpunct}{\mcitedefaultseppunct}\relax
\EndOfBibitem
\bibitem{Liu:2019tjn}
M.-Z. Liu {\em et~al.}, \ifthenelse{\boolean{articletitles}}{\emph{{Emergence
  of a complete heavy-quark spin symmetry multiplet: seven molecular
  pentaquarks in light of the latest LHCb analysis}},
  }{}\href{http://dx.doi.org/10.1103/PhysRevLett.122.242001}{Phys.\ Rev.\
  Lett.\  \textbf{122} (2019), no.~24 242001},
  \href{http://arxiv.org/abs/1903.11560}{{\normalfont\ttfamily
  arXiv:1903.11560}}\relax
\mciteBstWouldAddEndPuncttrue
\mciteSetBstMidEndSepPunct{\mcitedefaultmidpunct}
{\mcitedefaultendpunct}{\mcitedefaultseppunct}\relax
\EndOfBibitem
\bibitem{Yang:2011wz}
Z.-C. Yang {\em et~al.}, \ifthenelse{\boolean{articletitles}}{\emph{{The
  possible hidden-charm molecular baryons composed of anti-charmed meson and
  charmed baryon}},
  }{}\href{http://dx.doi.org/10.1088/1674-1137/36/1/002}{Chin.\ Phys.\ C
  \textbf{36} (2012) 6},
  \href{http://arxiv.org/abs/1105.2901}{{\normalfont\ttfamily
  arXiv:1105.2901}}\relax
\mciteBstWouldAddEndPuncttrue
\mciteSetBstMidEndSepPunct{\mcitedefaultmidpunct}
{\mcitedefaultendpunct}{\mcitedefaultseppunct}\relax
\EndOfBibitem
\bibitem{Wu:2010jy}
J.-J. Wu, R.~Molina, E.~Oset, and B.~S. Zou,
  \ifthenelse{\boolean{articletitles}}{\emph{{Prediction of narrow $N^*$ and
  $\Lambda^*$ resonances with hidden charm above 4 GeV}},
  }{}\href{http://dx.doi.org/10.1103/PhysRevLett.105.232001}{Phys.\ Rev.\
  Lett.\  \textbf{105} (2010) 232001},
  \href{http://arxiv.org/abs/1007.0573}{{\normalfont\ttfamily
  arXiv:1007.0573}}\relax
\mciteBstWouldAddEndPuncttrue
\mciteSetBstMidEndSepPunct{\mcitedefaultmidpunct}
{\mcitedefaultendpunct}{\mcitedefaultseppunct}\relax
\EndOfBibitem
\bibitem{Wu:2012md}
J.-J. Wu, T.-S.~H. Lee, and B.~S. Zou,
  \ifthenelse{\boolean{articletitles}}{\emph{{Nucleon Resonances with Hidden
  Charm in Coupled-Channel Models}},
  }{}\href{http://dx.doi.org/10.1103/PhysRevC.85.044002}{Phys.\ Rev.\ C
  \textbf{85} (2012) 044002},
  \href{http://arxiv.org/abs/1202.1036}{{\normalfont\ttfamily
  arXiv:1202.1036}}\relax
\mciteBstWouldAddEndPuncttrue
\mciteSetBstMidEndSepPunct{\mcitedefaultmidpunct}
{\mcitedefaultendpunct}{\mcitedefaultseppunct}\relax
\EndOfBibitem
\bibitem{Karliner:2015ina}
M.~Karliner and J.~L. Rosner, \ifthenelse{\boolean{articletitles}}{\emph{{New
  Exotic Meson and Baryon Resonances from Doubly-Heavy Hadronic Molecules}},
  }{}\href{http://dx.doi.org/10.1103/PhysRevLett.115.122001}{Phys.\ Rev.\
  Lett.\  \textbf{115} (2015) 122001},
  \href{http://arxiv.org/abs/1506.06386}{{\normalfont\ttfamily
  arXiv:1506.06386}}\relax
\mciteBstWouldAddEndPuncttrue
\mciteSetBstMidEndSepPunct{\mcitedefaultmidpunct}
{\mcitedefaultendpunct}{\mcitedefaultseppunct}\relax
\EndOfBibitem
\bibitem{Chen:2019asm}
R.~Chen, Z.-F. Sun, X.~Liu, and S.-L. Zhu,
  \ifthenelse{\boolean{articletitles}}{\emph{{Strong LHCb evidence supporting
  the existence of the hidden-charm molecular pentaquarks}},
  }{}\href{http://dx.doi.org/10.1103/PhysRevD.100.011502}{Phys.\ Rev.\ D
  \textbf{100} (2019), no.~1 011502},
  \href{http://arxiv.org/abs/1903.11013}{{\normalfont\ttfamily
  arXiv:1903.11013}}\relax
\mciteBstWouldAddEndPuncttrue
\mciteSetBstMidEndSepPunct{\mcitedefaultmidpunct}
{\mcitedefaultendpunct}{\mcitedefaultseppunct}\relax
\EndOfBibitem
\bibitem{Guo:2019fdo}
F.-K. Guo, H.-J. Jing, U.-G. Mei\ss{}ner, and S.~Sakai,
  \ifthenelse{\boolean{articletitles}}{\emph{{Isospin breaking decays as a
  diagnosis of the hadronic molecular structure of the $P_c(4457)$}},
  }{}\href{http://dx.doi.org/10.1103/PhysRevD.99.091501}{Phys.\ Rev.\ D
  \textbf{99} (2019), no.~9 091501},
  \href{http://arxiv.org/abs/1903.11503}{{\normalfont\ttfamily
  arXiv:1903.11503}}\relax
\mciteBstWouldAddEndPuncttrue
\mciteSetBstMidEndSepPunct{\mcitedefaultmidpunct}
{\mcitedefaultendpunct}{\mcitedefaultseppunct}\relax
\EndOfBibitem
\bibitem{He:2019ify}
J.~He, \ifthenelse{\boolean{articletitles}}{\emph{{Study of $P_c(4457)$,
  $P_c(4440)$, and $P_c(4312)$ in a quasipotential Bethe-Salpeter equation
  approach}}, }{}\href{http://dx.doi.org/10.1140/epjc/s10052-019-6906-1}{Eur.\
  Phys.\ J.\ C \textbf{79} (2019), no.~5 393},
  \href{http://arxiv.org/abs/1903.11872}{{\normalfont\ttfamily
  arXiv:1903.11872}}\relax
\mciteBstWouldAddEndPuncttrue
\mciteSetBstMidEndSepPunct{\mcitedefaultmidpunct}
{\mcitedefaultendpunct}{\mcitedefaultseppunct}\relax
\EndOfBibitem
\bibitem{Huang:2019jlf}
H.~Huang, J.~He, and J.~Ping,
  \ifthenelse{\boolean{articletitles}}{\emph{{Looking for the hidden-charm
  pentaquark resonances in $J/\psi p$ scattering}},
  }{}\href{http://arxiv.org/abs/1904.00221}{{\normalfont\ttfamily
  arXiv:1904.00221}}\relax
\mciteBstWouldAddEndPuncttrue
\mciteSetBstMidEndSepPunct{\mcitedefaultmidpunct}
{\mcitedefaultendpunct}{\mcitedefaultseppunct}\relax
\EndOfBibitem
\bibitem{Shimizu:2019ptd}
Y.~Shimizu, Y.~Yamaguchi, and M.~Harada,
  \ifthenelse{\boolean{articletitles}}{\emph{{Heavy quark spin multiplet
  structure of $P_c(4312)$, $P_c(4440)$, and $P_c(4457)$}},
  }{}\href{http://arxiv.org/abs/1904.00587}{{\normalfont\ttfamily
  arXiv:1904.00587}}\relax
\mciteBstWouldAddEndPuncttrue
\mciteSetBstMidEndSepPunct{\mcitedefaultmidpunct}
{\mcitedefaultendpunct}{\mcitedefaultseppunct}\relax
\EndOfBibitem
\bibitem{Guo:2019kdc}
Z.-H. Guo and J.~A. Oller, \ifthenelse{\boolean{articletitles}}{\emph{{Anatomy
  of the newly observed hidden-charm pentaquark states: $P_c(4312)$,
  $P_c(4440)$ and $P_c(4457)$}},
  }{}\href{http://dx.doi.org/10.1016/j.physletb.2019.04.053}{Phys.\ Lett.\ B
  \textbf{793} (2019) 144},
  \href{http://arxiv.org/abs/1904.00851}{{\normalfont\ttfamily
  arXiv:1904.00851}}\relax
\mciteBstWouldAddEndPuncttrue
\mciteSetBstMidEndSepPunct{\mcitedefaultmidpunct}
{\mcitedefaultendpunct}{\mcitedefaultseppunct}\relax
\EndOfBibitem
\bibitem{Xiao:2019mvs}
C.-J. Xiao {\em et~al.}, \ifthenelse{\boolean{articletitles}}{\emph{{Exploring
  the molecular scenario of Pc(4312) , Pc(4440) , and Pc(4457)}},
  }{}\href{http://dx.doi.org/10.1103/PhysRevD.100.014022}{Phys.\ Rev.\ D
  \textbf{100} (2019), no.~1 014022},
  \href{http://arxiv.org/abs/1904.00872}{{\normalfont\ttfamily
  arXiv:1904.00872}}\relax
\mciteBstWouldAddEndPuncttrue
\mciteSetBstMidEndSepPunct{\mcitedefaultmidpunct}
{\mcitedefaultendpunct}{\mcitedefaultseppunct}\relax
\EndOfBibitem
\bibitem{Meng:2019ilv}
L.~Meng, B.~Wang, G.-J. Wang, and S.-L. Zhu,
  \ifthenelse{\boolean{articletitles}}{\emph{{The hidden charm pentaquark
  states and $\Sigma_c\bar{D}^{(*)}$ interaction in chiral perturbation
  theory}}, }{}\href{http://dx.doi.org/10.1103/PhysRevD.100.014031}{Phys.\
  Rev.\ D \textbf{100} (2019), no.~1 014031},
  \href{http://arxiv.org/abs/1905.04113}{{\normalfont\ttfamily
  arXiv:1905.04113}}\relax
\mciteBstWouldAddEndPuncttrue
\mciteSetBstMidEndSepPunct{\mcitedefaultmidpunct}
{\mcitedefaultendpunct}{\mcitedefaultseppunct}\relax
\EndOfBibitem
\bibitem{Wu:2019rog}
Q.~Wu and D.-Y. Chen, \ifthenelse{\boolean{articletitles}}{\emph{{Production of
  $P_c $ states from $\Lambda_b$ decay}},
  }{}\href{http://dx.doi.org/10.1103/PhysRevD.100.114002}{Phys.\ Rev.\ D
  \textbf{100} (2019), no.~11 114002},
  \href{http://arxiv.org/abs/1906.02480}{{\normalfont\ttfamily
  arXiv:1906.02480}}\relax
\mciteBstWouldAddEndPuncttrue
\mciteSetBstMidEndSepPunct{\mcitedefaultmidpunct}
{\mcitedefaultendpunct}{\mcitedefaultseppunct}\relax
\EndOfBibitem
\bibitem{Shen:2019evi}
C.-W. Shen, J.-J. Wu, and B.-S. Zou,
  \ifthenelse{\boolean{articletitles}}{\emph{{Decay behaviors of possible
  $\Lambda_{c\bar{c}}$ states in hadronic molecule pictures}},
  }{}\href{http://dx.doi.org/10.1103/PhysRevD.100.056006}{Phys.\ Rev.\ D
  \textbf{100} (2019), no.~5 056006},
  \href{http://arxiv.org/abs/1906.03896}{{\normalfont\ttfamily
  arXiv:1906.03896}}\relax
\mciteBstWouldAddEndPuncttrue
\mciteSetBstMidEndSepPunct{\mcitedefaultmidpunct}
{\mcitedefaultendpunct}{\mcitedefaultseppunct}\relax
\EndOfBibitem
\bibitem{Xiao:2019gjd}
C.~W. Xiao, J.~Nieves, and E.~Oset,
  \ifthenelse{\boolean{articletitles}}{\emph{{Prediction of hidden charm
  strange molecular baryon states with heavy quark spin symmetry}},
  }{}\href{http://dx.doi.org/10.1016/j.physletb.2019.135051}{Phys.\ Lett.\ B
  \textbf{799} (2019) 135051},
  \href{http://arxiv.org/abs/1906.09010}{{\normalfont\ttfamily
  arXiv:1906.09010}}\relax
\mciteBstWouldAddEndPuncttrue
\mciteSetBstMidEndSepPunct{\mcitedefaultmidpunct}
{\mcitedefaultendpunct}{\mcitedefaultseppunct}\relax
\EndOfBibitem
\bibitem{Voloshin:2019aut}
M.~B. Voloshin, \ifthenelse{\boolean{articletitles}}{\emph{{Some decay
  properties of hidden-charm pentaquarks as baryon-meson molecules}},
  }{}\href{http://dx.doi.org/10.1103/PhysRevD.100.034020}{Phys.\ Rev.\ D
  \textbf{100} (2019), no.~3 034020},
  \href{http://arxiv.org/abs/1907.01476}{{\normalfont\ttfamily
  arXiv:1907.01476}}\relax
\mciteBstWouldAddEndPuncttrue
\mciteSetBstMidEndSepPunct{\mcitedefaultmidpunct}
{\mcitedefaultendpunct}{\mcitedefaultseppunct}\relax
\EndOfBibitem
\bibitem{Sakai:2019qph}
S.~Sakai, H.-J. Jing, and F.-K. Guo,
  \ifthenelse{\boolean{articletitles}}{\emph{{Decays of $P_c$ into $J/\psi N$
  and $\eta_cN$ with heavy quark spin symmetry}},
  }{}\href{http://dx.doi.org/10.1103/PhysRevD.100.074007}{Phys.\ Rev.\ D
  \textbf{100} (2019), no.~7 074007},
  \href{http://arxiv.org/abs/1907.03414}{{\normalfont\ttfamily
  arXiv:1907.03414}}\relax
\mciteBstWouldAddEndPuncttrue
\mciteSetBstMidEndSepPunct{\mcitedefaultmidpunct}
{\mcitedefaultendpunct}{\mcitedefaultseppunct}\relax
\EndOfBibitem
\bibitem{Wang:2019hyc}
Z.-G. Wang and X.~Wang, \ifthenelse{\boolean{articletitles}}{\emph{{Analysis of
  the strong decays of the $P_c(4312)$ as a pentaquark molecular state with QCD
  sum rules}}, }{}\href{http://dx.doi.org/10.1088/1674-1137/ababf7}{Chin.\
  Phys.\ C \textbf{44} (2020) 103102},
  \href{http://arxiv.org/abs/1907.04582}{{\normalfont\ttfamily
  arXiv:1907.04582}}\relax
\mciteBstWouldAddEndPuncttrue
\mciteSetBstMidEndSepPunct{\mcitedefaultmidpunct}
{\mcitedefaultendpunct}{\mcitedefaultseppunct}\relax
\EndOfBibitem
\bibitem{Yamaguchi:2019seo}
Y.~Yamaguchi {\em et~al.}, \ifthenelse{\boolean{articletitles}}{\emph{{$P_c$
  pentaquarks with chiral tensor and quark dynamics}},
  }{}\href{http://dx.doi.org/10.1103/PhysRevD.101.091502}{Phys.\ Rev.\ D
  \textbf{101} (2020), no.~9 091502},
  \href{http://arxiv.org/abs/1907.04684}{{\normalfont\ttfamily
  arXiv:1907.04684}}\relax
\mciteBstWouldAddEndPuncttrue
\mciteSetBstMidEndSepPunct{\mcitedefaultmidpunct}
{\mcitedefaultendpunct}{\mcitedefaultseppunct}\relax
\EndOfBibitem
\bibitem{Xu:2019zme}
Y.-J. Xu, C.-Y. Cui, Y.-L. Liu, and M.-Q. Huang,
  \ifthenelse{\boolean{articletitles}}{\emph{{Partial decay widths of
  $P_{c}(4312)$ as a $\bar{D}\Sigma_{c}$ molecular state}},
  }{}\href{http://dx.doi.org/10.1103/PhysRevD.102.034028}{Phys.\ Rev.\ D
  \textbf{102} (2020), no.~3 034028},
  \href{http://arxiv.org/abs/1907.05097}{{\normalfont\ttfamily
  arXiv:1907.05097}}\relax
\mciteBstWouldAddEndPuncttrue
\mciteSetBstMidEndSepPunct{\mcitedefaultmidpunct}
{\mcitedefaultendpunct}{\mcitedefaultseppunct}\relax
\EndOfBibitem
\bibitem{PavonValderrama:2019nbk}
M.~Pavon~Valderrama, \ifthenelse{\boolean{articletitles}}{\emph{{One pion
  exchange and the quantum numbers of the P$_c$(4440) and P$_c$(4457)
  pentaquarks}}, }{}\href{http://dx.doi.org/10.1103/PhysRevD.100.094028}{Phys.\
  Rev.\ D \textbf{100} (2019), no.~9 094028},
  \href{http://arxiv.org/abs/1907.05294}{{\normalfont\ttfamily
  arXiv:1907.05294}}\relax
\mciteBstWouldAddEndPuncttrue
\mciteSetBstMidEndSepPunct{\mcitedefaultmidpunct}
{\mcitedefaultendpunct}{\mcitedefaultseppunct}\relax
\EndOfBibitem
\bibitem{Peng:2019wys}
F.-Z. Peng {\em et~al.},
  \ifthenelse{\boolean{articletitles}}{\emph{{Five-flavor pentaquarks and other
  light- and heavy-flavor symmetry partners of the LHCb hidden-charm
  pentaquarks}},
  }{}\href{http://dx.doi.org/10.1016/j.nuclphysb.2022.115936}{Nucl.\ Phys.\ B
  \textbf{983} (2022) 115936},
  \href{http://arxiv.org/abs/1907.05322}{{\normalfont\ttfamily
  arXiv:1907.05322}}\relax
\mciteBstWouldAddEndPuncttrue
\mciteSetBstMidEndSepPunct{\mcitedefaultmidpunct}
{\mcitedefaultendpunct}{\mcitedefaultseppunct}\relax
\EndOfBibitem
\bibitem{Liu:2019zvb}
M.-Z. Liu {\em et~al.},
  \ifthenelse{\boolean{articletitles}}{\emph{{Spin-parities of the $P_c(4440)$
  and $P_c(4457)$ in the one-boson-exchange model}},
  }{}\href{http://dx.doi.org/10.1103/PhysRevD.103.054004}{Phys.\ Rev.\ D
  \textbf{103} (2021), no.~5 054004},
  \href{http://arxiv.org/abs/1907.06093}{{\normalfont\ttfamily
  arXiv:1907.06093}}\relax
\mciteBstWouldAddEndPuncttrue
\mciteSetBstMidEndSepPunct{\mcitedefaultmidpunct}
{\mcitedefaultendpunct}{\mcitedefaultseppunct}\relax
\EndOfBibitem
\bibitem{Pan:2019skd}
Y.-W. Pan {\em et~al.}, \ifthenelse{\boolean{articletitles}}{\emph{{Model
  independent determination of the spins of the $P_{c}$(4440) and $P_{c}$(4457)
  from the spectroscopy of the triply charmed dibaryons}},
  }{}\href{http://dx.doi.org/10.1103/PhysRevD.102.011504}{Phys.\ Rev.\ D
  \textbf{102} (2020), no.~1 011504},
  \href{http://arxiv.org/abs/1907.11220}{{\normalfont\ttfamily
  arXiv:1907.11220}}\relax
\mciteBstWouldAddEndPuncttrue
\mciteSetBstMidEndSepPunct{\mcitedefaultmidpunct}
{\mcitedefaultendpunct}{\mcitedefaultseppunct}\relax
\EndOfBibitem
\bibitem{Burns:2019iih}
T.~J. Burns and E.~S. Swanson,
  \ifthenelse{\boolean{articletitles}}{\emph{{Molecular interpretation of the
  $P_c$(4440) and $P_c$(4457) states}},
  }{}\href{http://dx.doi.org/10.1103/PhysRevD.100.114033}{Phys.\ Rev.\ D
  \textbf{100} (2019), no.~11 114033},
  \href{http://arxiv.org/abs/1908.03528}{{\normalfont\ttfamily
  arXiv:1908.03528}}\relax
\mciteBstWouldAddEndPuncttrue
\mciteSetBstMidEndSepPunct{\mcitedefaultmidpunct}
{\mcitedefaultendpunct}{\mcitedefaultseppunct}\relax
\EndOfBibitem
\bibitem{Guo:2015umn}
F.-K. Guo, U.-G. Mei{\ss}ner, W.~Wang, and Z.~Yang,
  \ifthenelse{\boolean{articletitles}}{\emph{{How to reveal the exotic nature
  of the P$_c$(4450)}},
  }{}\href{http://dx.doi.org/10.1103/PhysRevD.92.071502}{Phys.\ Rev.\
  \textbf{D92} (2015) 071502},
  \href{http://arxiv.org/abs/1507.04950}{{\normalfont\ttfamily
  arXiv:1507.04950}}\relax
\mciteBstWouldAddEndPuncttrue
\mciteSetBstMidEndSepPunct{\mcitedefaultmidpunct}
{\mcitedefaultendpunct}{\mcitedefaultseppunct}\relax
\EndOfBibitem
\bibitem{Meissner:2015mza}
U.-G. Mei{\ss}ner and J.~A. Oller,
  \ifthenelse{\boolean{articletitles}}{\emph{{Testing the $\chi_{c1}\, p$
  composite nature of the $P_c(4450)$}},
  }{}\href{http://dx.doi.org/10.1016/j.physletb.2015.10.015}{Phys.\ Lett.\
  \textbf{B751} (2015) 59},
  \href{http://arxiv.org/abs/1507.07478}{{\normalfont\ttfamily
  arXiv:1507.07478}}\relax
\mciteBstWouldAddEndPuncttrue
\mciteSetBstMidEndSepPunct{\mcitedefaultmidpunct}
{\mcitedefaultendpunct}{\mcitedefaultseppunct}\relax
\EndOfBibitem
\bibitem{Liu:2015fea}
X.-H. Liu, Q.~Wang, and Q.~Zhao,
  \ifthenelse{\boolean{articletitles}}{\emph{{Understanding the newly observed
  heavy pentaquark candidates}},
  }{}\href{http://dx.doi.org/10.1016/j.physletb.2016.03.089}{Phys.\ Lett.\
  \textbf{B757} (2016) 231},
  \href{http://arxiv.org/abs/1507.05359}{{\normalfont\ttfamily
  arXiv:1507.05359}}\relax
\mciteBstWouldAddEndPuncttrue
\mciteSetBstMidEndSepPunct{\mcitedefaultmidpunct}
{\mcitedefaultendpunct}{\mcitedefaultseppunct}\relax
\EndOfBibitem
\bibitem{Mikhasenko:2015vca}
M.~Mikhasenko, \ifthenelse{\boolean{articletitles}}{\emph{{A triangle
  singularity and the LHCb pentaquarks}},
  }{}\href{http://arxiv.org/abs/1507.06552}{{\normalfont\ttfamily
  arXiv:1507.06552}}\relax
\mciteBstWouldAddEndPuncttrue
\mciteSetBstMidEndSepPunct{\mcitedefaultmidpunct}
{\mcitedefaultendpunct}{\mcitedefaultseppunct}\relax
\EndOfBibitem
\bibitem{Eides:2019tgv}
M.~I. Eides, V.~Y. Petrov, and M.~V. Polyakov,
  \ifthenelse{\boolean{articletitles}}{\emph{{New LHCb pentaquarks as
  hadrocharmonium states}},
  }{}\href{http://dx.doi.org/10.1142/S0217732320501515}{Mod.\ Phys.\ Lett.\ A
  \textbf{35} (2020), no.~18 2050151},
  \href{http://arxiv.org/abs/1904.11616}{{\normalfont\ttfamily
  arXiv:1904.11616}}\relax
\mciteBstWouldAddEndPuncttrue
\mciteSetBstMidEndSepPunct{\mcitedefaultmidpunct}
{\mcitedefaultendpunct}{\mcitedefaultseppunct}\relax
\EndOfBibitem
\bibitem{Wang:2806799}
M.~Wang, {\em {Amplitude analysis of the $\Lambda_b^0 \to J/\psi p K^-$ decay
  and first observation of the $\Lambda_b^0 \to \eta_c(1S) p K^-$ decay}}, PhD
  thesis, Tsinghua U., 2021,
  \href{https://cds.cern.ch/record/2806799}{CERN-THESIS-2021-314}\relax
\mciteBstWouldAddEndPuncttrue
\mciteSetBstMidEndSepPunct{\mcitedefaultmidpunct}
{\mcitedefaultendpunct}{\mcitedefaultseppunct}\relax
\EndOfBibitem
\bibitem{Aaij:2016ymb}
LHCb, R.~Aaij {\em et~al.},
  \ifthenelse{\boolean{articletitles}}{\emph{{Evidence for exotic hadron
  contributions to $\Lambda_b^0 \to J/\psi p \pi^-$ decays}},
  }{}\href{http://dx.doi.org/10.1103/PhysRevLett.117.082003,
  10.1103/PhysRevLett.117.109902}{Phys.\ Rev.\ Lett.\  \textbf{117} (2016)
  082003}, \href{http://arxiv.org/abs/1606.06999}{{\normalfont\ttfamily
  arXiv:1606.06999}}, [Addendum: Phys. Rev. Lett.117 (2016) 109902]\relax
\mciteBstWouldAddEndPuncttrue
\mciteSetBstMidEndSepPunct{\mcitedefaultmidpunct}
{\mcitedefaultendpunct}{\mcitedefaultseppunct}\relax
\EndOfBibitem
\bibitem{LHCb:2019rmd}
LHCb, R.~Aaij {\em et~al.},
  \ifthenelse{\boolean{articletitles}}{\emph{{Observation of $B^0_{(s)} \to
  J/\psi p \overline{p}$ decays and precision measurements of the $B^0_{(s)}$
  masses}}, }{}\href{http://dx.doi.org/10.1103/PhysRevLett.122.191804}{Phys.\
  Rev.\ Lett.\  \textbf{122} (2019), no.~19 191804},
  \href{http://arxiv.org/abs/1902.05588}{{\normalfont\ttfamily
  arXiv:1902.05588}}\relax
\mciteBstWouldAddEndPuncttrue
\mciteSetBstMidEndSepPunct{\mcitedefaultmidpunct}
{\mcitedefaultendpunct}{\mcitedefaultseppunct}\relax
\EndOfBibitem
\bibitem{Hsiao:2014tda}
Y.~K. Hsiao and C.~Q. Geng,
  \ifthenelse{\boolean{articletitles}}{\emph{{$f_J(2220)$ and hadronic
  $\bar{B}^0_s$ decays}},
  }{}\href{http://dx.doi.org/10.1140/epjc/s10052-015-3317-9}{Eur.\ Phys.\ J.\ C
  \textbf{75} (2015), no.~3 101},
  \href{http://arxiv.org/abs/1412.4900}{{\normalfont\ttfamily
  arXiv:1412.4900}}\relax
\mciteBstWouldAddEndPuncttrue
\mciteSetBstMidEndSepPunct{\mcitedefaultmidpunct}
{\mcitedefaultendpunct}{\mcitedefaultseppunct}\relax
\EndOfBibitem
\bibitem{JPAC:2019ufm}
JPAC, M.~Mikhasenko {\em et~al.},
  \ifthenelse{\boolean{articletitles}}{\emph{{Dalitz-plot decomposition for
  three-body decays}},
  }{}\href{http://dx.doi.org/10.1103/PhysRevD.101.034033}{Phys.\ Rev.\ D
  \textbf{101} (2020), no.~3 034033},
  \href{http://arxiv.org/abs/1910.04566}{{\normalfont\ttfamily
  arXiv:1910.04566}}\relax
\mciteBstWouldAddEndPuncttrue
\mciteSetBstMidEndSepPunct{\mcitedefaultmidpunct}
{\mcitedefaultendpunct}{\mcitedefaultseppunct}\relax
\EndOfBibitem
\bibitem{Karliner:2022erb}
M.~Karliner and J.~L. Rosner, \ifthenelse{\boolean{articletitles}}{\emph{{New
  strange pentaquarks}},
  }{}\href{http://dx.doi.org/10.1103/PhysRevD.106.036024}{Phys.\ Rev.\ D
  \textbf{106} (2022), no.~3 036024},
  \href{http://arxiv.org/abs/2207.07581}{{\normalfont\ttfamily
  arXiv:2207.07581}}\relax
\mciteBstWouldAddEndPuncttrue
\mciteSetBstMidEndSepPunct{\mcitedefaultmidpunct}
{\mcitedefaultendpunct}{\mcitedefaultseppunct}\relax
\EndOfBibitem
\bibitem{Choi:2007wga}
Belle, S.~K. Choi {\em et~al.},
  \ifthenelse{\boolean{articletitles}}{\emph{{Observation of a resonance-like
  structure in the $\pi^\pm\psi'$ mass distribution in exclusive $B\to
  K\pi^\pm\psi'$ decays}},
  }{}\href{http://dx.doi.org/10.1103/PhysRevLett.100.142001}{Phys.\ Rev.\
  Lett.\  \textbf{100} (2008) 142001},
  \href{http://arxiv.org/abs/0708.1790}{{\normalfont\ttfamily
  arXiv:0708.1790}}\relax
\mciteBstWouldAddEndPuncttrue
\mciteSetBstMidEndSepPunct{\mcitedefaultmidpunct}
{\mcitedefaultendpunct}{\mcitedefaultseppunct}\relax
\EndOfBibitem
\bibitem{BaBar:2008bxw}
BaBar, B.~Aubert {\em et~al.},
  \ifthenelse{\boolean{articletitles}}{\emph{{Search for the $Z(4430)^-$ at
  BABAR}}, }{}\href{http://dx.doi.org/10.1103/PhysRevD.79.112001}{Phys.\ Rev.\
  D \textbf{79} (2009) 112001},
  \href{http://arxiv.org/abs/0811.0564}{{\normalfont\ttfamily
  arXiv:0811.0564}}\relax
\mciteBstWouldAddEndPuncttrue
\mciteSetBstMidEndSepPunct{\mcitedefaultmidpunct}
{\mcitedefaultendpunct}{\mcitedefaultseppunct}\relax
\EndOfBibitem
\bibitem{LHCb:2014zfx}
LHCb, R.~Aaij {\em et~al.},
  \ifthenelse{\boolean{articletitles}}{\emph{{Observation of the resonant
  character of the $Z(4430)^-$ state}},
  }{}\href{http://dx.doi.org/10.1103/PhysRevLett.112.222002}{Phys.\ Rev.\
  Lett.\  \textbf{112} (2014), no.~22 222002},
  \href{http://arxiv.org/abs/1404.1903}{{\normalfont\ttfamily
  arXiv:1404.1903}}\relax
\mciteBstWouldAddEndPuncttrue
\mciteSetBstMidEndSepPunct{\mcitedefaultmidpunct}
{\mcitedefaultendpunct}{\mcitedefaultseppunct}\relax
\EndOfBibitem
\bibitem{Chilikin:2013tch}
Belle, K.~Chilikin {\em et~al.},
  \ifthenelse{\boolean{articletitles}}{\emph{{Experimental constraints on the
  spin and parity of the $Z(4430)^+$}},
  }{}\href{http://dx.doi.org/10.1103/PhysRevD.88.074026}{Phys.\ Rev.\
  \textbf{D88} (2013) 074026},
  \href{http://arxiv.org/abs/1306.4894}{{\normalfont\ttfamily
  arXiv:1306.4894}}\relax
\mciteBstWouldAddEndPuncttrue
\mciteSetBstMidEndSepPunct{\mcitedefaultmidpunct}
{\mcitedefaultendpunct}{\mcitedefaultseppunct}\relax
\EndOfBibitem
\bibitem{Belle:2014nuw}
Belle, K.~Chilikin {\em et~al.},
  \ifthenelse{\boolean{articletitles}}{\emph{{Observation of a new charged
  charmoniumlike state in $\bar{B}^0\to\jpsi K^-\pi^+$ decays}},
  }{}\href{http://dx.doi.org/10.1103/PhysRevD.90.112009}{Phys.\ Rev.\ D
  \textbf{90} (2014), no.~11 112009},
  \href{http://arxiv.org/abs/1408.6457}{{\normalfont\ttfamily
  arXiv:1408.6457}}\relax
\mciteBstWouldAddEndPuncttrue
\mciteSetBstMidEndSepPunct{\mcitedefaultmidpunct}
{\mcitedefaultendpunct}{\mcitedefaultseppunct}\relax
\EndOfBibitem
\bibitem{LHCb:2015sqg}
LHCb, R.~Aaij {\em et~al.},
  \ifthenelse{\boolean{articletitles}}{\emph{{Model-independent confirmation of
  the $Z(4430)^-$ state}},
  }{}\href{http://dx.doi.org/10.1103/PhysRevD.92.112009}{Phys.\ Rev.\ D
  \textbf{92} (2015), no.~11 112009},
  \href{http://arxiv.org/abs/1510.01951}{{\normalfont\ttfamily
  arXiv:1510.01951}}\relax
\mciteBstWouldAddEndPuncttrue
\mciteSetBstMidEndSepPunct{\mcitedefaultmidpunct}
{\mcitedefaultendpunct}{\mcitedefaultseppunct}\relax
\EndOfBibitem
\bibitem{LHCb:2019maw}
LHCb, R.~Aaij {\em et~al.},
  \ifthenelse{\boolean{articletitles}}{\emph{{Model-Independent Observation of
  Exotic Contributions to $B^0\to J/\psi K^+\pi^-$ Decays}},
  }{}\href{http://dx.doi.org/10.1103/PhysRevLett.122.152002}{Phys.\ Rev.\
  Lett.\  \textbf{122} (2019), no.~15 152002},
  \href{http://arxiv.org/abs/1901.05745}{{\normalfont\ttfamily
  arXiv:1901.05745}}\relax
\mciteBstWouldAddEndPuncttrue
\mciteSetBstMidEndSepPunct{\mcitedefaultmidpunct}
{\mcitedefaultendpunct}{\mcitedefaultseppunct}\relax
\EndOfBibitem
\bibitem{Beiter:2854734}
A.~Beiter, {\em {Amplitude analysis of $B^0$ decays to $J/\psi\pi^-K^+$ and
  $\psi(2S)\pi^-K^+$}}, PhD thesis, Syracuse U., 2023,
  \href{http://cds.cern.ch/record/2854734}{CERN-THESIS-2023-027}\relax
\mciteBstWouldAddEndPuncttrue
\mciteSetBstMidEndSepPunct{\mcitedefaultmidpunct}
{\mcitedefaultendpunct}{\mcitedefaultseppunct}\relax
\EndOfBibitem
\bibitem{CDF:2009jgo}
CDF, T.~Aaltonen {\em et~al.},
  \ifthenelse{\boolean{articletitles}}{\emph{{Evidence for a Narrow
  Near-Threshold Structure in the $J/\psi\phi$ Mass Spectrum in $B^+\to
  J/\psi\phi K^+$ Decays}},
  }{}\href{http://dx.doi.org/10.1103/PhysRevLett.102.242002}{Phys.\ Rev.\
  Lett.\  \textbf{102} (2009) 242002},
  \href{http://arxiv.org/abs/0903.2229}{{\normalfont\ttfamily
  arXiv:0903.2229}}\relax
\mciteBstWouldAddEndPuncttrue
\mciteSetBstMidEndSepPunct{\mcitedefaultmidpunct}
{\mcitedefaultendpunct}{\mcitedefaultseppunct}\relax
\EndOfBibitem
\bibitem{Cheng-Ping:2009sgk}
Belle, S.~Cheng-Ping, \ifthenelse{\boolean{articletitles}}{\emph{{XYZ particles
  at Belle}}, }{}\href{http://dx.doi.org/10.1088/1674-1137/34/6/001}{Chin.\
  Phys.\ C \textbf{34} (2010) 615},
  \href{http://arxiv.org/abs/0912.2386}{{\normalfont\ttfamily
  arXiv:0912.2386}}\relax
\mciteBstWouldAddEndPuncttrue
\mciteSetBstMidEndSepPunct{\mcitedefaultmidpunct}
{\mcitedefaultendpunct}{\mcitedefaultseppunct}\relax
\EndOfBibitem
\bibitem{CDF:2011pep}
CDF, T.~Aaltonen {\em et~al.},
  \ifthenelse{\boolean{articletitles}}{\emph{{Observation of the $Y(4140)$
  Structure in the $J/\psi\phi$ Mass Spectrum in $B^\pm\to J/\psi\phi K^\pm$
  Decays}}, }{}\href{http://dx.doi.org/10.1142/S0217732317501395}{Mod.\ Phys.\
  Lett.\ A \textbf{32} (2017), no.~26 1750139},
  \href{http://arxiv.org/abs/1101.6058}{{\normalfont\ttfamily
  arXiv:1101.6058}}\relax
\mciteBstWouldAddEndPuncttrue
\mciteSetBstMidEndSepPunct{\mcitedefaultmidpunct}
{\mcitedefaultendpunct}{\mcitedefaultseppunct}\relax
\EndOfBibitem
\bibitem{LHCb:2012wyi}
LHCb, R.~Aaij {\em et~al.}, \ifthenelse{\boolean{articletitles}}{\emph{{Search
  for the $X(4140)$ state in $B^+ \to J/\psi \phi K^+$ decays}},
  }{}\href{http://dx.doi.org/10.1103/PhysRevD.85.091103}{Phys.\ Rev.\ D
  \textbf{85} (2012) 091103},
  \href{http://arxiv.org/abs/1202.5087}{{\normalfont\ttfamily
  arXiv:1202.5087}}\relax
\mciteBstWouldAddEndPuncttrue
\mciteSetBstMidEndSepPunct{\mcitedefaultmidpunct}
{\mcitedefaultendpunct}{\mcitedefaultseppunct}\relax
\EndOfBibitem
\bibitem{BaBar:2014wwp}
BaBar, J.~P. Lees {\em et~al.},
  \ifthenelse{\boolean{articletitles}}{\emph{{Study of $B^{\pm,0} \to J/\psi
  K^+ K^- K^{\pm,0}$ and search for $B^0 \to J/\psi\phi$ at BABAR}},
  }{}\href{http://dx.doi.org/10.1103/PhysRevD.91.012003}{Phys.\ Rev.\ D
  \textbf{91} (2015), no.~1 012003},
  \href{http://arxiv.org/abs/1407.7244}{{\normalfont\ttfamily
  arXiv:1407.7244}}\relax
\mciteBstWouldAddEndPuncttrue
\mciteSetBstMidEndSepPunct{\mcitedefaultmidpunct}
{\mcitedefaultendpunct}{\mcitedefaultseppunct}\relax
\EndOfBibitem
\bibitem{CMS:2013jru}
CMS, S.~Chatrchyan {\em et~al.},
  \ifthenelse{\boolean{articletitles}}{\emph{{Observation of a Peaking
  Structure in the $J/\psi \phi$ Mass Spectrum from $B^{\pm} \to J/\psi \phi
  K^{\pm}$ Decays}},
  }{}\href{http://dx.doi.org/10.1016/j.physletb.2014.05.055}{Phys.\ Lett.\ B
  \textbf{734} (2014) 261},
  \href{http://arxiv.org/abs/1309.6920}{{\normalfont\ttfamily
  arXiv:1309.6920}}\relax
\mciteBstWouldAddEndPuncttrue
\mciteSetBstMidEndSepPunct{\mcitedefaultmidpunct}
{\mcitedefaultendpunct}{\mcitedefaultseppunct}\relax
\EndOfBibitem
\bibitem{D0:2015nxw}
D0, V.~M. Abazov {\em et~al.},
  \ifthenelse{\boolean{articletitles}}{\emph{{Inclusive Production of the
  X(4140) State in $p \overline p$ Collisions at D0}},
  }{}\href{http://dx.doi.org/10.1103/PhysRevLett.115.232001}{Phys.\ Rev.\
  Lett.\  \textbf{115} (2015), no.~23 232001},
  \href{http://arxiv.org/abs/1508.07846}{{\normalfont\ttfamily
  arXiv:1508.07846}}\relax
\mciteBstWouldAddEndPuncttrue
\mciteSetBstMidEndSepPunct{\mcitedefaultmidpunct}
{\mcitedefaultendpunct}{\mcitedefaultseppunct}\relax
\EndOfBibitem
\bibitem{D0:2013jvp}
D0, V.~M. Abazov {\em et~al.},
  \ifthenelse{\boolean{articletitles}}{\emph{{Search for the $X$(4140) state in
  $\Bp\to\jpsi\phi\Kp$ decays with the D0 Detector}},
  }{}\href{http://dx.doi.org/10.1103/PhysRevD.89.012004}{Phys.\ Rev.\ D
  \textbf{89} (2014), no.~1 012004},
  \href{http://arxiv.org/abs/1309.6580}{{\normalfont\ttfamily
  arXiv:1309.6580}}\relax
\mciteBstWouldAddEndPuncttrue
\mciteSetBstMidEndSepPunct{\mcitedefaultmidpunct}
{\mcitedefaultendpunct}{\mcitedefaultseppunct}\relax
\EndOfBibitem
\bibitem{LHCb:2016nsl}
LHCb, R.~Aaij {\em et~al.},
  \ifthenelse{\boolean{articletitles}}{\emph{{Amplitude analysis of $B^+\to
  J/\psi \phi K^+$ decays}},
  }{}\href{http://dx.doi.org/10.1103/PhysRevD.95.012002}{Phys.\ Rev.\ D
  \textbf{95} (2017), no.~1 012002},
  \href{http://arxiv.org/abs/1606.07898}{{\normalfont\ttfamily
  arXiv:1606.07898}}\relax
\mciteBstWouldAddEndPuncttrue
\mciteSetBstMidEndSepPunct{\mcitedefaultmidpunct}
{\mcitedefaultendpunct}{\mcitedefaultseppunct}\relax
\EndOfBibitem
\bibitem{LHCb:2016axx}
LHCb, R.~Aaij {\em et~al.},
  \ifthenelse{\boolean{articletitles}}{\emph{{Observation of $J/\psi\phi$
  structures consistent with exotic states from amplitude analysis of $B^+\to
  J/\psi \phi K^+$ decays}},
  }{}\href{http://dx.doi.org/10.1103/PhysRevLett.118.022003}{Phys.\ Rev.\
  Lett.\  \textbf{118} (2017), no.~2 022003},
  \href{http://arxiv.org/abs/1606.07895}{{\normalfont\ttfamily
  arXiv:1606.07895}}\relax
\mciteBstWouldAddEndPuncttrue
\mciteSetBstMidEndSepPunct{\mcitedefaultmidpunct}
{\mcitedefaultendpunct}{\mcitedefaultseppunct}\relax
\EndOfBibitem
\bibitem{LHCb:2021uow}
LHCb, R.~Aaij {\em et~al.},
  \ifthenelse{\boolean{articletitles}}{\emph{{Observation of New Resonances
  Decaying to $\jpsi\Kp$ and $\jpsi\phi$}},
  }{}\href{http://dx.doi.org/10.1103/PhysRevLett.127.082001}{Phys.\ Rev.\
  Lett.\  \textbf{127} (2021), no.~8 082001},
  \href{http://arxiv.org/abs/2103.01803}{{\normalfont\ttfamily
  arXiv:2103.01803}}\relax
\mciteBstWouldAddEndPuncttrue
\mciteSetBstMidEndSepPunct{\mcitedefaultmidpunct}
{\mcitedefaultendpunct}{\mcitedefaultseppunct}\relax
\EndOfBibitem
\bibitem{LHCb:2023hxg}
LHCb, R.~Aaij {\em et~al.},
  \ifthenelse{\boolean{articletitles}}{\emph{{Evidence of a $\jpsi\KS$
  Structure in $\Bd\to\jpsi\phi\KS$ Decays}},
  }{}\href{http://dx.doi.org/10.1103/PhysRevLett.131.131901}{Phys.\ Rev.\
  Lett.\  \textbf{131} (2023), no.~13 131901},
  \href{http://arxiv.org/abs/2301.04899}{{\normalfont\ttfamily
  arXiv:2301.04899}}\relax
\mciteBstWouldAddEndPuncttrue
\mciteSetBstMidEndSepPunct{\mcitedefaultmidpunct}
{\mcitedefaultendpunct}{\mcitedefaultseppunct}\relax
\EndOfBibitem
\bibitem{LHCb:2022dvn}
LHCb, R.~Aaij {\em et~al.}, \ifthenelse{\boolean{articletitles}}{\emph{{First
  observation of the $\Bp\to\Dsp\Dsm\Kp$ decay}},
  }{}\href{http://dx.doi.org/10.1103/PhysRevD.108.034012}{Phys.\ Rev.\ D
  \textbf{108} (2023) 034012},
  \href{http://arxiv.org/abs/2211.05034}{{\normalfont\ttfamily
  arXiv:2211.05034}}\relax
\mciteBstWouldAddEndPuncttrue
\mciteSetBstMidEndSepPunct{\mcitedefaultmidpunct}
{\mcitedefaultendpunct}{\mcitedefaultseppunct}\relax
\EndOfBibitem
\bibitem{LHCb:2022aki}
LHCb, R.~Aaij {\em et~al.},
  \ifthenelse{\boolean{articletitles}}{\emph{{Observation of a Resonant
  Structure near the $\Dsp\Dsm$ Threshold in the $\Bp\to\Dsp\Dsm\Kp$ Decay}},
  }{}\href{http://dx.doi.org/10.1103/PhysRevLett.131.071901}{Phys.\ Rev.\
  Lett.\  \textbf{131} (2023), no.~7 071901},
  \href{http://arxiv.org/abs/2210.15153}{{\normalfont\ttfamily
  arXiv:2210.15153}}\relax
\mciteBstWouldAddEndPuncttrue
\mciteSetBstMidEndSepPunct{\mcitedefaultmidpunct}
{\mcitedefaultendpunct}{\mcitedefaultseppunct}\relax
\EndOfBibitem
\bibitem{Molina:2010tx}
R.~Molina, T.~Branz, and E.~Oset, \ifthenelse{\boolean{articletitles}}{\emph{{A
  new interpretation for the $D^*_{s2}(2573)$ and the prediction of novel
  exotic charmed mesons}},
  }{}\href{http://dx.doi.org/10.1103/PhysRevD.82.014010}{Phys.\ Rev.\ D
  \textbf{82} (2010) 014010},
  \href{http://arxiv.org/abs/1005.0335}{{\normalfont\ttfamily
  arXiv:1005.0335}}\relax
\mciteBstWouldAddEndPuncttrue
\mciteSetBstMidEndSepPunct{\mcitedefaultmidpunct}
{\mcitedefaultendpunct}{\mcitedefaultseppunct}\relax
\EndOfBibitem
\bibitem{Liu:2016ogz}
Y.-R. Liu, X.~Liu, and S.-L. Zhu,
  \ifthenelse{\boolean{articletitles}}{\emph{{$X(5568)$ and and its partner
  states}}, }{}\href{http://dx.doi.org/10.1103/PhysRevD.93.074023}{Phys.\ Rev.\
  D \textbf{93} (2016), no.~7 074023},
  \href{http://arxiv.org/abs/1603.01131}{{\normalfont\ttfamily
  arXiv:1603.01131}}\relax
\mciteBstWouldAddEndPuncttrue
\mciteSetBstMidEndSepPunct{\mcitedefaultmidpunct}
{\mcitedefaultendpunct}{\mcitedefaultseppunct}\relax
\EndOfBibitem
\bibitem{Cheng:2020nho}
J.-B. Cheng {\em et~al.}, \ifthenelse{\boolean{articletitles}}{\emph{{Spectrum
  and rearrangement decays of tetraquark states with four different flavors}},
  }{}\href{http://dx.doi.org/10.1103/PhysRevD.101.114017}{Phys.\ Rev.\ D
  \textbf{101} (2020), no.~11 114017},
  \href{http://arxiv.org/abs/2001.05287}{{\normalfont\ttfamily
  arXiv:2001.05287}}\relax
\mciteBstWouldAddEndPuncttrue
\mciteSetBstMidEndSepPunct{\mcitedefaultmidpunct}
{\mcitedefaultendpunct}{\mcitedefaultseppunct}\relax
\EndOfBibitem
\bibitem{LHCb:2020bls}
LHCb, R.~Aaij {\em et~al.}, \ifthenelse{\boolean{articletitles}}{\emph{{A
  model-independent study of resonant structure in ${\Bp\to\Dp\Dm\Kp}$
  decays}}, }{}\href{http://dx.doi.org/10.1103/PhysRevLett.125.242001}{Phys.\
  Rev.\ Lett.\  \textbf{125} (2020) 242001},
  \href{http://arxiv.org/abs/2009.00025}{{\normalfont\ttfamily
  arXiv:2009.00025}}\relax
\mciteBstWouldAddEndPuncttrue
\mciteSetBstMidEndSepPunct{\mcitedefaultmidpunct}
{\mcitedefaultendpunct}{\mcitedefaultseppunct}\relax
\EndOfBibitem
\bibitem{LHCb:2020pxc}
LHCb, R.~Aaij {\em et~al.},
  \ifthenelse{\boolean{articletitles}}{\emph{{Amplitude analysis of the
  ${\Bp\to\Dp\Dm\Kp}$ decay}},
  }{}\href{http://dx.doi.org/10.1103/PhysRevD.102.112003}{Phys.\ Rev.\ D
  \textbf{102} (2020) 112003},
  \href{http://arxiv.org/abs/2009.00026}{{\normalfont\ttfamily
  arXiv:2009.00026}}\relax
\mciteBstWouldAddEndPuncttrue
\mciteSetBstMidEndSepPunct{\mcitedefaultmidpunct}
{\mcitedefaultendpunct}{\mcitedefaultseppunct}\relax
\EndOfBibitem
\bibitem{Burns:2020xne}
T.~J. Burns and E.~S. Swanson,
  \ifthenelse{\boolean{articletitles}}{\emph{{Discriminating among
  interpretations for the $X(2900)$ states}},
  }{}\href{http://dx.doi.org/10.1103/PhysRevD.103.014004}{Phys.\ Rev.\ D
  \textbf{103} (2021), no.~1 014004},
  \href{http://arxiv.org/abs/2009.05352}{{\normalfont\ttfamily
  arXiv:2009.05352}}\relax
\mciteBstWouldAddEndPuncttrue
\mciteSetBstMidEndSepPunct{\mcitedefaultmidpunct}
{\mcitedefaultendpunct}{\mcitedefaultseppunct}\relax
\EndOfBibitem
\bibitem{Guo:2021mja}
T.~Guo, J.~Li, J.~Zhao, and L.~He,
  \ifthenelse{\boolean{articletitles}}{\emph{{Mass spectra and decays of
  open-heavy tetraquark states}},
  }{}\href{http://dx.doi.org/10.1103/PhysRevD.105.054018}{Phys.\ Rev.\ D
  \textbf{105} (2022), no.~5 054018},
  \href{http://arxiv.org/abs/2108.06222}{{\normalfont\ttfamily
  arXiv:2108.06222}}\relax
\mciteBstWouldAddEndPuncttrue
\mciteSetBstMidEndSepPunct{\mcitedefaultmidpunct}
{\mcitedefaultendpunct}{\mcitedefaultseppunct}\relax
\EndOfBibitem
\bibitem{Molina:2022jcd}
R.~Molina and E.~Oset,
  \ifthenelse{\boolean{articletitles}}{\emph{{$T_{cs}(2900)^{-}$ as a threshold
  effect from the interaction of the $\Dstar\Kstar$, $\Dstars\rho$ channels}},
  }{}\href{http://dx.doi.org/10.1103/PhysRevD.107.056015}{Phys.\ Rev.\ D
  \textbf{107} (2023), no.~5 056015},
  \href{http://arxiv.org/abs/2211.01302}{{\normalfont\ttfamily
  arXiv:2211.01302}}\relax
\mciteBstWouldAddEndPuncttrue
\mciteSetBstMidEndSepPunct{\mcitedefaultmidpunct}
{\mcitedefaultendpunct}{\mcitedefaultseppunct}\relax
\EndOfBibitem
\bibitem{Liu:2020orv}
X.-H. Liu {\em et~al.}, \ifthenelse{\boolean{articletitles}}{\emph{{Triangle
  singularity as the origin of $X_0(2900)$ and $X_1(2900)$ observed in $B^+\to
  D^+ D^- K^+$}},
  }{}\href{http://dx.doi.org/10.1140/epjc/s10052-020-08762-6}{Eur.\ Phys.\ J.\
  C \textbf{80} (2020), no.~12 1178},
  \href{http://arxiv.org/abs/2008.07190}{{\normalfont\ttfamily
  arXiv:2008.07190}}\relax
\mciteBstWouldAddEndPuncttrue
\mciteSetBstMidEndSepPunct{\mcitedefaultmidpunct}
{\mcitedefaultendpunct}{\mcitedefaultseppunct}\relax
\EndOfBibitem
\bibitem{LHCb:2022sfr}
LHCb, R.~Aaij {\em et~al.}, \ifthenelse{\boolean{articletitles}}{\emph{{First
  Observation of a Doubly Charged Tetraquark and Its Neutral Partner}},
  }{}\href{http://dx.doi.org/10.1103/PhysRevLett.131.041902}{Phys.\ Rev.\
  Lett.\  \textbf{131} (2023), no.~4 041902},
  \href{http://arxiv.org/abs/2212.02716}{{\normalfont\ttfamily
  arXiv:2212.02716}}\relax
\mciteBstWouldAddEndPuncttrue
\mciteSetBstMidEndSepPunct{\mcitedefaultmidpunct}
{\mcitedefaultendpunct}{\mcitedefaultseppunct}\relax
\EndOfBibitem
\bibitem{LHCb:2022lzp}
LHCb, R.~Aaij {\em et~al.},
  \ifthenelse{\boolean{articletitles}}{\emph{{Amplitude analysis of
  $\Bd\to\Dzb\Dsp\pim$ and $\Bp\to\Dm\Dsp\pip$ decays}},
  }{}\href{http://dx.doi.org/10.1103/PhysRevD.108.012017}{Phys.\ Rev.\ D
  \textbf{108} (2023), no.~1 012017},
  \href{http://arxiv.org/abs/2212.02717}{{\normalfont\ttfamily
  arXiv:2212.02717}}\relax
\mciteBstWouldAddEndPuncttrue
\mciteSetBstMidEndSepPunct{\mcitedefaultmidpunct}
{\mcitedefaultendpunct}{\mcitedefaultseppunct}\relax
\EndOfBibitem
\bibitem{Lian:2023cgs}
D.-K. Lian {\em et~al.}, \ifthenelse{\boolean{articletitles}}{\emph{{Strong
  decays of $T_{c\bar s0}(2900)^{++/0}$ as a fully open-flavor tetraquark
  state}}, }{}\href{http://arxiv.org/abs/2302.01167}{{\normalfont\ttfamily
  arXiv:2302.01167}}\relax
\mciteBstWouldAddEndPuncttrue
\mciteSetBstMidEndSepPunct{\mcitedefaultmidpunct}
{\mcitedefaultendpunct}{\mcitedefaultseppunct}\relax
\EndOfBibitem
\bibitem{Agaev:2022eyk}
S.~S. Agaev, K.~Azizi, and H.~Sundu,
  \ifthenelse{\boolean{articletitles}}{\emph{{Modeling the resonance
  $T_{cs0}^{a}(2900)^{++}$ as a hadronic molecule $\Dstarp\Kstar^{+}$}},
  }{}\href{http://dx.doi.org/10.1103/PhysRevD.107.094019}{Phys.\ Rev.\ D
  \textbf{107} (2023), no.~9 094019},
  \href{http://arxiv.org/abs/2212.12001}{{\normalfont\ttfamily
  arXiv:2212.12001}}\relax
\mciteBstWouldAddEndPuncttrue
\mciteSetBstMidEndSepPunct{\mcitedefaultmidpunct}
{\mcitedefaultendpunct}{\mcitedefaultseppunct}\relax
\EndOfBibitem
\bibitem{Dmitrasinovic:2023eei}
V.~Dmitra\v{s}inovi\'c, \ifthenelse{\boolean{articletitles}}{\emph{{Are LHCb
  exotics $T_{c\bar{s}0}(2900)^0$, $T_{c\bar{s}0}(2900)^{++}$ and
  $\overline{X}_0(2900)$ members of an $SU_F(3)$ 6-plet?}},
  }{}\href{http://arxiv.org/abs/2301.05471}{{\normalfont\ttfamily
  arXiv:2301.05471}}\relax
\mciteBstWouldAddEndPuncttrue
\mciteSetBstMidEndSepPunct{\mcitedefaultmidpunct}
{\mcitedefaultendpunct}{\mcitedefaultseppunct}\relax
\EndOfBibitem
\bibitem{LHCb:2020bwg}
LHCb, R.~Aaij {\em et~al.},
  \ifthenelse{\boolean{articletitles}}{\emph{{Observation of structure in the
  $J /\psi$ -pair mass spectrum}},
  }{}\href{http://dx.doi.org/10.1016/j.scib.2020.08.032}{Sci.\ Bull.\
  \textbf{65} (2020), no.~23 1983},
  \href{http://arxiv.org/abs/2006.16957}{{\normalfont\ttfamily
  arXiv:2006.16957}}\relax
\mciteBstWouldAddEndPuncttrue
\mciteSetBstMidEndSepPunct{\mcitedefaultmidpunct}
{\mcitedefaultendpunct}{\mcitedefaultseppunct}\relax
\EndOfBibitem
\bibitem{LHCb:2018uwm}
LHCb, R.~Aaij {\em et~al.}, \ifthenelse{\boolean{articletitles}}{\emph{{Search
  for beautiful tetraquarks in the $\Upsilon(1S)\mu^+\mu^-$ invariant-mass
  spectrum}}, }{}\href{http://dx.doi.org/10.1007/JHEP10(2018)086}{JHEP
  \textbf{10} (2018) 086},
  \href{http://arxiv.org/abs/1806.09707}{{\normalfont\ttfamily
  arXiv:1806.09707}}\relax
\mciteBstWouldAddEndPuncttrue
\mciteSetBstMidEndSepPunct{\mcitedefaultmidpunct}
{\mcitedefaultendpunct}{\mcitedefaultseppunct}\relax
\EndOfBibitem
\bibitem{CMS:2020qwa}
CMS, A.~M. Sirunyan {\em et~al.},
  \ifthenelse{\boolean{articletitles}}{\emph{{Measurement of the $\Upsilon$(1S)
  pair production cross section and search for resonances decaying to
  $\Upsilon$(1S)$\mu^+\mu^-$ in proton-proton collisions at $\sqrt{s} =$
  13\tev}}, }{}\href{http://dx.doi.org/10.1016/j.physletb.2020.135578}{Phys.\
  Lett.\ B \textbf{808} (2020) 135578},
  \href{http://arxiv.org/abs/2002.06393}{{\normalfont\ttfamily
  arXiv:2002.06393}}\relax
\mciteBstWouldAddEndPuncttrue
\mciteSetBstMidEndSepPunct{\mcitedefaultmidpunct}
{\mcitedefaultendpunct}{\mcitedefaultseppunct}\relax
\EndOfBibitem
\bibitem{Iwasaki:1976cn}
Y.~Iwasaki, \ifthenelse{\boolean{articletitles}}{\emph{{Is a State c anti-c c
  anti-c Found at 6.0-GeV?}},
  }{}\href{http://dx.doi.org/10.1103/PhysRevLett.36.1266}{Phys.\ Rev.\ Lett.\
  \textbf{36} (1976) 1266}\relax
\mciteBstWouldAddEndPuncttrue
\mciteSetBstMidEndSepPunct{\mcitedefaultmidpunct}
{\mcitedefaultendpunct}{\mcitedefaultseppunct}\relax
\EndOfBibitem
\bibitem{Chao:1980dv}
K.-T. Chao, \ifthenelse{\boolean{articletitles}}{\emph{{The (cc) - ($\bar{cc}$)
  (Diquark - Anti-Diquark) States in $e^+ e^-$ Annihilation}},
  }{}\href{http://dx.doi.org/10.1007/BF01431564}{Z.\ Phys.\ C \textbf{7} (1981)
  317}\relax
\mciteBstWouldAddEndPuncttrue
\mciteSetBstMidEndSepPunct{\mcitedefaultmidpunct}
{\mcitedefaultendpunct}{\mcitedefaultseppunct}\relax
\EndOfBibitem
\bibitem{Ader:1981db}
J.~P. Ader, J.~M. Richard, and P.~Taxil,
  \ifthenelse{\boolean{articletitles}}{\emph{Do narrow heavy multiquark states
  exist?}, }{}\href{http://dx.doi.org/10.1103/PhysRevD.25.2370}{Phys.\ Rev.\ D
  \textbf{25} (1982) 2370}\relax
\mciteBstWouldAddEndPuncttrue
\mciteSetBstMidEndSepPunct{\mcitedefaultmidpunct}
{\mcitedefaultendpunct}{\mcitedefaultseppunct}\relax
\EndOfBibitem
\bibitem{Li:1983ru}
B.-A. Li and K.-F. Liu, \ifthenelse{\boolean{articletitles}}{\emph{{$J/\psi$
  Pair Production in Hadronic Collisions}},
  }{}\href{http://dx.doi.org/10.1103/PhysRevD.29.426}{Phys.\ Rev.\ D
  \textbf{29} (1984) 426}\relax
\mciteBstWouldAddEndPuncttrue
\mciteSetBstMidEndSepPunct{\mcitedefaultmidpunct}
{\mcitedefaultendpunct}{\mcitedefaultseppunct}\relax
\EndOfBibitem
\bibitem{Berezhnoy:2011xn}
A.~V. Berezhnoy, A.~V. Luchinsky, and A.~A. Novoselov,
  \ifthenelse{\boolean{articletitles}}{\emph{{Tetraquarks Composed of 4 Heavy
  Quarks}}, }{}\href{http://dx.doi.org/10.1103/PhysRevD.86.034004}{Phys.\ Rev.\
  D \textbf{86} (2012) 034004},
  \href{http://arxiv.org/abs/1111.1867}{{\normalfont\ttfamily
  arXiv:1111.1867}}\relax
\mciteBstWouldAddEndPuncttrue
\mciteSetBstMidEndSepPunct{\mcitedefaultmidpunct}
{\mcitedefaultendpunct}{\mcitedefaultseppunct}\relax
\EndOfBibitem
\bibitem{Wu:2016vtq}
J.~Wu {\em et~al.}, \ifthenelse{\boolean{articletitles}}{\emph{{Heavy-flavored
  tetraquark states with the $QQ\bar{Q}\bar{Q}$ configuration}},
  }{}\href{http://dx.doi.org/10.1103/PhysRevD.97.094015}{Phys.\ Rev.\ D
  \textbf{97} (2018), no.~9 094015},
  \href{http://arxiv.org/abs/1605.01134}{{\normalfont\ttfamily
  arXiv:1605.01134}}\relax
\mciteBstWouldAddEndPuncttrue
\mciteSetBstMidEndSepPunct{\mcitedefaultmidpunct}
{\mcitedefaultendpunct}{\mcitedefaultseppunct}\relax
\EndOfBibitem
\bibitem{Karliner:2016zzc}
M.~Karliner, S.~Nussinov, and J.~L. Rosner,
  \ifthenelse{\boolean{articletitles}}{\emph{{$Q Q \bar Q \bar Q$ states:
  masses, production, and decays}},
  }{}\href{http://dx.doi.org/10.1103/PhysRevD.95.034011}{Phys.\ Rev.\
  \textbf{D95} (2017) 034011},
  \href{http://arxiv.org/abs/1611.00348}{{\normalfont\ttfamily
  arXiv:1611.00348}}\relax
\mciteBstWouldAddEndPuncttrue
\mciteSetBstMidEndSepPunct{\mcitedefaultmidpunct}
{\mcitedefaultendpunct}{\mcitedefaultseppunct}\relax
\EndOfBibitem
\bibitem{Barnea:2006sd}
N.~Barnea, J.~Vijande, and A.~Valcarce,
  \ifthenelse{\boolean{articletitles}}{\emph{{Four-quark spectroscopy within
  the hyperspherical formalism}},
  }{}\href{http://dx.doi.org/10.1103/PhysRevD.73.054004}{Phys.\ Rev.\ D
  \textbf{73} (2006) 054004},
  \href{http://arxiv.org/abs/hep-ph/0604010}{{\normalfont\ttfamily
  arXiv:hep-ph/0604010}}\relax
\mciteBstWouldAddEndPuncttrue
\mciteSetBstMidEndSepPunct{\mcitedefaultmidpunct}
{\mcitedefaultendpunct}{\mcitedefaultseppunct}\relax
\EndOfBibitem
\bibitem{Debastiani:2017msn}
V.~R. Debastiani and F.~S. Navarra,
  \ifthenelse{\boolean{articletitles}}{\emph{{A non-relativistic model for the
  $[cc][\bar{c}\bar{c}]$ tetraquark}},
  }{}\href{http://arxiv.org/abs/1706.07553}{{\normalfont\ttfamily
  arXiv:1706.07553}}\relax
\mciteBstWouldAddEndPuncttrue
\mciteSetBstMidEndSepPunct{\mcitedefaultmidpunct}
{\mcitedefaultendpunct}{\mcitedefaultseppunct}\relax
\EndOfBibitem
\bibitem{Liu:2019zuc}
M.-S. Liu, Q.-F. L\"u, X.-H. Zhong, and Q.~Zhao,
  \ifthenelse{\boolean{articletitles}}{\emph{{All-heavy tetraquarks}},
  }{}\href{http://dx.doi.org/10.1103/PhysRevD.100.016006}{Phys.\ Rev.\ D
  \textbf{100} (2019), no.~1 016006},
  \href{http://arxiv.org/abs/1901.02564}{{\normalfont\ttfamily
  arXiv:1901.02564}}\relax
\mciteBstWouldAddEndPuncttrue
\mciteSetBstMidEndSepPunct{\mcitedefaultmidpunct}
{\mcitedefaultendpunct}{\mcitedefaultseppunct}\relax
\EndOfBibitem
\bibitem{Chen:2016jxd}
W.~Chen {\em et~al.}, \ifthenelse{\boolean{articletitles}}{\emph{{Hunting for
  exotic doubly hidden-charm/bottom tetraquark states}},
  }{}\href{http://dx.doi.org/10.1016/j.physletb.2017.08.034}{Phys.\ Lett.\ B
  \textbf{773} (2017) 247},
  \href{http://arxiv.org/abs/1605.01647}{{\normalfont\ttfamily
  arXiv:1605.01647}}\relax
\mciteBstWouldAddEndPuncttrue
\mciteSetBstMidEndSepPunct{\mcitedefaultmidpunct}
{\mcitedefaultendpunct}{\mcitedefaultseppunct}\relax
\EndOfBibitem
\bibitem{Wang:2019rdo}
G.-J. Wang, L.~Meng, and S.-L. Zhu,
  \ifthenelse{\boolean{articletitles}}{\emph{{Spectrum of the fully-heavy
  tetraquark state $QQ\bar Q' \bar Q'$}},
  }{}\href{http://dx.doi.org/10.1103/PhysRevD.100.096013}{Phys.\ Rev.\ D
  \textbf{100} (2019), no.~9 096013},
  \href{http://arxiv.org/abs/1907.05177}{{\normalfont\ttfamily
  arXiv:1907.05177}}\relax
\mciteBstWouldAddEndPuncttrue
\mciteSetBstMidEndSepPunct{\mcitedefaultmidpunct}
{\mcitedefaultendpunct}{\mcitedefaultseppunct}\relax
\EndOfBibitem
\bibitem{Bedolla:2019zwg}
M.~A. Bedolla, J.~Ferretti, C.~D. Roberts, and E.~Santopinto,
  \ifthenelse{\boolean{articletitles}}{\emph{{Spectrum of fully-heavy
  tetraquarks from a diquark+antidiquark perspective}},
  }{}\href{http://dx.doi.org/10.1140/epjc/s10052-020-08579-3}{Eur.\ Phys.\ J.\
  C \textbf{80} (2020), no.~11 1004},
  \href{http://arxiv.org/abs/1911.00960}{{\normalfont\ttfamily
  arXiv:1911.00960}}\relax
\mciteBstWouldAddEndPuncttrue
\mciteSetBstMidEndSepPunct{\mcitedefaultmidpunct}
{\mcitedefaultendpunct}{\mcitedefaultseppunct}\relax
\EndOfBibitem
\bibitem{Lloyd:2003yc}
R.~J. Lloyd and J.~P. Vary, \ifthenelse{\boolean{articletitles}}{\emph{{All
  charm tetraquarks}},
  }{}\href{http://dx.doi.org/10.1103/PhysRevD.70.014009}{Phys.\ Rev.\ D
  \textbf{70} (2004) 014009},
  \href{http://arxiv.org/abs/hep-ph/0311179}{{\normalfont\ttfamily
  arXiv:hep-ph/0311179}}\relax
\mciteBstWouldAddEndPuncttrue
\mciteSetBstMidEndSepPunct{\mcitedefaultmidpunct}
{\mcitedefaultendpunct}{\mcitedefaultseppunct}\relax
\EndOfBibitem
\bibitem{Chen:2020lgj}
X.~Chen, \ifthenelse{\boolean{articletitles}}{\emph{{Fully-charm tetraquarks:
  $cc\bar{c}\bar{c}$}},
  }{}\href{http://arxiv.org/abs/2001.06755}{{\normalfont\ttfamily
  arXiv:2001.06755}}\relax
\mciteBstWouldAddEndPuncttrue
\mciteSetBstMidEndSepPunct{\mcitedefaultmidpunct}
{\mcitedefaultendpunct}{\mcitedefaultseppunct}\relax
\EndOfBibitem
\bibitem{Wang:2018poa}
Z.-G. Wang and Z.-Y. Di, \ifthenelse{\boolean{articletitles}}{\emph{{Analysis
  of the vector and axialvector $QQ\bar{Q}\bar{Q}$ tetraquark states with QCD
  sum rules}}, }{}\href{http://dx.doi.org/10.5506/APhysPolB.50.1335}{Acta
  Phys.\ Polon.\ B \textbf{50} (2019) 1335},
  \href{http://arxiv.org/abs/1807.08520}{{\normalfont\ttfamily
  arXiv:1807.08520}}\relax
\mciteBstWouldAddEndPuncttrue
\mciteSetBstMidEndSepPunct{\mcitedefaultmidpunct}
{\mcitedefaultendpunct}{\mcitedefaultseppunct}\relax
\EndOfBibitem
\bibitem{Anwar:2017toa}
M.~N. Anwar {\em et~al.},
  \ifthenelse{\boolean{articletitles}}{\emph{{Spectroscopy and decays of the
  fully-heavy tetraquarks}},
  }{}\href{http://arxiv.org/abs/1710.02540}{{\normalfont\ttfamily
  arXiv:1710.02540}}\relax
\mciteBstWouldAddEndPuncttrue
\mciteSetBstMidEndSepPunct{\mcitedefaultmidpunct}
{\mcitedefaultendpunct}{\mcitedefaultseppunct}\relax
\EndOfBibitem
\bibitem{ATLAS:2023bft}
ATLAS, G.~Aad {\em et~al.},
  \ifthenelse{\boolean{articletitles}}{\emph{{Observation of an Excess of
  Dicharmonium Events in the Four-Muon Final State with the ATLAS Detector}},
  }{}\href{http://dx.doi.org/10.1103/PhysRevLett.131.151902}{Phys.\ Rev.\
  Lett.\  \textbf{131} (2023), no.~15 151902},
  \href{http://arxiv.org/abs/2304.08962}{{\normalfont\ttfamily
  arXiv:2304.08962}}\relax
\mciteBstWouldAddEndPuncttrue
\mciteSetBstMidEndSepPunct{\mcitedefaultmidpunct}
{\mcitedefaultendpunct}{\mcitedefaultseppunct}\relax
\EndOfBibitem
\bibitem{CMS:2023owd}
CMS, A.~Hayrapetyan {\em et~al.},
  \ifthenelse{\boolean{articletitles}}{\emph{{Observation of new structure in
  the J/$\psi$J/$\psi$ mass spectrum in proton-proton collisions at $\sqrt{s}$
  = 13\tev}}, }{}\href{http://arxiv.org/abs/2306.07164}{{\normalfont\ttfamily
  arXiv:2306.07164}}\relax
\mciteBstWouldAddEndPuncttrue
\mciteSetBstMidEndSepPunct{\mcitedefaultmidpunct}
{\mcitedefaultendpunct}{\mcitedefaultseppunct}\relax
\EndOfBibitem
\bibitem{Dong:2020nwy}
X.-K. Dong {\em et~al.},
  \ifthenelse{\boolean{articletitles}}{\emph{{Coupled-Channel Interpretation of
  the LHCb Double-~$J/\psi$~Spectrum and Hints of a New State Near the~ $J/\psi
  J/\psi$~~Threshold}},
  }{}\href{http://dx.doi.org/10.1103/PhysRevLett.127.119901}{Phys.\ Rev.\
  Lett.\  \textbf{126} (2021), no.~13 132001},
  \href{http://arxiv.org/abs/2009.07795}{{\normalfont\ttfamily
  arXiv:2009.07795}}, [Erratum: Phys.Rev.Lett. 127, 119901 (2021)]\relax
\mciteBstWouldAddEndPuncttrue
\mciteSetBstMidEndSepPunct{\mcitedefaultmidpunct}
{\mcitedefaultendpunct}{\mcitedefaultseppunct}\relax
\EndOfBibitem
\bibitem{LHCb:2017iph}
LHCb, R.~Aaij {\em et~al.},
  \ifthenelse{\boolean{articletitles}}{\emph{{Observation of the doubly charmed
  baryon $\Xi_{cc}^{++}$}},
  }{}\href{http://dx.doi.org/10.1103/PhysRevLett.119.112001}{Phys.\ Rev.\
  Lett.\  \textbf{119} (2017), no.~11 112001},
  \href{http://arxiv.org/abs/1707.01621}{{\normalfont\ttfamily
  arXiv:1707.01621}}\relax
\mciteBstWouldAddEndPuncttrue
\mciteSetBstMidEndSepPunct{\mcitedefaultmidpunct}
{\mcitedefaultendpunct}{\mcitedefaultseppunct}\relax
\EndOfBibitem
\bibitem{LHCb:2019epo}
LHCb, R.~Aaij {\em et~al.},
  \ifthenelse{\boolean{articletitles}}{\emph{{Precision measurement of the
  $\Xi_{cc}^{++}$ mass}},
  }{}\href{http://dx.doi.org/10.1007/JHEP02(2020)049}{JHEP \textbf{02} (2020)
  049}, \href{http://arxiv.org/abs/1911.08594}{{\normalfont\ttfamily
  arXiv:1911.08594}}\relax
\mciteBstWouldAddEndPuncttrue
\mciteSetBstMidEndSepPunct{\mcitedefaultmidpunct}
{\mcitedefaultendpunct}{\mcitedefaultseppunct}\relax
\EndOfBibitem
\bibitem{BALLOT1983449}
J.~L. Ballot and J.~M. Richard, \ifthenelse{\boolean{articletitles}}{\emph{Four
  quark states in additive potentials},
  }{}\href{http://dx.doi.org/https://doi.org/10.1016/0370-2693(83)90991-7}{Physics
  Letters B \textbf{123} (1983), no.~6 449}\relax
\mciteBstWouldAddEndPuncttrue
\mciteSetBstMidEndSepPunct{\mcitedefaultmidpunct}
{\mcitedefaultendpunct}{\mcitedefaultseppunct}\relax
\EndOfBibitem
\bibitem{Karliner:2017qjm}
M.~Karliner and J.~L. Rosner,
  \ifthenelse{\boolean{articletitles}}{\emph{{Discovery of doubly-charmed
  $\Xi_{cc}$ baryon implies a stable ($b b \bar u \bar d$) tetraquark}},
  }{}\href{http://dx.doi.org/10.1103/PhysRevLett.119.202001}{Phys.\ Rev.\
  Lett.\  \textbf{119} (2017) 202001},
  \href{http://arxiv.org/abs/1707.07666}{{\normalfont\ttfamily
  arXiv:1707.07666}}\relax
\mciteBstWouldAddEndPuncttrue
\mciteSetBstMidEndSepPunct{\mcitedefaultmidpunct}
{\mcitedefaultendpunct}{\mcitedefaultseppunct}\relax
\EndOfBibitem
\bibitem{LHCb:2021vvq}
LHCb, R.~Aaij {\em et~al.},
  \ifthenelse{\boolean{articletitles}}{\emph{{Observation of an exotic narrow
  doubly charmed tetraquark}},
  }{}\href{http://dx.doi.org/10.1038/s41567-022-01614-y}{Nature Phys.\
  \textbf{18} (2022), no.~7 751},
  \href{http://arxiv.org/abs/2109.01038}{{\normalfont\ttfamily
  arXiv:2109.01038}}\relax
\mciteBstWouldAddEndPuncttrue
\mciteSetBstMidEndSepPunct{\mcitedefaultmidpunct}
{\mcitedefaultendpunct}{\mcitedefaultseppunct}\relax
\EndOfBibitem
\bibitem{LHCb:2021auc}
LHCb, R.~Aaij {\em et~al.}, \ifthenelse{\boolean{articletitles}}{\emph{{Study
  of the doubly charmed tetraquark $T_{cc}^{+}$}},
  }{}\href{http://dx.doi.org/10.1038/s41467-022-30206-w}{Nature Commun.\
  \textbf{13} (2022), no.~1 3351},
  \href{http://arxiv.org/abs/2109.01056}{{\normalfont\ttfamily
  arXiv:2109.01056}}\relax
\mciteBstWouldAddEndPuncttrue
\mciteSetBstMidEndSepPunct{\mcitedefaultmidpunct}
{\mcitedefaultendpunct}{\mcitedefaultseppunct}\relax
\EndOfBibitem
\bibitem{Janc:2004qn}
D.~Janc and M.~Rosina, \ifthenelse{\boolean{articletitles}}{\emph{{The $T_{cc}
  = DD^*$ molecular state}},
  }{}\href{http://dx.doi.org/10.1007/s00601-004-0068-9}{Few Body Syst.\
  \textbf{35} (2004) 175},
  \href{http://arxiv.org/abs/hep-ph/0405208}{{\normalfont\ttfamily
  arXiv:hep-ph/0405208}}\relax
\mciteBstWouldAddEndPuncttrue
\mciteSetBstMidEndSepPunct{\mcitedefaultmidpunct}
{\mcitedefaultendpunct}{\mcitedefaultseppunct}\relax
\EndOfBibitem
\bibitem{LHCb:2023hlw}
LHCb, R.~Aaij {\em et~al.}, \ifthenelse{\boolean{articletitles}}{\emph{{The
  LHCb upgrade I}},
  }{}\href{http://arxiv.org/abs/2305.10515}{{\normalfont\ttfamily
  arXiv:2305.10515}}\relax
\mciteBstWouldAddEndPuncttrue
\mciteSetBstMidEndSepPunct{\mcitedefaultmidpunct}
{\mcitedefaultendpunct}{\mcitedefaultseppunct}\relax
\EndOfBibitem
\bibitem{SMOG}
C.~Barschel, {\em {Precision luminosity measurement at LHCb with beam-gas
  imaging}}, PhD thesis, RWTH Aachen U., 2014,
  \href{https://cds.cern.ch/record/1693671}{CERN-THESIS-2013-301}\relax
\mciteBstWouldAddEndPuncttrue
\mciteSetBstMidEndSepPunct{\mcitedefaultmidpunct}
{\mcitedefaultendpunct}{\mcitedefaultseppunct}\relax
\EndOfBibitem
\bibitem{LHCbCollaboration:2776420}
LHCb, \ifthenelse{\boolean{articletitles}}{\emph{{Framework TDR for the LHCb
  Upgrade II: Opportunities in flavour physics, and beyond, in the HL-LHC
  era}}, }{} tech. rep., CERN, Geneva, 2021.
\newblock \href{https://cds.cern.ch/record/2776420}{CERN-LHCC-2021-012,
  LHCB-TDR-023}\relax
\mciteBstWouldAddEndPuncttrue
\mciteSetBstMidEndSepPunct{\mcitedefaultmidpunct}
{\mcitedefaultendpunct}{\mcitedefaultseppunct}\relax
\EndOfBibitem
\bibitem{Lebed:2022vfu}
R.~F. Lebed {\em et~al.}, \ifthenelse{\boolean{articletitles}}{\emph{{Summary
  of Topical Group on Hadron Spectroscopy (RF07) Rare Processes and Precision
  Frontier of Snowmass 2021}}, }{} in {\em {Snowmass 2021}} (R.~F. Lebed and
  T.~Skwarnicki, eds.), 7, 2022.
\newblock \href{http://arxiv.org/abs/2207.14594}{{\normalfont\ttfamily
  arXiv:2207.14594}}\relax
\mciteBstWouldAddEndPuncttrue
\mciteSetBstMidEndSepPunct{\mcitedefaultmidpunct}
{\mcitedefaultendpunct}{\mcitedefaultseppunct}\relax
\EndOfBibitem
\end{mcitethebibliography}
\end{document}